\documentclass[11pt]{article}

\RequirePackage{fix-cm}
\usepackage{amsmath, amssymb}
\usepackage{verbatim}
\usepackage[dvipdfm,a4paper,left=30mm,right=30mm,top=35mm,bottom=35mm]{geometry}
\usepackage{color}
\usepackage{bm}
\usepackage{graphicx}
\usepackage{framed}
\usepackage[inline]{showlabels}
\usepackage{hyperref}
\usepackage{tcolorbox}	
\usepackage{subcaption}
\usepackage{mathtools}

\usepackage{accents}
\usepackage[]{algorithm2e}
\usepackage{algpseudocode}
\usepackage[page]{appendix}

\newtheorem{proof}{Proof}

\def\phi{\varphi}
\def\C{\hbox{\rlap{\kern.24em\raise.1ex\hbox
      {\vrule height1.3ex width.9pt}}C}}
\def\R{\mathbb{R}}
\def\S{{\rm I\kern-.2em S}}
\def\E{{\rm I\kern-.2em E}} 
\def\P{\hbox{\rlap{I}\kern.16em P}}

\def\N{\hbox{\rlap{I}\kern.16em\rlap{I}N}}
\def\Z{\hbox{\rlap{Z}\kern.20em Z}}
\def\({\begin{eqnarray}}
\def\){\end{eqnarray}}
\def\[{\begin{eqnarray*}}
\def\]{\end{eqnarray*}}
\def\part#1#2{{\partial #1\over\partial #2}}

\def\dx{\nabla_x}

\def\d{\, \text{d}}

\def\pmb#1{\setbox0=\hbox{$#1$}
  \kern-.025em\copy0\kern-\wd0
  \kern-.05em\copy0\kern-\wd0
  \kern-.025em\raise.0433em\box0 }
\def\bar{\overline}

\DeclarePairedDelimiter\floor{\lfloor}{\rfloor}

\begin{document}

\title{A new model for the emergence of blood capillary networks}

\author{
P. Aceves-S\'anchez%
\thanks{Department of Mathematics, Imperial College London, London SW7 2AZ, UK. {\tt p.aceves-sanchez@imperial.ac.uk}},
B.~Aymard%
\thanks{MathNeuro Team, Inria Sophia Antipolis M\'editerran\'ee, 2004 Route
des Lucioles, BP93, 06902 Valbonne cedex, France. 
{\tt benjamin.aymard@inria.fr}
}, 
D. Peurichard%
\thanks{INRIA Paris, 2, rue Simone Iff, 75589 Paris Cedex 12, France. {\tt diane.a.peurichard@inria.fr}
}, 
P. Kennel%
\thanks{Universit\'e de Toulouse-INPT-UPS, Institut de M\'ecanique des Fluides, 31000 Toulouse, France. {\tt pol.kennel@gmail.com} 
},
A. Lorsignol%
\thanks{STROMALab, Universit\'e de Toulouse, Inserm U1031, EFS, INP-ENVT, UPS, CNRS ERL5311, Toulouse, France. {\tt anne.lorsignol@inserm.fr}
}, \\
F. Plourabou\'e%
\thanks{Universit\'e de Toulouse-INPT-UPS, Institut de M\'ecanique des Fluides, 31000 Toulouse, France. {\tt fplourab@imft.fr}
},
L. Casteilla%
\thanks{STROMALab, Universit\'e de Toulouse, Inserm U1031, EFS, INP-ENVT, UPS, CNRS ERL5311, Toulouse, France. {\tt louis.casteilla@inserm.fr}
},
\, \& P. Degond%
\thanks{Department of Mathematics, Imperial College London, London SW7 2AZ, UK. {\tt p.degond@imperial.ac.uk} 
}
}

\date{}

\maketitle

\noindent{\bf Abstract.} 
We propose a new model for the emergence of blood capillary networks. We assimilate the tissue and extra cellular matrix as a porous medium, using Darcy's law for describing both blood and intersticial fluid flows. Oxygen obeys a convection-diffusion-reaction equation describing advection by the blood, diffusion and consumption by the tissue. Discrete agents named capillary elements and modelling groups of endothelial cells are created or deleted according to different rules involving the oxygen concentration gradient, the blood velocity, the sheer stress or the capillary element density. Once created, a capillary element locally enhances the hydraulic conductivity matrix,  contributing to a local increase of the blood velocity and oxygen flow. No connectivity between the capillary elements is imposed. The coupling between blood, oxygen flow and capillary elements provides a positive feedback mechanism which triggers the emergence of a network of channels of high hydraulic conductivity which we identify as new blood capillaries. We provide two different, biologically relevant geometrical settings and numerically analyze the influence of each of the capillary creation mechanism in detail. All mechanisms seem to concur towards a harmonious network but the most important ones are those involving oxygen gradient and sheer stress. A detailed discussion of this model with respect to the literature and its potential future developments concludes the paper. 

\vskip 0.2cm

\medskip


\noindent{\bf Acknowledgment:} This project was supported by the R\'egion Midi-Pyr\'en\'ees, France, under contract ref. APRTCN 14050455. PD acknowledges support by the Engineering and Physical Sciences Research Council (EPSRC) under grants no. EP/M006883/1 and EP/N014529/1, by the Royal Society and the Wolfson Foundation through a Royal Society Wolfson Research Merit Award no. WM130048 and by the National Science Foundation (NSF) under grant no. RNMS11-07444 (KI-Net). PD is on leave  from CNRS, Institut de Math\'ematiques de Toulouse, France. BA gratefully acknowledges the hospitality of the Department of Mathematics, Imperial College London, where part of this research was conducted. PAS was supported by EPSRC under grant no. EP/M006883/1.

\medskip

\noindent{\bf Data statement:} no new data were collected in the course of this research. 
\medskip

\noindent{\bf Link to video:} \href{https://doi.org/10.6084/m9.figshare.c.4287575.v1}{https://doi.org/10.6084/m9.figshare.c.4287575.v1}
\medskip



\section{Introduction}

Networks appear ubiquitously in nature and in man-made structures. 
Examples of networks in living systems are the circulatory, respiratory and nervous systems of vertebrates, the branches, roots and leaves of trees, coral formations, bacterial colonies. In nature, river networks, erosion patterns, lightnings of thunder are other example of networks. The literature on network modelling is enormous and growing at an impressive speed. Yet, most works assume a given topological structure, possibly random or dynamic, but which is, for a given realization or a given time, well-defined. However, in many cases, networks are fuzzy objects which, according to the observed scale, adopt a different structure. This is particularly the case of emergent networks, where a network pattern emerges from a continuum. One can think for instance of erosion gradually sculpting an originally flat landscape into gullies, channels and ultimately valleys. Many networks in nature emerge in this way. Current network modelling methodologies are unadapted to capture this emergence phase because they need an already established network structure from the start. In this paper, we propose a novel methodology to model emergent networks and apply it to the emergence of blood capillary networks, the so-called vasculogenesis or angiogenesis phenomena. However, the approach is fairly versatile and could easily be adapted - modulo appropriate changes - to other emergent networks. In particular, we refer to earlier works developing similar ideas for ant trail formation \cite{BDM2013} and tissue self-organization \cite{P+2017}. 

Vasculogenesis (the de novo formation of new blood vessels) or angiogenesis (the development of new blood vessels from existing ones, see e.g. \cite{R1997}) play a fundamental role in living systems by influencing growth, regeneration and reparation through an increase of blood flow and tissue oxygenation and consequently by boosting the transport of nutrients as well as the disposal of waste. Angiogenesis has been the subject of intense study over the last decades for its crucial role in the development of cancerous tumours. Once a tumour has reached certain level of maturation, the surrounding tissue starts to suffer from hypoxia, or lack of oxygen, which induces the accumulation of hypoxia-inducible factors. As a consequence, hypoxic cells start to secrete tumour angiogenic growth factors (TAFs) such as the vascular endothelial growth factor (VEGF) \cite{K+2005, X+2014}. The TAFs diffuse through the tissue until reaching the nearby blood vessels, triggering the formation of a new vascular network to supply the tumour with nutrients and oxygen. Angiogenesis is not unique to tumour growth but is also a key part of wound healing, the menstrual cycle, embryonic development and of several diseases such as rheumatoid arthritis \cite{CJ2000, F+2005, F1995}. New blood vessels grow through the Extracellular Matrix (ECM), which is a collection of collagen fibers, interstitial cells, proteoglycans and matrix binding molecules among many other components. The ECM provides support for cells which gather together to form organs. It also provides mechanical support for cells and offers an environment for the transmission of chemical cues.

There are many different approaches to model vasculogenesis and angiogenesis in the literature.  Many of them are two-dimensional cell-based models, describing the migration of endothetial cells to form new blood vessels and including various aspects of the tissue environment \cite{BJJ2009, DM2013, MC+2006, OAMB2009, T+2014}). Evolving network models have been developed in \cite{S+2012, S+2013}. Other models use a continuum approach for the endothelial cell (EC) density \cite{balding1985mathematical, byrne1995mathematical, T+2011} and the link between cell-based and continuum approaches is investigated e.g. in \cite{pillay2017modeling}. In most of these models, blood flow plays no or little role in determining the geometry of the network. For instance, the efficiency of blood perfusion in a predetermined network is studied in \cite{GP2000}. For a review of recent models, we refer to \cite{SBP2013}. One important characteristics of our model is that blood and oxygen flows have the leading role in shaping the network structure. Another feature is that the network structure is not imposed a priori but emerges as an outcome of the evolution rules ascribed to the agents. 

In the present work, we describe the emergence of vascular networks by considering agents (EC or group of such cells referred to as capillary elements) obeying elementary heuristic rules. Indeed, the biology, biophysics and biochemistry involved are incredibly complex and still incompletely known \cite{PH2009,SF2007}. However, by their combined influence cells are likely to respond in a determinable (if not deterministic) way to environmental cues, such as blood flow, oxygen concentration, etc. In other words, cells behave like social agents with a behavior determined by their own state and their environment. This is the viewpoint developed here. Moreover, cell social behavior is not fully understood yet and the model aims to provide a platform to test the validity of such behavioral hypotheses. Once elucidated, cell behavior can be traced back to biochemistry through experiments whose design is facilitated by the intuition gained by the model. 

In this model, unlike most models of the literature we put the flow of blood, interstitial fluid and oxygen at the heart of the capillary creation process. We assimilate the tissue and ECM as a porous medium in which blood and interstitial fluid (we make no distinction between those) flow under the influence of a pressure gradient (powered by the pressure difference between arteries and veins). We assume blood velocity follows a Darcy law, i.e. the fluid is incompressible and its velocity is proportional to the pressure gradient through a proportionality matrix called the hydraulic conductivity tensor. Oxygen is supplied by e.g. an oxygenated tissue, transported by the blood velocity and diffuses through and is consumed by the tissue. If there are no capillaries, due to low hydraulic conductivity of avascular tissue flow is very slow and oxygen is mostly transported by diffusion. Thus, oxygen concentration falls quickly off even at short distance from the supply due to quick consumption. This is where capillary creation comes at play. So far, the treatment of blood flow and oxygen transport are through a continuum model. We now introduce an agent-based description of the formation of the capillary network, making our model a hybrid model coupling continuum and agent-based descriptions of the system.  

We assume that capillary elements (representing individual EC or group of such cells) can be created in response to a gradient of oxygen concentration. The creation of this capillary element could either correspond to a de novo blood vessel creation or to the recruitment of EC previously at a different location. We do not make any difference between the two processes as, for the sake of simplicity, we do not follow the motion of these EC prior to their recruitment. For the same reason, we assume that once created, a capillary element stays in place without moving, until it is eventually destroyed. A capillary element is modelled by a rod-shape particle with finite area (we assume a 2-dimensional model). Once created, it locally enhances the hydraulic conductivity in the direction of the rod, contributing to locally increase the blood velocity and the flow of oxygen in this direction. This increase is limited to the area of the particle. The process is time-dynamic: capillary elements are created following a spatio-temporal Poisson process the intensity of which is large when and where the gradient of oxygen concentration is large. Each time a capillary element is created, we assume an instantaneous adaptation of the blood velocity through solving Darcy's equation with the new value of the hydraulic conductivity. This new velocity is fed into the convection-diffusion-reaction equation satisfied by oxygen concentration and the process goes on. It leads to the appearance of more capillary elements. 

Importantly, we do not assume any connectivity between the capillary elements. They could appear anywhere following the Poisson process and do not need to intersect. The influence of each of them is summed up at each update of the hydraulic conductivity tensor. In spite of the absence of connectivity constraint between capillary elements, simulations show the creation of streets of capillary elements forming what we could assimilate as a new blood capillary. Connectivity emerges as an outcome of the model but does not need to be prescribed. This is an important advantage of the model as keeping track of connectivity requires the use of adequate data structures that considerably increase the complexity of code development and its computational cost. 

In our model, the creation of capillary elements in response to a gradient of oxygen concentration is a proxy for the influence of growth factors such as VEGF, which we do not include in the model. Indeed, both gradients of VEGF and of oxygen concentration carry information about the direction to be followed by new capillaries. The VEGF signaling is likely to amplify this information for improved sensing by the cells or to be a chemical intermediate that triggers a stronger reaction than mere oxygen gradient. However, both encode similar information and ignoring VEGF allows us to reduce the model complexity without distorting the phenomenology too much. We do assume several other capillary creation mechanisms. One is by reinforcement of existing branches. Another one is through wall shear stress which has proven to play a key role in the creation of new blood vessels (see for instance \cite{G+2014, K+2008}). We also consider pruning mechanisms when the capillary element density is too high.

The outline of the article if the following. In Section \ref{sect:model}, we provide a complete description of the model although the details of the numerical methods are deferred to the appendices. Results of the model are given in Section \ref{sect:results}. A discussion is then developed in Section \ref{sect:discussion}.


\section{Model}
\label{sect:model}

\subsection{Presentation}

We consider a two-dimensional model. The tissue occupies a spatial domain $\Omega \subset {\mathbb R}^2$. We will specifically consider two different geometries described in Section \ref{subsec:geom}. 
As outlined above, the model considers four different entities: 
\begin{itemize}
\item The flow of blood or interstitial fluid described by its pressure $p(x,t)$ and its velocity $u(x,t)$ where $x \in \Omega$ and $t \geq 0$ and where $p \in {\mathbb R}$ and $u \in {\mathbb R}^2$. The quantities $u$ and $p$ are continuum variables which satisfy Partial Differential Equations (PDE), namely Darcy's equations, described in Section \ref{subsec:blood_flow}, with boundary conditions given, according to the geometrical case studied, in Section \ref{subsec:geom}.
\item The flow of oxygen, described by the oxygen concentration $\rho(x,t) \in {\mathbb R}$. 
The quantity $\rho$ is also a continuum variable satisfying a Partial Differential Equation (PDE), namely a convection-diffusion-reaction equation described in Section \ref{subsec:oxygen} with initial and boundary conditions given in Section \ref{subsec:geom}.
\item Capillary elements in turn are discrete quantities (particles) carrying a direction and their dynamics is given by an agent-based model described in Section \ref{subsec:capillaries}. 
\item The tissue and the ECM mediate the interaction between the capillaries on the one hand and the blood and oxygen flows on the other hand. The tissue characteristics determine the coefficients of the PDE for the blood and the oxygen and those are dependent on the position and direction of the capillaries. The computation of these coefficients is detailed in Section \ref{subsec:tissue}. 
\end{itemize}
We now describe the various elements of the model in detail. The numerical values of all the parameters are summarized in Table \ref{TableOfParameters}.


\subsection{Blood and intersticial fluid flow}
\label{subsec:blood_flow} 

We lump blood and interstitial fluid in one and single fluid which we will call 'blood' for simplicity. We assume that blood is an incompressible Newtonian fluid described by its pressure $p(x,t)$ and its velocity $u(x,t)$. The tissue viewed as a mixture of cells, ECM and blood vessels is assimilated to a single porous medium with hydraulic conductivity tensor $\textbf{K}(x,t)$, which is a symmetric uniformly positive-definite matrix. Because of the presence of capillaries and the possible creation/removal of capillaries, $\textbf{K}$ is dependent on both position and time. In a porous medium velocity and pressure are linked by Darcy's law 
\begin{align}
 \textbf{u} (x,t) = -\textbf{K} (x,t)\, \nabla_x p(x,t), \quad x \in \Omega, \quad t \geq0,    \label{eq:bloodPDE1}
\end{align} 
where $\nabla_x p$ denotes the gradient of $p$. This equation asserts that in spite of the time variations of the conductivity matrix $\textbf{K}$, pressure and velocity instantaneously adjust one with each other according to (\ref{eq:bloodPDE1}). This is consistent with the assumptions leading to Darcy's law, i.e. viscous terms dominate inertial ones, resulting in a quasi-steady flow. Darcy's law is complemented with the incompressibility condition:
\begin{align}
\nabla_x \cdot \textbf{u}  = 0 \, \label{eq:bloodPDE2} , 
\end{align} 
where $\nabla_x \cdot \textbf{u}$ stands for the divergence of the vector field $u$. Substituting \eqref{eq:bloodPDE1} into \eqref{eq:bloodPDE2} we obtain $p$ as a solution of the following elliptic problem:
\begin{equation}\label{eq:elliptic} 
- \nabla_x \cdot (\textbf{K} \, \nabla_x p) = 0 \qquad \text{in } \Omega \times [0,T],  
\end{equation}
where $T$ is the total simulation time. The expression of $\textbf{K}$ will be given in Section \ref{subsec:tissue}. This equation is complemented by boundary conditions detailed in Section \ref{subsec:geom}.


\subsection{Oxygen flow}
\label{subsec:oxygen} 

We recall that the oxygen concentration at a point $x \in \Omega$ and time $t \geq 0$ is denoted by $\rho ( x, t)$. The oxygen may be either convected by the blood flow, diffused through the tissue or consumed. Thus, we assume that $\rho$ is described by the following convection-diffusion-reaction equation
\begin{equation}
\partial_t \rho + \nabla_x  \cdot (  \textbf{v} \rho) = -\beta(\rho) \rho \, ,
\label{O2PDE}
\end{equation}
where
\begin{align}
\textbf{v}(x,t)  & = \textbf{u}(x,t) - \textbf{D}(x,t) \frac{\nabla_x \rho(x,t)}{\rho(x,t) + \widetilde{\rho}} \, , \label{eq:v}\\
\beta(\rho) & = \frac{\beta_{\mbox{sat}}}{\rho + K_m} \, . \label{eq:beta} 
\end{align}
The left-hand side of  \eqref{O2PDE} together with \eqref{eq:v} is the convection-diffusion part. The first term of \eqref{eq:v} is the blood velocity $\textbf{u}$ and models the convection of oxygen by the blood flow. The second term is a nonlinear diffusion with diffusivity matrix $\textbf{D}(x,t) \frac{\rho(x,t)}{\rho(x,t) + \widetilde{\rho}}$. For small values of the oxygen concentration $\rho \ll \widetilde{\rho}$, this nonlinear diffusion behaves like $- \nabla_x \cdot ( \frac{\textbf{D}}{2 \widetilde{\rho}} \nabla_x \rho^2)$ and leads to the porous medium equation. In particular, it maintains $\textbf{v}$ finite (and actually equal to $0$) where $\rho = 0$. On the other hand, for high oxygen concentrations $\rho \gg \widetilde{\rho}$, it behaves like $- \nabla_x (\textbf{D} \nabla_x \rho)$, which leads to the standard linear diffusion equation. The diffusivity $\textbf{D}(x,t)$ is a symmetric nonnegative matrix which, like the hydraulic conductivity $\textbf{K}$, depends on the presence of capillaries and consequently, on space and time. It will be specified in Section \ref{subsec:tissue}. The right-hand side of \eqref{O2PDE} is a reaction term that accounts for oxygen consumption by the tissue. Formula (\ref{eq:beta}) is the classical Menten-Michaelis reaction term which is widely used for biological reactions. The consumption term $\beta(\rho) \rho$ is a linear function of the oxygen concentration at low oxygen concentration and models the increase of the number of cells able to consume oxygen (through for instance, cell proliferation). The consumption term $\beta(\rho) \rho$ saturates at large oxygen concentration because the number of cells and reaction sites able to consume oxygen cannot grow indefinitely and saturate to a maximal value. The Michaelis constant $K_m$ is the oxygen concentration at which the consumption rate reaches half of its maximum value and $\beta_{\text{sat}}$ is the maximal consumption rate. Eq. \eqref{O2PDE} must be supplemented with initial and boundary conditions which will be given in Section \ref{subsec:geom}.

We comment on the chosen values for the parameters $\widetilde{\rho}$, $\beta_{\mbox{sat}}$ and $K_m$ (see Table \ref{TableOfParameters}). All quantities relating to oxygen densities are expressed relative to the oxygen density in the oxygenated tissue $\rho_0$ which is a boundary condition for \eqref{O2PDE} (see Section \ref{subsec:geom}). The particular value of $\rho_0$ is irrelevant. Indeed, since the model is two-dimensional, it can be viewed as representing a three dimensional volume of certain thickness where all quantities are independent of the third coordinate. Thus, $\rho_0$ is equal to the actual volumetric oxygen density in the oxygenated tissue times the thickness of the considered volume and that thickness is arbitrary. The chosen value of $\rho_0$ is thus purely arbitrary. The value $\widetilde{\rho}$ is chosen to be $10\%$ of $\rho_0$ meaning that we switch to nonlinear diffusion when the oxygen density is $10\%$ lower than that of the oxygenated tissue. We could not find any documentation of this effect in the literature and this value appeared as a sensible choice. As for the oxygen consumption, we refer to \cite{takahashi1966oxygen} where the rate of consumption of oxygen was measured in a rabbit eye. This rate was measured to be of the order of $10^{-5}$ ml(O$_2$) (ml(tissue) s)$^{-1}$, while the O$_2$ volume fraction was of the order of $4 \, 10^{-3}$ ml(O$_2$) (ml(tissue))$^{-1}$ which gives a consumption rate of $14\%$ min$^{-1}$. However, we have chosen the much lower value $2\%$ min$^{-1}$. If we match this value to the maximal consumption rate of formula \eqref{eq:beta}, this gives us the value $\frac{\beta_{\mbox{{\scriptsize sat}}}}{K_m} = 0.02$ min$^{-1}$ which we can deduce from Table \ref{TableOfParameters}. The reason for choosing this low value lies in the approximations made in the model. Indeed, as already mentioned above, we use the oxygen gradient as a proxy for the VEGF gradient, which we do not include in the model. However, for a physiological value of the consumption rate, oxygen is consumed too quickly and there is not enough oxygen left to trigger the formation of capillaries (see Section \ref{subsec:capillaries}). By lowering the value of oxygen consumption, we allow for enough oxygen to remain and fuel capillary growth. A physiologically realistic value of the oxygen consumption should be usable when we include VEGF in future developments of the model. Now, we have chosen $K_m$, the oxygen concentration which halves the consumption rate, to be equal to half the oxygen concentration in the oxygenated tissue. This is consistent with values given in \cite{curcio2014kinetics} which suggest that the Michaelis constant is of the same order of magnitude as the oxygen concentration in the oxygenated tissue.

\subsection{Capillaries}
\label{subsec:capillaries}

\subsubsection{Capillary element}
\label{subsubsec_capillary_element}

\begin{figure}[ht!]
  \begin{center}
    \includegraphics[width=5cm]{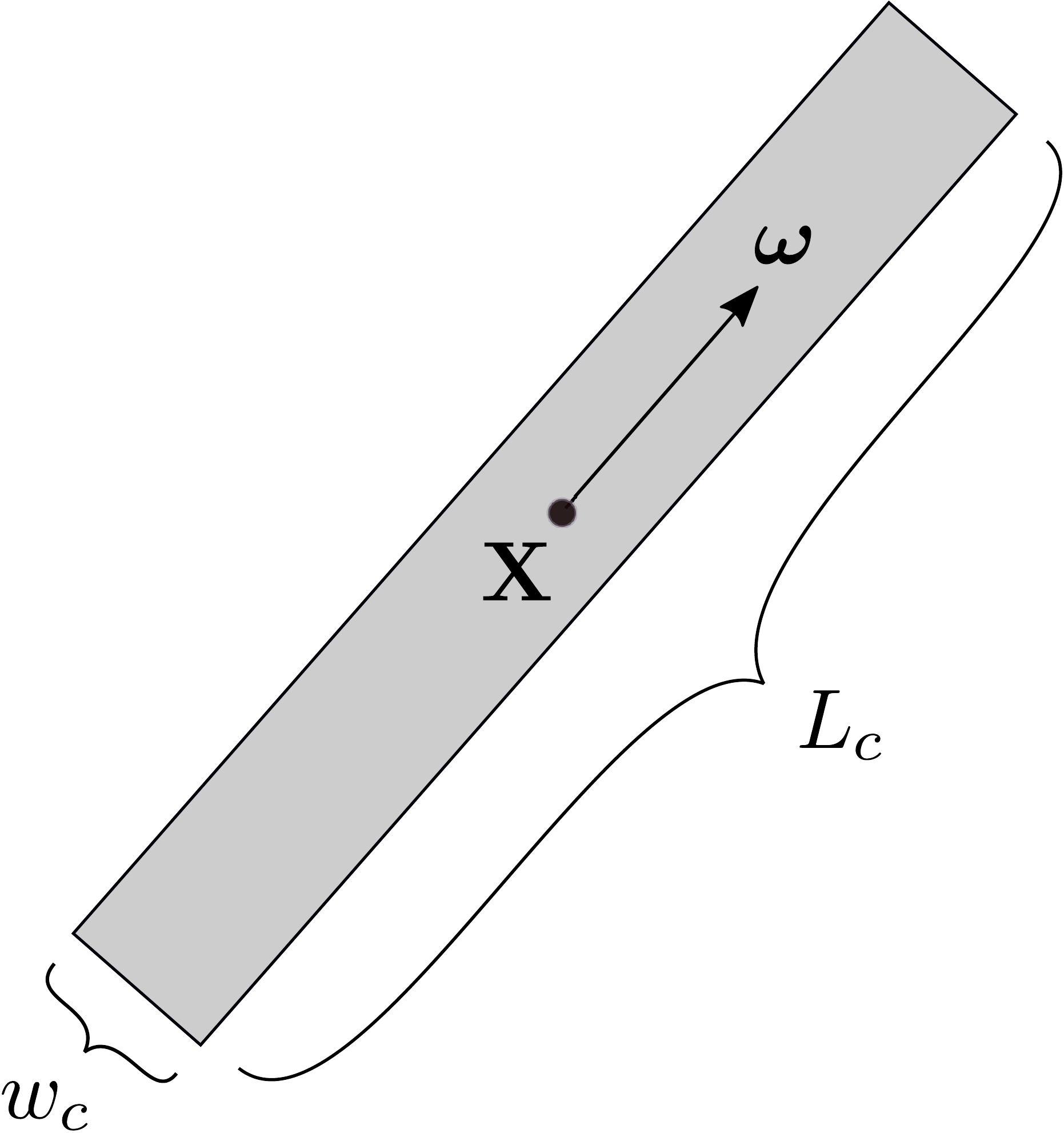} 
    \caption{A capillary element of length $L_c$ and width $w_c$ with center at $\textbf{X}$ and direction~$\boldsymbol{\omega}$.}
    \label{fig:rod}
  \end{center}
\end{figure}

The fundamental assumption of the model is to consider capillary elements as directional information of the flow, rather than describing the exact geometry of the network. Biologically capillary elements could represent one or a group of EC depending on the size chosen for the capillary element. We represent each capillary element as a rod of fixed length $L_c$ and width $w_c$, defined by the position of its centre  $\textbf{X} \in \Omega$ and direction $\boldsymbol{\omega} = ( \cos \theta, \sin \theta)$ where $\theta \in [0, 2 \pi)$ (see Fig. \ref{fig:rod}). We assume that each capillary element influences the local direction of the blood flow as well as the diffusivity properties of the tissue (see Sect. \ref{subsec:tissue}). The capillary network is formed by superposing all the capillary elements and its connectivity is not imposed at the particle level but is recovered through an averaging process at the continuum level detailed in Section \ref{subsec:tissue}. We have taken the values $L_c = 15 \, \mu$m and $w_c = 4 \, \mu$m (see Table \ref{TableOfParameters}) as these correspond to the typical size of human EC \cite{garipcan2011image}. 

We assume that capillary elements are created in response to environmental cues. This process could represent both de novo formation of new capillaries as in vasculogenesis or recruitment of existing EC that are attracted to the location where this creation happens. We do not model the process of EC migration (by contrast to many other works in the literature) for the sake of modelling simplicity. Likewise, once created, capillary elements remain immobile, until they are eventually destroyed. Thus capillary element dynamics reduces to creation and destruction processes. We assume three different creation processes and one removal process according to the nature of the environmental cues considered. They are all described by spatio-temporal Poisson processes whose intensities depend on these environmental cues. These processes are:
\begin{itemize}
\item Creation by oxygen concentration gradient (detailed in Section \ref{Subsec:creation_gradient}): when the norm of the gradient of oxygen concentration at some location exceeds a certain threshold, this signals a fast drop of oxygen concentration and a potentially hypoxic region at some distance along the direction of the gradient. Then, the probability of capillary creation is increased. The newly created capillary element, directed along the direction of the gradient, will contribute to bring more blood and eventually more oxygen, feeding the hypoxic region. However, if the oxygen concentration is already high, there is no need for more blood inflow and the capillary creation is inhibited. As already noted, this mechanism is a proxy for the creation of new capillaries in response to VEGF gradients, while we ignore VEGF in this model. 
\item Creation by reinforcement of small capillaries (detailed in Section \ref{Subsec:reinforcement}): to stabilize the newly created capillary elements, we have introduced a reinforcement mechanism. As long as the norm of the blood velocity is smaller than a threshold value and the oxygen concentration lies between two bounds, we trigger the formation of new capillary elements directed in the direction of the blood velocity. The bound on the velocity is because we want to avoid the blood flow going too fast, and 'missing' some nearby hypoxic regions. The bounds on the oxygen concentration is because there is no point in reinforcing capillaries if no or too much oxygen is already flowing in them. 
\item Creation by Wall Shear Stress (WSS) (detailed in Section \ref{Subsec:wss}): WSS is known to play a fundamental role in the development of new capillaries \cite{I+1997,K+2008,K+2011,M2015,SP1996}). A fluid flowing in a duct exerts a shear stress on its walls, called WSS. There is evidence of a mechanosensing mechanism which triggers capillary sprouting when WSS exceeds a certain value. WSS is expressed as the maximal eigenvalue of the viscous stress tensor. In the model, when at a given location WSS exceeds a threshold value, the creation of a new capillary element in the normal direction to the velocity is favored. The choice of this direction is motivated by experimental observations \cite{K+2011} where it seems favored. But this point will be scrutinized in future work. 
\item Capillary pruning (detailed in Section \ref{Subsec:cap_pruning}): we assume that capillary elements can be destroyed if the tissue hydraulic conductivity is already large. Indeed, the creation of new capillaries increases the tissue hydraulic conductivity (see Section \ref{subsec:tissue}) but on the other hand, the conductivity should not be locally too large, to avoid channeling all the blood flow in a localized region and leaving the rest of the domain hypoxic. We use the Frobenius norm $\gamma:=\| \textbf{K}\|$ (defined in Section \ref{Subsec:cap_pruning}) of the hydraulic conductivity tensor $\textbf{K}$  to sense the conductivity of the tissue. We trigger the removal of capillary elements when $\gamma$ exceeds a threshold value $\gamma^*$ and the removal probability is then a quadratic function of $\gamma-\gamma^*$. The choice of the quadratic function is motivated by \cite{H+2016} where it is assumed that there is a cost for the maintenance of the network which is a quadratic function of a quantity which plays a similar role as the hydraulic conductivity used here. In \cite{H+2016}, such a nonlinear cost is needed for the appearance of capillaries and we wanted to test if it is similar here. 
\end{itemize}

In the following sections, these four processes are described in detail.


\subsubsection{Capillary creation triggered by oxygen gradient}
\label{Subsec:creation_gradient}

\begin{figure}
\begin{center}
\hspace{-1in}
\begin{minipage}[t]{0.25\linewidth}
\vspace{0pt}
\centering
\includegraphics[width=2.5in]{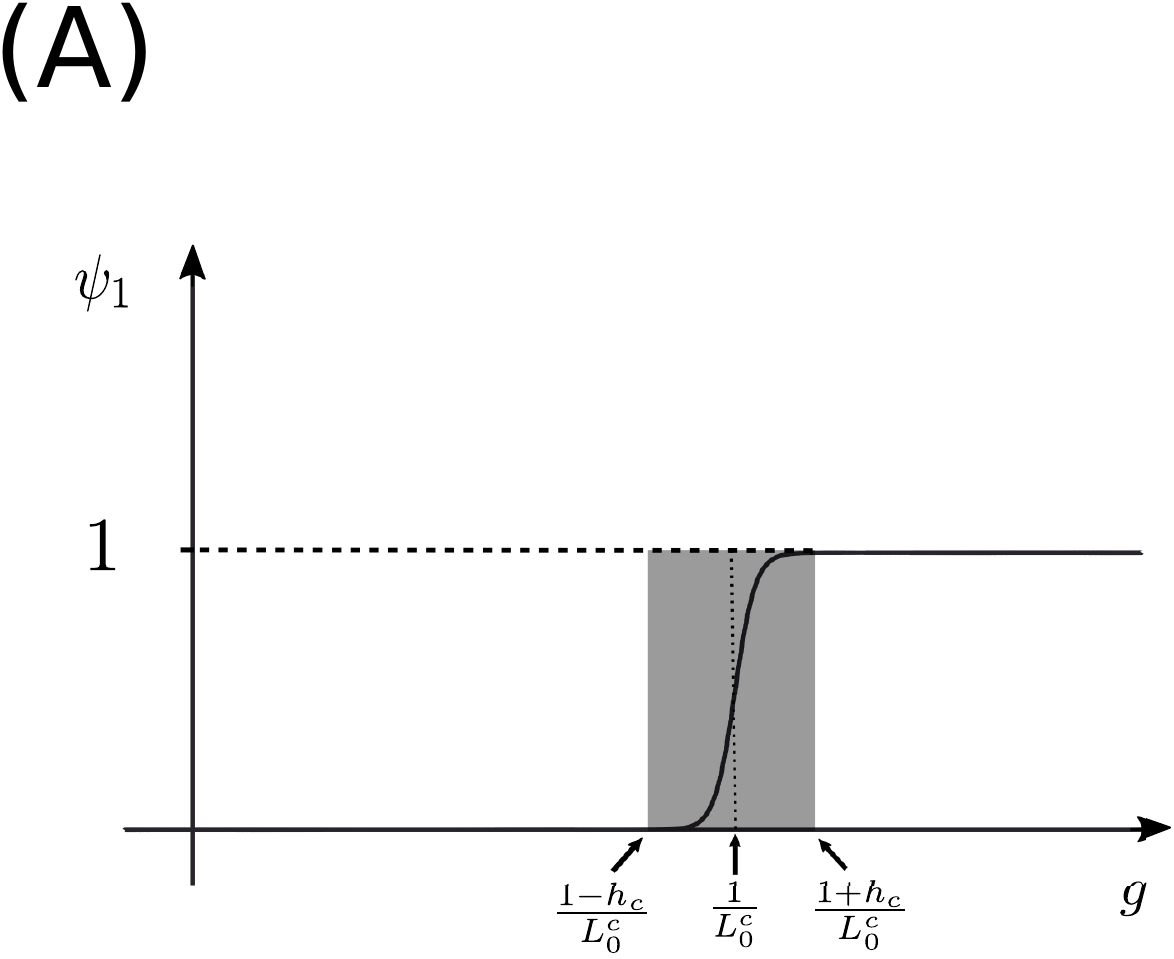}
\end{minipage}%
\hspace{1.3in}
\begin{minipage}[t]{0.25\linewidth}
\vspace{0pt}
\centering
\includegraphics[width=2.5in]{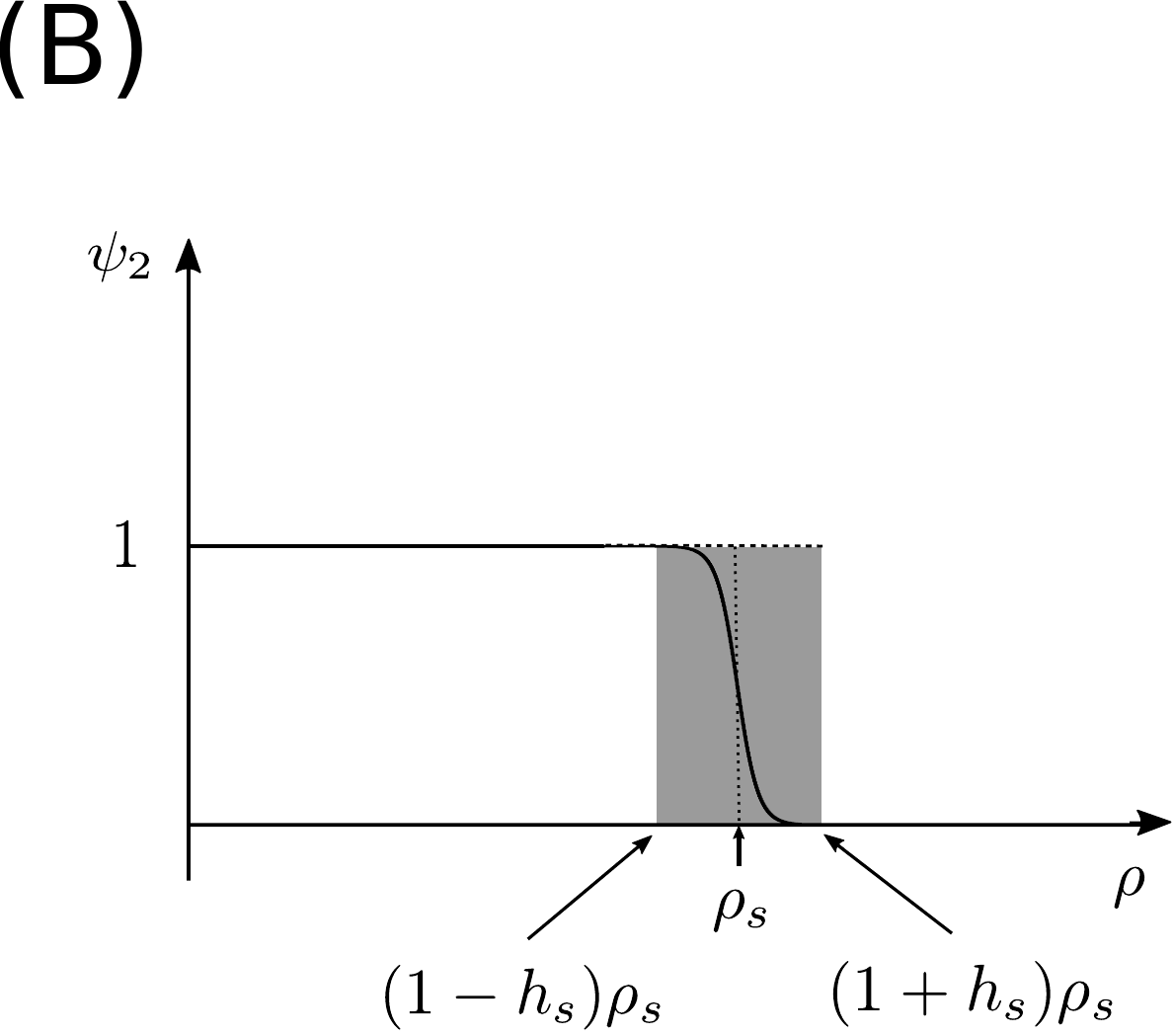}
\end{minipage}
\caption{(A) The function $g \mapsto \psi_1 ( g) = \psi ( ( L_0^c g - 1)/h_c)$ where $\psi$ is defined in \eqref{eq:tanh} models an on/off switch. Its fuzziness region is shadowed in gray. On its left-hand-side the switch is off whereas on its right-hand-side it is on. (B) The function $\rho \mapsto \psi_2 ( \rho) = \psi ( (1 - \rho / \rho_s) / h_s)$ with fuzzy region shadowed in gray. As opposed to (A) the switch is on at the left-hand-side of the shadowed region and it is off on the right-hand-side. }  
    \label{fig:fuzziness}
\end{center}
\end{figure}

We recall that $\rho$ denotes the oxygen concentration in the tissue. We assume that the creation of capillary elements is given by a spatio-temporal Poisson process with intensity
\begin{equation}
\nu_c = \nu_c^*
\psi
\left(
\frac{L_0^c \, g( \rho, \dx \rho) - 1}{h_c}
\right)
\psi
\left(
\frac{1 - \frac{\rho}{\rho_s}}{h_s}
\right),
\label{nuc}
\end{equation}
where
\begin{align}
\psi(z) & = \displaystyle \frac{1}{2} ( 1 + \tanh(z)) \, , \label{eq:tanh} \\
g( \rho, \dx \rho) & = \displaystyle\frac{| \dx \rho |}{\rho + \rho^*}  \, , \label{eq:gradient_length}
\end{align}
and $\nu_c^*$, $L_0^c$, $h_c$, $\rho_s$, $h_s$, $\rho^*$ are positive parameters. The function $g = g( \rho, \dx \rho)$ defined in~\eqref{eq:gradient_length} represents a smoothed logarithmic sensing function of the oxygen gradient $\nabla_x \rho$ where the parameter $\rho^*$ is introduced in order to avoid singularities where $\rho = 0$.  The function $\psi$ defined in \eqref{eq:tanh} is a sigmoid. It operates as a switch which is turned on or off whether $z$ is positive or negative respectively. It smoothly transitions from  0 to 1 which introduces uncertainty in the turn on/off of the switch, with a mushy zone located between the values $z=-1$ and $z=1$.  For instance, consider the first activation function
\begin{equation}
g \mapsto \psi_1 ( g) = \psi \bigg( \frac{ L_0^c g - 1}{ h_c} \bigg) \quad \text{ where } \quad g \in \mathbb{R},
\label{eq:phis1}
\end{equation}
on the right-hand-side of \eqref{nuc}. The parameter $h_c$ prescribes the width of the fuzziness region of the activation function. We set $h_c = 0.1$ which represents a $\pm 10\%$ uncertainty around the value $1/L_0^c$ (see Fig. \ref{fig:fuzziness}(A)). Therefore, \eqref{eq:phis1} turns on the creation process when the logarithmic sensing $g( \rho, \dx \rho)$ of the oxygen gradient exceeds a threshold $1/L_0^c \pm 10\%$. The second sigmoid on the right-hand-side of \eqref{nuc} follows the same rationale. It activates capillary creation when $\rho$ is less than the threshold value $\rho_s$  up to a fuzziness of order $\pm 10 \%$  (taking $h_s = 0.1$) (see Fig. \ref{fig:fuzziness}(B)). This factor prevents the formation of more capillary elements if the oxygen concentration is high enough. The parameter $\nu_c^*$ is the intensity of the creation rate when both switches in \eqref{nuc} are totally on. If a capillary element is created at the position $\textbf{X} \in \Omega$ then it is oriented towards the oxygen gradient
\[
\boldsymbol{\omega} = \frac{\nabla_x \rho(\textbf{X})}{| \nabla_x \rho(\textbf{X}) |},  
\]
which represents the optimal direction for the spreading of the oxygen.

We have already commented on the choice of $h_c=h_s=0.1$ corresponding to $\pm 10 \%$ uncertainties on the threshold values. The constant $\nu_c^*$ is the parameter of the Poisson process where all the switches are on i.e. where the two factors involving the function $\psi$ in \eqref{nuc} are equal to $1$. It corresponds to the fastest rate of capillary creation where large oxygen concentration gradient and low oxygen concentration signal a strongly hypoxic region. It is difficult to obtain such data from the experimental literature and we have resorted to trial and error with the model. The choice made, $\nu_c^* = 0.05$ min$^{-1}$ $\mu$m$^{-1}$ can be best understood when relative to a capillary element area $S_c=L_c \, w_c = 60 \mu$m$^2$. Then $\nu_c^* S_c = 3$ min$^{-1}$ meaning the maximal creation rate of capillary elements due to oxygen concentration gradient is one new capillary element every $20$ seconds per surface of a capillary element. This value sounds too high and was chosen on the basis of numerical constraints. This time-scale issue will be discussed in greater details in Sect. \ref{sect:results}. We have taken the oxygen gradient length threshold $L_0^c$ which triggers the formation of new capillaries as half the size of an EC ($L_0^c = 8 \, \mu$m) as concentration sensing should be done at the cell level. The quantity $\rho^*$ is just a regularizing constant and has been chosen as $10\%$ of the oxygen concentration in the oxygenated tissue $\rho_0$ consistently with the choice done for the other regularizing constant $\widetilde{\rho}$. The threshold oxygen concentration beyond which the capillary creation is turned off has been chosen equal to the oxygen concentration in the oxygenated tissue $\rho_0$ which means that for concentrations above $\rho_0$ the tissue is assumed fully oxygenated. The numerical values of parameters $\nu_c^*$, $L_0^c$, $h_c$, $\rho_s$, $h_s$, $\rho^*$ are summarized in Table~\ref{TableOfParameters}.


\subsubsection{Capillary creation by reinforcement}
\label{Subsec:reinforcement}

In order to enhance the creation of small branches we introduce a spatio-temporal Poisson process with intensity function
\begin{equation}
\nu_f = \nu_f^{*} \,
\psi
\left(
\frac{1 - \frac{| \textbf{u}|}{ \bar{u}}}{h_f}
\right)
\psi
\left(
\frac{\frac{\rho}{\underline{\rho}} - 1}{h_f}
\right)
\psi
\left(
\frac{1 - \frac{\rho}{\bar{\rho}}}{h_f}
\right),
\label{nuf}
\end{equation}
where $\nu_f^{*}$, $\bar{u}$, $h_f$, $\underline{\rho}$ and $\bar{\rho}$ are positive parameters and $\textbf{u}$ is the blood velocity. The activation function $\psi$ is defined in \eqref{eq:tanh}. Following the same rationale as in Sect. \ref{Subsec:creation_gradient}, the intensity \eqref{nuf} promotes the creation of capillary elements when the magnitude of the blood velocity $|\textbf{u}|$ is below the threshold parameter $\bar{u}$ up to a fuzziness region of $\pm 10\%$ set by the parameter $h_f = 0.1$. We also impose a cut-off when the oxygen concentration $\rho$ is below $\underline{\rho}$ or above $\bar{\rho}$ up to an uncertainty of $\pm 10\%$ set by the same parameter $h_f$. If a capillary element is created at a point $\textbf{X}$ its direction is set to be the normalized velocity vector at~$\textbf{X}$, i.e.
\[
 \boldsymbol{\omega} = \frac{ \textbf{u} ( \textbf{X})}{ | \textbf{u} ( \textbf{X})|} \, .
\]
Indeed, this choice is best to reinforce the flow in the direction of the blood velocity $\textbf{u} ( \textbf{X})$.

The parameter $\bar{u}$ was estimated by taking it slightly above the typical blood velocity in the tissue in the absence of capillaries. This velocity can be estimated through Darcy's law using the hydraulic conductivity of the tissue in the absence of capillaries $\kappa_h = 400 \, \mu$m$^2$ min$^{-1}$ mmHg$^{-1}$ (see Section \ref{subsec:tissue} for a justification of this value). With a pressure difference of about $20$ mmHg and a typical tissue length of $1$ mm (see section \ref{subsec:geom}), such velocity is $8 \, \mu$m/min. We have chosen $\bar u = 20 \, \mu$m/min. 
The values that we chose for $\underline{\rho}$ and $\bar{\rho}$ are, respectively, $10\%$ and $50\%$ of the the oxygen concentration $\rho_0$ injected to the tissue (see Sect. \ref{subsec:geom}). Indeed below $0.1 \, \rho_0$ the oxygen density is too small and capillaries must first be created by other creation mechanisms. Above $0.5 \, \rho_0$ the oxygen density is close to that of the oxygenated region and capillaries do not need any reinforcement. The parameter of the Poisson process $\nu_f^{*}$ corresponding to the maximum intensity when all switches are on is set to be $1/5$ of the corresponding intensity for the creation process by oxygen gradient. Indeed, once capillaries are created, the threat posed by hypoxia on the tissue is reduced and the reinforcement mechanism can be slower, but still within an order of magnitude of the faster process of creation by oxygen gradient. All these parameters are summarized in Table~\ref{TableOfParameters}.


\subsubsection{Capillary creation by Wall-Shear-Stress}
\label{Subsec:wss}

For a given point $x$ we consider the deviatoric stress tensor at $x$ which, thanks to the incompressibility 
condition for the blood flow \eqref{eq:bloodPDE2}, reduces to
\begin{equation}
\sigma (x) = \mu \big( \nabla_x \textbf{u} ( x) + (\nabla_x \textbf{u} ( x))^T \big) \, , \label{eq:sigma} 
\end{equation}
where $( \nabla_x \textbf{u})_{ij} = \partial \textbf{u}_j / \partial x_i$ is the tensor gradient of the vector field $\textbf{u}$, $A^T$, where $A$ is a $2 \times 2$  matrix, denotes the transpose of $A$ and $\mu$ is the dynamic viscosity of the blood. The $2 \times 2$ tensor $\sigma (x)$ is nonnegative symmetric and traceless. Hence it has eigenvalues $\pm \lambda$ with $\lambda \geq 0$. 
The creation by WSS is given by the spatio-temporal Poisson process with intensity function
\begin{equation}\label{eq:wall}
 \nu_w = \nu_w^* \, \psi \bigg( \frac{ \frac{ \lambda}{ \lambda^*} - 1 }{ h_w} \bigg) \, ,
\end{equation}
where $\nu_w^*$, $\lambda^*$ and $h_w$ are positive parameters and the activation function $\psi$ is defined in~\eqref{eq:tanh}. Following the same rationale as in Sect. \ref{Subsec:creation_gradient}, capillary elements are created if $\lambda$ is bigger than the threshold parameter $\lambda^*$ up to an uncertainty given by $h_w$ (also set to be $0.1$) and $\nu_w^*$ is the intensity when the switch is totally on. If a capillary element is created at a point $\textbf{X}$ its orientation is taken to be
\[
 \boldsymbol{\omega} = \frac{ \textbf{u}^\perp ( \textbf{X})}{ | \textbf{u}^\perp ( \textbf{X})|}
\]
where $\textbf{u}^\perp ( \textbf{X})$ denotes the rotation of $\textbf{u} ( \textbf{X})$ by $90^\circ$ in the counterclockwise direction. We chose this branching angle following the experimental data presented in \cite{K+2008} and \cite{K+2011}. The value of $\lambda^*$ was taken from \cite{P+2011} and the value of the dynamic viscosity of the blood $\mu$ at  $37^\circ$C was taken from \cite{F2017}. Following a set of numerical trials, and owing to the literature documenting the importance of this mechanism, the value of $\nu_w^*$ was chosen large and set $6$~times larger that the rate of creation by oxygen gradient. See Table \ref{TableOfParameters} for a summary of the numerical  values of these parameters.


\subsubsection{Capillary pruning}
\label{Subsec:cap_pruning}

In order to impede excessive concentrations of capillary elements we remove them following a spatio-temporal Poisson process with intensity defined by 
\begin{equation}
\nu_r = \nu_r^* \, 
\Big(
\big(\frac{ \gamma}{ \gamma^*} - 1 \big)_+
\Big)^2,
\label{nur}
\end{equation}
where $\nu_r^*$, $\gamma^*$ are positive parameters, $\gamma$ is the Frobenius norm of $\textbf{K}$ and $z_+ = \max \{ 0, z \}$ for $z$ in $\mathbb{R}$. The Frobenius norm of a $2 \times 2$ matrix with real valued entries $\{a_{ij}\}_{1 \leq i,j \leq 2}$ is given by $\| A \| = (\sum_{1 \leq i,j\leq 2} (a_{ij})^2)^{1/2}$. Formula \eqref{nur} states that the removal mechanism is only turned on when $\gamma$ exceeds the threshold value $\gamma^*$. Then for $\gamma \geq \gamma^*$ the removal intensity  increases quadratically. The quantity $\nu_r^*$ is the intensity of the process for $\gamma = 2 \gamma^*$. The choice of the parameter $\gamma^*$ was estimated to be $5$ times the hydraulic conductivity of individual capillary elements (estimated in Section \ref{subsec:tissue}) meaning that we do not allow more than $5$ individual capillary elements to superimpose at the same place (superposition of capillary elements modelling broader capillaries). Indeed, larger number of superposed capillaries would model larger vessels for which the Darcy approximation and consequently the whole model would lose validity. We want to promote a fast removal rate when $\gamma$ exceeds the threshold $\gamma^*$. So, the value of $\nu_r^*$ is chosen to be such that the lifetime of a capillary element when $\gamma = 2 \gamma^*$ is equal to 2 seconds which is comparable to the maximal creation rate of capillaries by WSS, equal 1 every 3 seconds per capillary element area (see Section \ref{Subsec:wss}). Finally, the pruning rate being a quadratic function of $\gamma - \gamma^*$ reflects the nonlinear loss rate in the continuum model of network formation of \cite{H+2016}. This feature will be commented in Section \ref{sect:discussion}. We refer to Table \ref{TableOfParameters} for a summary of the parameter values.


\subsection{Tissue}
\label{subsec:tissue} 

The tissue is a mixture of cells, ECM and capillary vessels. It behaves like a porous medium (see \cite{SF2007}) for the blood and oxygen fluids. It is characterized by its hydraulic conductivity tensor $\textbf{K}$ for the blood flow and its diffusivity tensor $\textbf{D}$ for the oxygen. We assume that in the absence of capillaries, the tissue is an isotropic homogeneous medium characterized by scalar background hydraulic conductivity $k_h$ and oxygen diffusivity $\Delta_h$. Each capillary element $( \textbf{X}_k, \boldsymbol{\omega}_k)$ induces a local increase of the tensor $\textbf{K}$ by the quantity $\kappa \, (\boldsymbol{\omega}_k \otimes \boldsymbol{\omega}_k) \, \chi_{S_k}(\textbf{X})$. The positive scalar $\kappa$ is the conductivity of a capillary element, computed below under the assumption of Poiseuille flow in the capillary. The symbol $\otimes$ denotes the tensor product of two vectors, i.e. if $A=(A_i)_{i=1,2}$ and $B=(B_i)_{i=1,2}$ are two vectors in ${\mathbb R}^2$, then $A \otimes B$ is the $2 \times 2$ matrix whose entries are $(A \otimes B)_{ij} = A_i \, B_j$. We have the formula $(\boldsymbol{\omega}_k \otimes \boldsymbol{\omega}_k) \nabla_x p = (\boldsymbol{\omega}_k \cdot \nabla_x p) \, \boldsymbol{\omega}_k$ giving that the flow velocity through a capillary element is proportional to the directional derivative of $p$ in the direction of $\boldsymbol{\omega}_k$ and points in the direction of $\boldsymbol{\omega}_k$. This is the feature of a porous medium whose pores are all oriented in the direction of $\boldsymbol{\omega}_k$ and which is impermeable in the orthogonal direction $\boldsymbol{\omega}_k^\bot$ (where $\boldsymbol{\omega}_k^\perp$ represents the counterclockwise rotation of $\boldsymbol{\omega}_k$ by $90^\circ$). The increase of hydraulic conductivity is limited to the domain occupied by the capillary element, which is reflected by the factor $\chi_{S_k}(X)$ where $S_k$ is the capillary element domain, i.e. the rectangle defined by 
$$ S_k = \{ \textbf{X} \in {\mathbb R}^2 \, \, \big| \, \, |(\textbf{X}_k - \textbf{X}) \cdot \boldsymbol{\omega}_k| \leq \frac{L_c}{2}, \, \,  |(\textbf{X}_k - \textbf{X}) \cdot \boldsymbol{\omega}_k^{\perp}|\leq \frac{w_c}{2} \}. 
$$ 
For a subset $S$ of ${\mathbb R}^2$, the function $\chi_S(\textbf{X})$ is the indicator function of $S$ i.e. it takes the value $1$ if $\textbf{X} \in S$ and $0$ otherwise. The contribution of each capillary element is summed up with those of the other capillary elements and with that of the bulk tissue given by $ k_h \textbf{I}_2$ (where $\textbf{I}_2$ is the $2 \times 2$ identity matrix and reflects the fact that the bulk conductivity is isotropic). The same phenomenology holds for the oxygen diffusivity tensor $\textbf{D}$ with a local increase of the diffusivity due to the $k$-th capillary element by the quantity $\Delta (\boldsymbol{\omega}_k \otimes \boldsymbol{\omega}_k) \chi_{S_k}(\textbf{X})$ where $\Delta$ will be estimated below. 

\begin{figure}[ht!]
  \begin{center}
    \includegraphics[trim={0 13cm 0 0},clip,width=10cm]{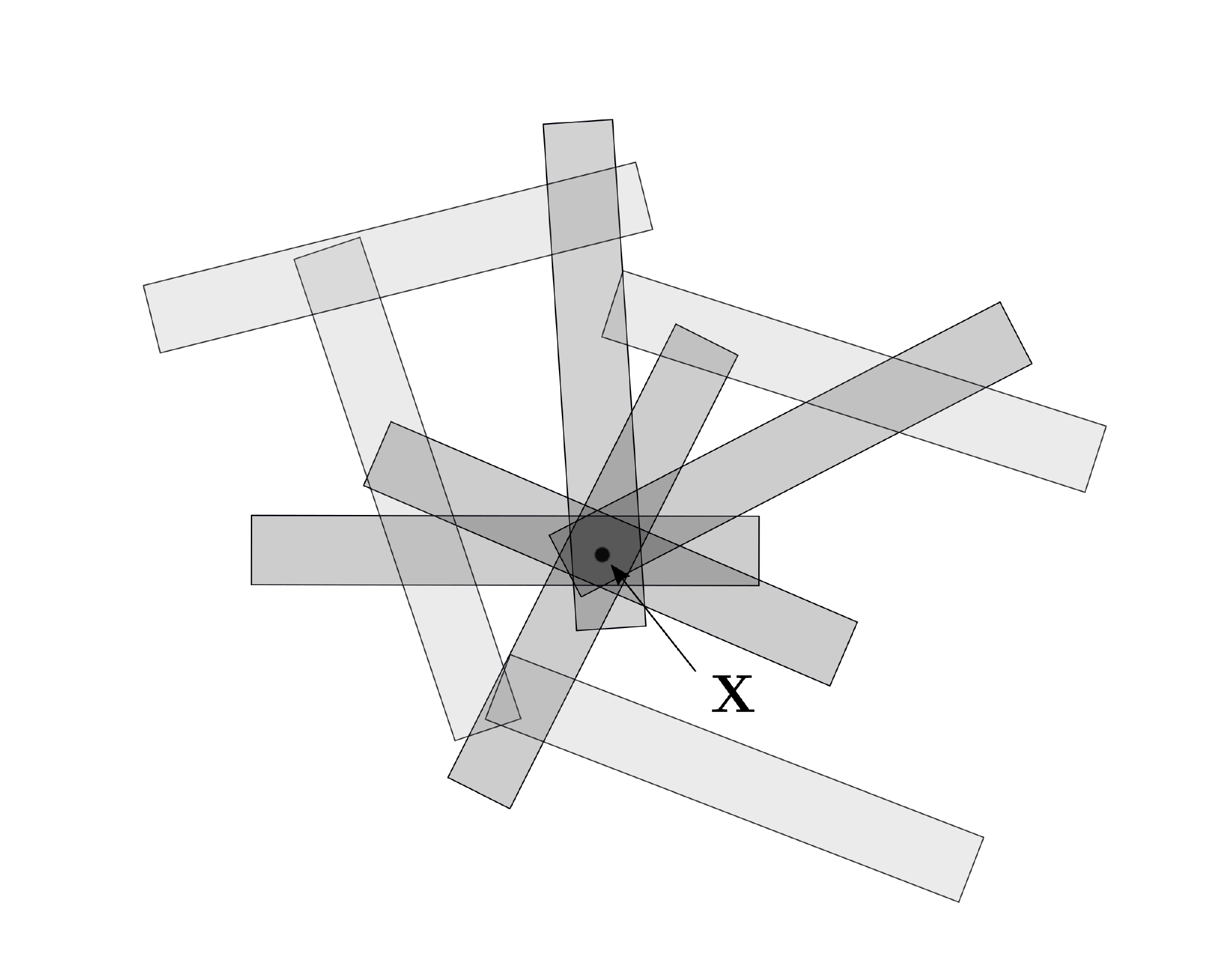} 
    \caption{Given a point 
		$\textbf{X}$ in the tissue, the second term of the right-hand-side of the tensors $\textbf{K}$ and $\textbf{D}$ defined in \eqref{eq:tissue1} and \eqref{eq:tissue2} are computed by summing the tensors $\boldsymbol{\omega}_k \otimes \boldsymbol{\omega}_k$ over all capillary elements $k$ that contain $\textbf{X}$ in their domain $S_k$. For instance, in this sketch, only five (dark-shadowed rods) out of the nine capillary elements are combined to form tensors $\textbf{K}$ and $\textbf{D}$ at $\textbf{X}$.} 
    \label{fig:rods}
  \end{center}
\end{figure}

The resulting formula for the hydraulic conductivity and diffusivity tensors of the tissue are
\begin{align}
\textbf{K}(\textbf{X}) = k_h \textbf{I}_2 + \sum_{\textbf{X} \in S_k} \kappa \, (\boldsymbol{\omega}_k \otimes \boldsymbol{\omega}_k) \, , \label{eq:tissue1} \\
\textbf{D}(\textbf{X}) = \Delta_h \textbf{I}_2 + 
\sum_{\textbf{X} \in S_k} \Delta \ (\boldsymbol{\omega}_k \otimes \boldsymbol{\omega}_k) \, , \label{eq:tissue2}
\end{align}
The first component on the right-hand-side of \eqref{eq:tissue1} is the background conductivity of the tissue in the absence of capillaries and the second one combines the influence of all capillary elements whose domains contain $\textbf{X}$ (see Fig. \ref{fig:rods}). So, each capillary element $(\textbf{X}_k, \boldsymbol{\omega}_k)$ containing $\textbf{X}$ in its domain exerts a local bias by  facilitating blood flow along $\boldsymbol{\omega}_k$. The same phenomenology holds for the oxygen diffusivity given by \eqref{eq:tissue2}.

The value of $\kappa_h$ in \eqref{eq:tissue1} was taken from \cite{SF2007}, whereas $\kappa$ was estimated supposing that the flow within a capillary follows Poiseuille's law. It relates the pressure drop $\Delta p$ in a duct of length $L_c$ and radius $w_c/2$ to the velocity $|u|$ of the flow by the relation $\frac{\Delta p}{L_c} = \frac{8 \mu |u|}{(w_c/2)^2}$ where $\mu$ is the dynamic viscosity of the fluid. With Darcy's law within a capillary element $|u| = \kappa  \frac{\Delta p}{L_c}$, we get $\kappa = \frac{w_c^2}{32 \mu}$. With the values of $w_c$ and $\mu$ from Table \ref{TableOfParameters}, we get $\kappa \approx 10^6 \, \mu$m$^2$ (mmHg min)$^{-1}$. In fact this value  brings too large stiffness to the elliptic problem \eqref{eq:elliptic} and we settle to a value about $10$ times smaller equal to $0.8 \, 10^5 \mu$m$^2$ (mmHg min)$^{-1}$. The value $\Delta_h$ deduced from from \cite{T+2014} is about $0.5 \, \mu$m$^2$ min$^{-1}$. We take a slightly larger value of $10 \, \mu$m$^2$ min$^{-1}$ owing to the large variability of this parameter according to the type of tissue. This larger value allows for a more uniform distribution of oxygen in the tissue. The diffusivity of oxygen in a vessel is enhanced but not as much as the hydraulic conductivity because the fluid molecules still constitute obstacles against the diffusion of the gaseous oxygen. We assume for our model that $\Delta = 20 \, \Delta_h$. Again, the entire set of parameter values is reminded in Table \ref{TableOfParameters}.


\subsection{Geometry and boundary conditions}
\label{subsec:geom}

\subsubsection{Geometry}
\label{subsubsect:geometry}

For the sake of computational efficiency, we consider a 2D rectangular domain $\Omega = [0,L_x] \times [0,L_y]$. Given a capillary-free tissue at initial time, we want to explore how the network appears. We assume that there is a highly oxygenated tissue on the left border with high blood pressure bringing oxygenated blood into the tissue. We analyze two different geometrical conditions distinguished by their boundary conditions. 

\begin{figure}
\begin{center}
\hspace{-12.5cm}
\begin{minipage}[t]{0.25\linewidth}
\vspace{-2cm}
\centering
\includegraphics[trim={0cm 3cm 5cm 2cm}, width=4.4\textwidth, clip]{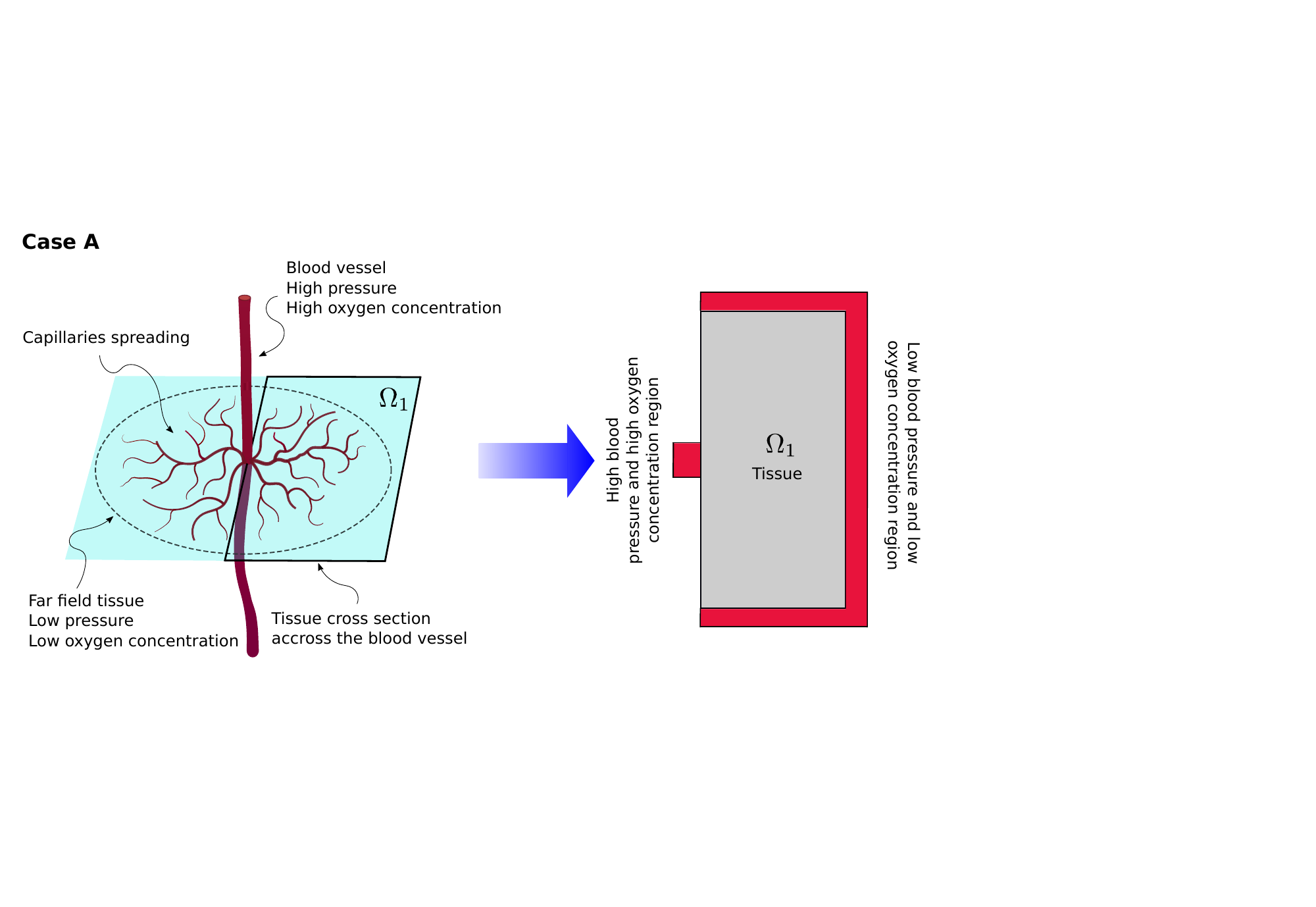}
\end{minipage}%
\\
\hspace{-13.8cm}
\begin{minipage}[t]{0.25\linewidth}
\vspace{-1cm}
\centering
\includegraphics[trim={0cm 3cm 1cm 0cm}, width=4.4\textwidth, clip]{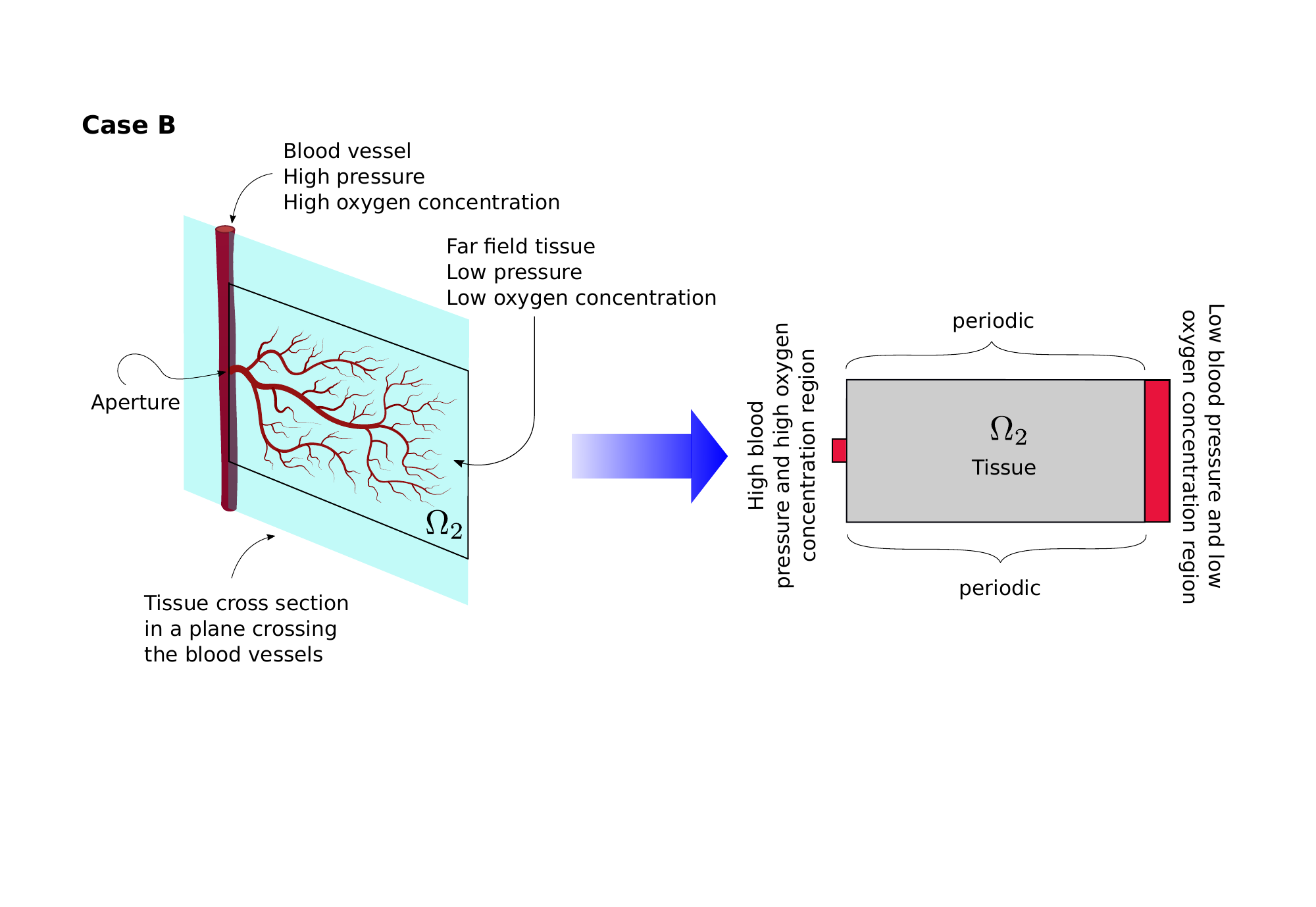}
\end{minipage}
\caption{(A) Geometrical setting for $\Omega_1$, which mimics a cross-section of the tissue in the direction normal to a blood vessel. (B) Geometrical setting for $\Omega_2$ which mimics a cross-section in a plane containing the blood vessel. The dimensions of $\Omega_1$ and $\Omega_2$ are given in Table. \ref{TableOfParameters}.}  
\label{fig:bio_geometries}
\end{center}
\end{figure}

\paragraph{Case 1} In the first domain $\Omega_1$ depicted in Fig \ref{fig:bio_geometries} (A), the blood inflow region on the left-hand boundary is surrounded by a lower blood pressure region, such as a region of uptake of blood by the venous system, modelled by the top, bottom and right-hand-side boundaries. The blood source is located in the middle of the left boundary and has small extent. The rest of the left boundary is impermeable. This case models the spread of capillaries around a blood vessel in a cross-section of the tissue normal to the blood vessel. Only half of the environment of the blood vessel is taken into account since, at least in a homogeneous tissue, the capillaries will spread in a similar fashion on the other side of the left boundary. We take domain sizes equal to $L_x = 1$  mm and $L_y=2$ mm. This corresponds roughly to a domain which extends up to a distance of $1$ mm to the blood source, which is a typical distance covered by capillaries in wound healing for instance. The extent of the blood and oxygen sources in the middle of the left boundary is $100 \, \mu$m which corresponds to the diameter of a small blood vessel. 

\paragraph{Case 2} In the second one denoted by $\Omega_2$ and shown in Fig. \ref{fig:bio_geometries} (B), a highly oxygenated high blood pressure region stands on the left-hand boundary. It faces a low blood pressure region corresponding to the right boundary. The top and bottom boundaries are assumed periodic: they represent a tissue that extends in the vertical direction on large distances in a somewhat similar fashion and for which only a representative fraction is simulated. The creation of a new capillary is triggered by an inflow of oxygen on a small portion of the left boundary. This case depicts the spread of a blood capillary longitudinally, i.e. when a cross-section of the tissue is made in a plane containing the blood vessel making the blood source. This situation also mimics the geometry of a corneal micropocket angiogenesis assay \cite{grogan2018importance}. We take domain sizes equal to $L_x=2$ mm and $L_y = 1$ mm, which is the order of magnitude encountered in such experiments. The oxygen sources in the middle of the left boundary extends along $100 \, \mu$m similar to the previous case. 

\medskip
\noindent
In both cases, the difference of pressures between the high and low blood pressure regions produces a flow of the interstitial fluid which triggers the formation of a new vascular network. The bulk of the domains are constituted of a tissue assimilated to a porous medium as described in Sect. \ref{subsec:tissue}.

\subsubsection{Boundary conditions on the pressure and boundary/initial conditions on the oxygen concentration}
\label{subsubsect:boundary_conditions}

Since we initially suppose a capillary-free tissue, we impose for consistency that the oxygen concentration is initially zero, i.e. for both geometries $\Omega_1$ and $\Omega_2$ we impose $\rho ( x, 0)=0$ for all $x$. For the boundary conditions, we distinguish between the two cases depicted in the previous section.

\begin{figure}[ht!]
  \begin{center}
  \includegraphics[trim={1.5cm 7cm 0 3cm}, clip, width=10cm]{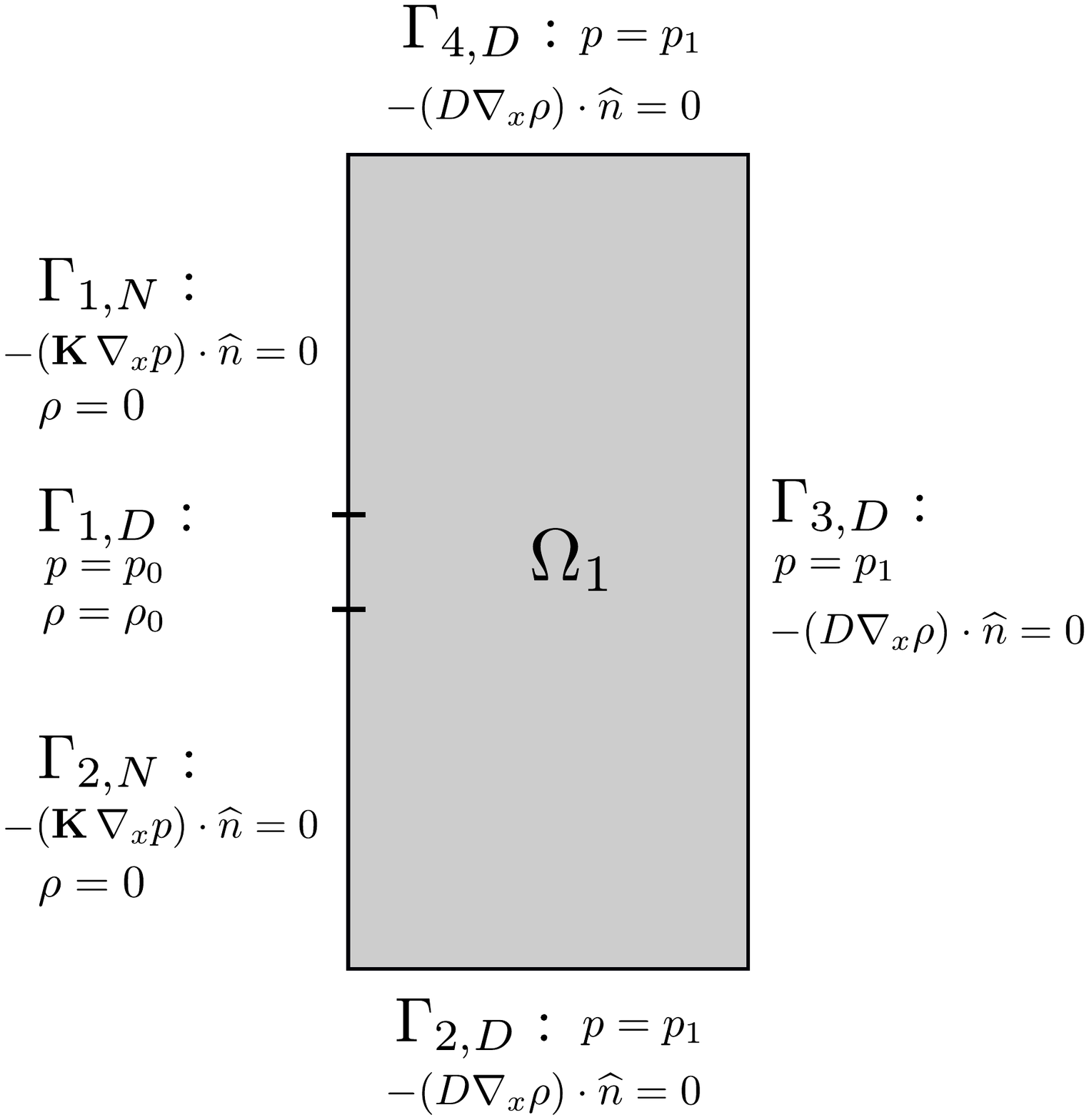} 
    \caption{Labeling of boundaries and boundary conditions for the pressure $p$ and oxygen density $\rho$ in $\Omega_1$.}
    \label{fig:stress}
  \end{center}
\end{figure}

\paragraph{Case 1} The labeling of the different parts of the boundary of $\Omega_1$ is given in Fig. \ref{fig:stress}. We impose the following boundary conditions for the pressure: 
\begin{align}
p & = p_0 \text{ on } \Gamma_{1,D} \, , \label{eq:dir1} \\
p & = p_1 \text{ on } \Gamma_{2,D} \cup \Gamma_{3,D} \cup \Gamma_{4,D} \, , \label{eq:dir2} \\
 - ( \textbf{K} \, \nabla_x p) \cdot \widehat{n} & = 0 \text{ on } \Gamma_{1,N} \cup \Gamma_{2,N} \, , \label{eq:dir3}
\end{align}
and for the oxygen concentration:
\begin{align}
\rho & = \rho_0 \text{ on } \Gamma_{1,D} \, , \label{eq:dir1_rho} \\
- ( \textbf{D} \, \nabla_x \rho) \cdot \widehat{n} & = 0 \text{ on } \Gamma_{2,D} \cup \Gamma_{3,D} \cup \Gamma_{4,D} \, , \label{eq:dir2_rho} \\
\rho & = 0 \text{ on } \Gamma_{1,N} \cup \Gamma_{2,N} \, , \label{eq:dir3_rho}
\end{align}
where $\widehat{n}$ is the outward unit normal vector to the boundary. Boundary $\Gamma_{1,D}$ has length $100 \, \mu$m and is located at the centre of the left boundary of the rectangle $\Omega_1$ i.e. it starts at the height $L_{min}= \,950 \, \mu$m and ends at $L_{max}= \, 1050 \, \mu$m. We set $p_0 > p_1$. Condition \eqref{eq:dir1} expresses that $\Gamma_{1,D}$ is the high blood pressure region with pressure $p_0$ (for instance a blood vessel). $\Gamma_{1,D}$ is also the source of oxygen and $\rho_0$ stands for the oxygen concentration in this highly oxygenated region, hence \eqref{eq:dir1_rho}. The subset $\Gamma_{2,D} \cup \Gamma_{3,D} \cup \Gamma_{4,D} $ of the boundary is the low blood pressure region, hence the Dirichlet condition \eqref{eq:dir2} with pressure $p_1$. It surrounds the high pressure source to mimic the conditions depicted in Fig~\ref{fig:bio_geometries} (A). The uptake of blood by the venous system takes place in this region and the oxygen flows out of the domain. Along this boundary, we assume a zero normal gradient of the oxygen concentration \eqref{eq:dir2_rho}, meaning (with \eqref{eq:v}) that the normal oxygen flow velocity $v \cdot \widehat{n}$ is equal to the normal blood flow velocity $u \cdot \widehat{n}$. We expect the latter to be large and positive (i.e. leaving the domain). Thus, oxygen will exit the domain freely without any possible inflow. So,~\eqref{eq:dir2_rho} expresses that the incoming flux of oxygen is zero. The Neumann boundary condition \eqref{eq:dir3} for the pressure along $\Gamma_{1,N} \cup \Gamma_{2,N}$ is a rigid wall condition stating that the normal blood velocity to the wall is zero. We also assume that there is no oxygen, leading to \eqref{eq:dir3_rho}. Because the problem is two-dimensional, the quantity $\rho_0$ is a surface density.  As explained in Section \ref{subsec:oxygen}, its numerical value is irrelevant because it depends on the height of the tissue in the third dimension, which is arbitrary. The boundary conditions on the pressure are those that prevail between the inflow and outflow of the capillary system as measured in \cite{williams1988dynamic}. The parameters are given in Table \ref{TableOfParameters}. 

\begin{figure}[ht!]
  \begin{center} 
    \includegraphics[trim={2cm 19.3cm 0 3cm}, clip, width=16.5cm]{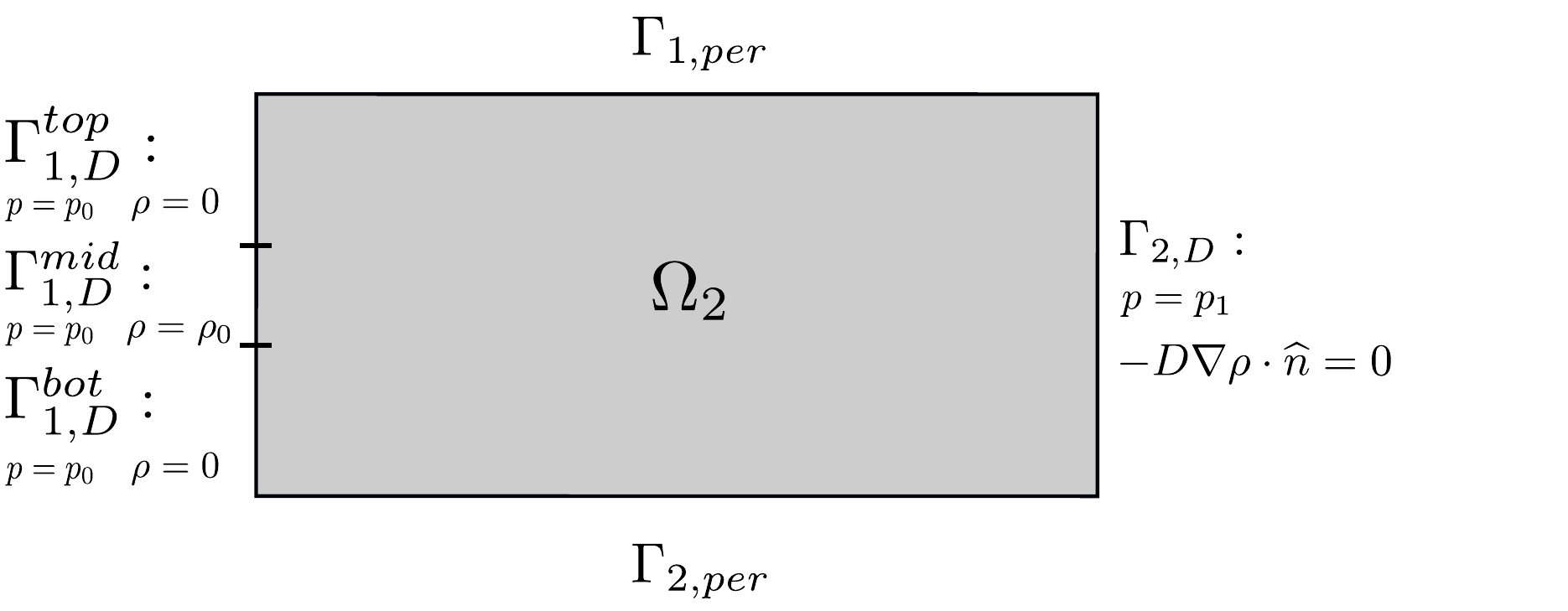} 
    \caption{Labeling of boundaries and boundary conditions for the pressure $p$ and oxygen density $\rho$ for $\Omega_2$.}
    \label{fig:boundary}
  \end{center}
\end{figure}

\paragraph{Case 2} We refer to Fig. \ref{fig:boundary} for the labeling of the various parts of the boundary of $\Omega_2$ and consider the following boundary conditions for the pressure:
\begin{align}
p & = p_0 \text{ on } \Gamma_{1,D}^{\mbox{\scriptsize top}} \cup \Gamma_{1,D}^{\mbox{\scriptsize mid}} \cup \Gamma_{1,D}^{\mbox{\scriptsize bot}}  \, , \label{eq:C2_p0} \\
p & = p_1 \text{ on } \Gamma_{2,D} \, , \label{eq:C2_p1} \\
p(x) & = p(x - L_y e_2) \text{ for all $x$ on }   \Gamma_{1,\mbox{\scriptsize per}} \label{eq:C2_p-per}
\end{align}
where $e_2$ is the unit vector in the vertical direction i.e. $e_2=(0,1)$. For the oxygen concentration, the boundary conditions are set to:
\begin{align}
\rho & = \rho_0 \text{ on } \Gamma_{1,D}^{\mbox{\scriptsize mid}} \, , \label{eq:C2_rho0} \\
\rho & = 0 \text{ on } \Gamma_{1,D}^{\mbox{\scriptsize top}} \cup \Gamma_{1,D}^{\mbox{\scriptsize bot}}  \, , \label{eq:C2_rho=0} \\
- ( \textbf{D} \, \nabla_x \rho) \cdot \widehat{n} & = 0 \text{ on } \Gamma_{2,D} \, , \label{eq:C2_rho_outflow} \\
\rho(x) & = \rho(x - L_y e_2) \text{ for all $x$ on }  \Gamma_{1,\mbox{\scriptsize per}} \label{eq:C2_rho-per}
\end{align}
Again, we impose $p_0 > p_1$ and the Dirichlet condition \eqref{eq:C2_p0} sets the pressure along the left boundary $\Gamma_{1,D}^{\mbox{\scriptsize top}} \cup \Gamma_{1,D}^{\mbox{\scriptsize mid}} \cup \Gamma_{1,D}^{\mbox{\scriptsize bot}}$ to the high pressure value $p_0$, while the Dirichlet condition~\eqref{eq:C2_p1} sets it to the low value one $p_1$ on the right boundary $\Gamma_{2,D}$. Condition \eqref{eq:C2_p-per} simply states that the values of the pressure on $\Gamma_{1,\mbox{\scriptsize per}}$ must be equal to those of $\Gamma_{2,\mbox{\scriptsize per}}$ expressing the periodicity of the solution in the vertical direction. To trigger the formation of a capillary we assume a large oxygen concentration along $\Gamma_{1,D}^{\mbox{\scriptsize mid}}$, hence the Dirichlet condition~\eqref{eq:C2_rho0} setting the oxygen concentration to $\rho_0$ on this boundary. Along the remaining part of the left boundary $\Gamma_{1,D}^{\mbox{\scriptsize top}} \cup \Gamma_{1,D}^{\mbox{\scriptsize bot}}$, we assume no oxygen is present, hence the Dirichlet condition~\eqref{eq:C2_rho=0}. We assume that $\Gamma_{1,D}^{\mbox{\scriptsize mid}}$ has length $100 \, \mu$m and is located at the centre of the left boundary i.e. it starts at the height $L_{min}= \,450 \, \mu$m and ends at $L_{max}= \, 550 \, \mu$m. Like for the pressure, condition \eqref{eq:C2_rho-per} along $\Gamma_{1,\mbox{\scriptsize per}}$ expresses the periodicity constraint on $\rho$ in the vertical direction. Finally, \eqref{eq:C2_rho_outflow} is an outflow condition for the oxygen concentration along the right boundary $\Gamma_{2,D}$. Its interpretation is the same as for \eqref{eq:dir2_rho} and we refer to the previous case for a detailed explanation. 

\begin{table}{}
{\scriptsize
\hspace{-1.cm}
\begin{tabular}{|l|l|l|l|l|}
\hline
{\bf Quantity}                                            &{\bf Symbol}        &  {\bf Value}              & {\bf Units}                        & {\bf Source} \\
\hline
{\bf Geometry 1 (Sect. \ref{subsec:geom})}                &                    &                           &                                    & \\
\hline
Domain size in $x$-direction                              &$L_x$               & $1000$                    &$  \mu \mbox{m} $                   & estim.\\
Domain size in $y$-direction                              &$L_y$               & $2000$                    &$  \mu \mbox{m} $                   & estim.\\
Oxygen injection region: $y$-coordinate of lower end      &$L_{\min}$          & $950$                     &$  \mu \mbox{m} $                   & estim.\\
Oxygen injection region: $y$-coordinate of  upper end     &$L_{\max}$          & $1050$                    &$  \mu \mbox{m} $                   & estim.\\
\hline
{\bf Geometry 2 (Sect. \ref{subsec:geom})}                &                     &                           && \\
\hline
Domain size in $x$-direction                              &$L_x$               & $2000$                    &$  \mu \mbox{m} $                   & estim.\\
Domain size in $y$-direction                              &$L_y$               & $1000$                    &$  \mu \mbox{m} $                   & estim.\\
Oxygen injection region: $y$-coordinate of lower end      &$L_{\min}$          & $450$                     &$  \mu \mbox{m} $                   & estim.\\
Oxygen injection region: $y$-coordinate of upper end      &$L_{\max}$          & $550$                     &$  \mu \mbox{m} $                   & estim.\\
\hline
{\bf Blood (Sect. \ref{subsec:blood_flow} \& \ref{Subsec:wss})}&               &                           &                                    & \\
\hline
Pressure at high pressure boundary                        &$p_0$               & $37.7$                    &$  \mbox{mmHg} $                    & \cite{williams1988dynamic} \\
Pressure at low pressure boundary                         &$p_1$               & $14.6$                    &$  \mbox{mmHg} $                    & \cite{williams1988dynamic} \\
Dynamic viscosity                                         &$\mu$               & $3.75 \times 10^{-7}$     &$  \mbox{mmHg min} $                & \cite{F2017} \\
\hline
{\bf Oxygen and oxygen dynamics (Sect. \ref{subsec:oxygen})}&                  &                           &                                    & \\
\hline
Concentration at injection boundary                       &$\rho_0$            & $0.025$                   &$  \mu \mbox{m}^{-2} $              & estim. \\
Concentration for linear/nonlinear diffusion shift        &$\widetilde{\rho}$  & $0.1 \times \rho_0$       &$   \mu \mbox{m}^{-2} $             & estim. \\
Maximum consumption rate                                  &$\beta_{\mbox{sat}}$& $ 0.01 \times \rho_0$     &$  \mbox{min}^{-1}\mu \mbox{m}^{-2}$& estim. from \cite{curcio2014kinetics, takahashi1966oxygen} \\
Michaelis constant                                        &$K_m$              & $0.5 \times \rho_0$       &$  \mu \mbox{m}^{-2} $              & estim. from \cite{curcio2014kinetics, takahashi1966oxygen} \\
\hline
{\bf Capillary elements (Sect. \ref{subsubsec_capillary_element} \& \ref{subsec:tissue})}&   &            &                                    & \\
\hline
Length                                                    &$L_c$              & $ 15 $                    &$ \mu \mbox{m} $                    & \cite{garipcan2011image}\\
Width                                                     &$w_c$              & $ 4 $                     &$ \mu \mbox{m}$                     & \cite{garipcan2011image}\\
Hydraulic conductivity                                    &$\kappa$           & $80000 $                  &$ \mu \mbox{m}^2 \mbox{min}^{-1}\mbox{mmHg}^{-1} $& estim.\\
Oxygen diffusivity                                        &$\Delta$           & $200 $                    &$ \mu \mbox{m}^2\mbox{min}^{-1} $& estim. \\
\hline
{\bf Capillary creation: oxygen gradient (Sect. \ref{Subsec:creation_gradient})}&   &                     &                                    & \\
\hline
Maximal creation rate                                     &$\nu_c^*$          & $0.05 $                   & $ \mu \mbox{m}^{-2}\mbox{min}^{-1} $& estim. \\
Oxygen concentration gradient length threshold            &$L_0^c$            & $8 $                      & $ \mu \mbox{m}$                   & estim. \\
Concentration for regularization of logarithmic sensing   &$\rho^*$           & $0.1 \times \rho_0 $      & $  \mu \mbox{m}^{-2}$             & estim. \\
Width of sigmoid: oxygen gradient                     &$h_c$              & $0.1 $                    & $ - $                              & estim. \\
Oxygen concentration threshold                            &$\rho_s$           & $\rho_0$                  & $   \mu \mbox{m}^{-2} $            & estim. \\
Width of sigmoid: oxygen concentration                &$h_s$              & $0.1 $                    & $ - $                              & estim. \\
\hline
{\bf Capillary creation: reinforcement (Sect. \ref{Subsec:reinforcement})}&   &                           &                                    & \\
\hline
Maximal creation rate                                     &$\nu_f^*$          & $0.01$                    &$ \mu \mbox{m}^{-2}\mbox{min}^{-1} $& estim. \\
Blood velocity threshold                                  &$\bar{u}$          & $20$                      &$ \mu \mbox{m } \mbox{min}^{-1} $   & estim. \\
Lower oxygen concentration threshold                      &$\underline{\rho}$ & $0.1 \times \rho_0$       & $ \mu \mbox{m}^{-2} $              & estim.\\
Upper oxygen concentration threshold                      &$\bar{\rho}$       & $0.5 \times \rho_0$       & $ \mu \mbox{m}^{-2} $              & estim.  \\
Width of sigmoids                                     &$h_f$              & $0.1 $                    &$ - $                               & estim. \\
\hline
{\bf Capillary creation: WSS (Sect. \ref{Subsec:wss})}    &                   &                           &                                    & \\
\hline
Maximal creation rate                                     &$\nu_w^*$          & $0.3$                     & $ \mu \mbox{m}^{-2}\mbox{min}^{-1} $ & estim. \\
Width of sigmoid                                      &$h_w$              & $0.1 $                    &$ - $                               & estim. \\
WSS threshold                                             &$\lambda^*$        & $3.75 \times 10^{-8}$     & \mbox{mmHg}                        & estim. from \cite{P+2011} \\
\hline
{\bf Capillary removal (Sect. \ref{Subsec:cap_pruning})}  &                   &                           &                                    & \\
\hline
Removal rate at twice threshold                           &$\nu_r^*$          & $30.0 $                   &$\mbox{min}^{-1}$                 & estim. \\
Hydraulic conductivity threshold                          &$\gamma^*$         & $400000$                  &  $  \mu \mbox{m}^2 \, \mbox{min}^{-1} \mbox{mmHg}^{-1} $  & estim. \\
\hline
{\bf Tissue (Sect. \ref{subsec:tissue})}                  &                   &                           &                                    & \\
\hline
Hydraulic conductivity                                    &$k_h$              & $400 $                    &$ \mu \mbox{m}^2 \mbox{min}^{-1}\mbox{mmHg}^{-1} $& \cite{SF2007} \\
Oxygen diffusivity                                        &$\Delta_h$         & $10 $                     &$ \mu \mbox{m}^2\mbox{min}^{-1} $& \cite{T+2014} \\
\hline
\end{tabular}
\caption{\it Parameters of the model. In case of estimated parameters (``estim.'' in the last column), we refer to the corresponding section (indicated in the first column) for the details of this estimation.}
\label{TableOfParameters}
}
\end{table}


\subsection{Numerical method}
\label{subsec:numerics}

In this section, we provide a brief summary of the numerical methods used and refer to the Appendices \ref{appendix:fem}, \ref{appendix:sph}, \ref{appendix:acceptance} for a detailed description. A synopsis of the numerical treatment of the problem is given in Algorithm \ref{alg:alg1} below and the choices for the numerical parameters are summarized in Table \ref{numericalParameters}. 

The elliptic equation describing blood flow \eqref{eq:bloodPDE1}-\eqref{eq:bloodPDE2} is approximated using a  $\mathcal{Q}_1$ finite element method on a rectangular spatial grid \cite{BS2007} of uniform mesh sizes $\Delta x$ and $\Delta y$ in the horizontal and vertical directions respectively. Finite-element methods are regarded as methods of choice for the resolution of elliptic problems with complex boundary conditions. A $\mathcal{Q}_1$ finite element method provides an approximation of the solution by polynomials linear in each variable on quadrangular numerical cells. We favored $\mathcal{Q}_1$ elements over more classical $\mathcal{P}_1$ elements (which provide approximations by globally linear polynomials on triangular numerical cells) to avoid too much numerical directional bias. For all numerical simulations shown in this paper, otherwise stated, we set $\Delta x = \Delta y = 1.25 \, \mu$m, which resolves the scale of capillary elements. If we take finer mesh sizes we do not note any significant changes in the network formation. 

The diffusion-advection-reaction equation \eqref{O2PDE} describing the evolution of the oxygen concentration $\rho$ is discretized using the Smoothed Particle Hydrodynamics (SPH) methodology \cite{monaghan1992smoothed}. The choice of a SPH method is motivated by the initially zero oxygen concentration. Dealing with regions of zero concentration with classical grid methods, such as finite volume methods, is delicate due to the high risk of appearance of negative values, which often leads to a breakdown of the simulation. The SPH method is not subject to this risk and is the method of choice for problems involving jets or injections of gases in vacuum. It consists of approximating the function $\rho(x,t)$ by a sum of $N$ Dirac deltas located at positions $\{\textbf{Y}_\ell(t)\}_{\ell = 1, \ldots, N}$ and of equal and time-constant masses $m$. Here, since the oxygen concentration scale is irrelevant (see Sect. \ref{subsec:oxygen}), we take $m=1$. These particles move in space with a velocity $\textbf{V}_\ell(t)$ given by formula \eqref{eq:v} evaluated at $\textbf{Y}_\ell(t)$. However since \eqref{eq:v} involves $\nabla_x \rho$, it does not make sense with $\rho$ equal to a sum of Dirac deltas. To make sense of it, we approximate all Dirac deltas by smooth regularizations. Their role is to spread the Dirac deltas over regions of finite extent $\eta$. We take $\eta = 5 \,  \mu$m which spreads the Dirac deltas over $10 \, \mu$m, i.e. approximately the scale of a capillary. The process of representing diffusion of a cloud of particles by their convection along their regularized concentration gradient is known as the diffusion velocity method, introduced in \cite{DM1990}. It stands as an alternative to stochastic methods for the treatment of diffusion which introduces lower noise uncertainty. The regularization kernel is taken to be the poly6 kernel (see Appendix) which has  well documented reliability \cite{M2003}. To account for the consumption term (right-hand side of \eqref{O2PDE}), we use a splitting method. We first solve for the left-hand-side of \eqref{O2PDE} by the SPH method and then we use a stochastic death process to discretize the Michaelis-Menten reaction term. 

The simulations of the spatio-temporal Poisson point processes (\ref{nuc}), (\ref{nuf}), \eqref{eq:wall} and (\ref{nur}) are performed using a classical acceptance-rejection method (see for instance \cite{G+2016}). For each creation process and each discrete time $t$, we pick $N_c$ points $x$ in the domain independently and uniformly and compute the value of the corresponding Poisson parameter $\nu(x,t)$.  We can then estimate the probability of creation of a capillary element by this process. The value $N_c = 10^5$ was selected by trial and error. Larger values of $N_c$ do not improve the quality of the solution but require longer computer time. Lower values of $N_c$ deteriorate the quality of the results. For the removal process, we just loop over all the existing capillaries and compute the value of the removal Poisson rate at their location. Again, this allows us to estimate the probability of removal of the considered capillary. We could have accelerated the treatment of these processes by using Metropolis-Hastings or Monte-carlo Markov Chain methods but this step was so quick compared with the resolution of the elliptic equation that it did not seem to be worth the effort. 

We implemented the code in Fortran 95 and used the Mersenne twister generator \cite{MN1998} for the generation of random numbers. The plots were generated in Gnuplot. The simulations were conducted on two workstations, one with two Intel Xeon E5-2637v3 quad core processors clocked at 3.5 GHz with 12 MB of on-CPU cache and the other one with two Intel Xeon Gold 5122 quad core processors clocked at 3.6 GHz with 16 MB of on-CPU cache. The time taken for each simulation is in average less than 72 hours for Geometry 1 and for Geometry 2 it takes in average 6 days depending on the creation/deletion mechanisms involved. We summarize the choice of numerical parameters in Table \ref{numericalParameters}.

\begin{algorithm}[ht!]
 $\textbf{Input:}$ Final time $T$ and parameters in Tables \ref{TableOfParameters} and \ref{numericalParameters} \\
 $\textbf{Initialisation:}$ For every node $\textbf{X}$ in the mesh of the FEM we set $\textbf{K}(\textbf{X}) = k_h \textbf{I}_2$, $\textbf{D}(\textbf{X}) = \Delta_h \textbf{I}_2$
 and set $t = 0$ \\
 Update the blood pressure $p$ (solve the linear system \eqref{eq:linsys}) \\
 Update the blood velocity $\textbf{u}$ (Eqs. \eqref{centeredFDM} \& \eqref{forwardFDM}) \\
 Oxygen particles injection (see Sect. \ref{sec:artery_boundary}) \\
 $\textbf{While } t \leq T \textbf{ Do}$ \\
 \qquad $\bullet$ Acceptance-rejection sampling method for the capillary element dynamics \\
 \qquad \qquad $\ast$ Add capillary elements (Algorithms \ref{algonuc}, \ref{algonuf} or \ref{algonuwss}) \\
 \qquad \qquad $\ast$ Remove capillary elements (Algorithm \ref{algonur}) \\
 \qquad \qquad $\ast$ Update tensors \textbf{K} \& \textbf{D} on finite element grid (Eqs. \eqref{eq:tissue1} \& \eqref{eq:tissue2}) \\
 \qquad $\bullet$ Finite Element Method for blood flow \\
 \qquad \qquad $\ast$ Update the blood pressure $p$ (solve the linear system \eqref{eq:linsys}) \\
 \qquad \qquad $\ast$ Update the blood velocity $\textbf{u}$ (Eqs. \eqref{centeredFDM} \& \eqref{forwardFDM}) \\
 \qquad $\bullet$ Smoothed Particles Hydrodynamics for oxygen flow \\
 \qquad \qquad $\ast$ First loop over all particles $\ell$ for the SPH method at time $t$ \\  
 \qquad \qquad \qquad $\cdot$ Compute $\textbf{u}_\ell$, $\textbf{D}_\ell$ by interpolation from their grid values\\
 \qquad \qquad \qquad $\cdot$ Compute $\textbf{V}^n_\ell$ by \eqref{eq:O2_particle} \\
 \qquad \qquad \qquad $\cdot$ Compute the time-step $\Delta t^n$ by \eqref{CFL} \\
 \qquad \qquad \qquad $\cdot$ Impose $\Delta t^n \leq 0.01$ for accuracy of Acceptance-Rejection process \\
 \qquad \qquad $\ast$ Second loop over all particles $\ell$ for the SPH method at time $t$ \\
 \qquad \qquad \qquad $\cdot$ Update particle positions from $\textbf{Y}_\ell^{n}$ to $\textbf{Y}_\ell^{n+1}$ through \eqref{updatePosition} \\
 \qquad \qquad \qquad $\cdot$ For $\Omega_2$ apply the periodic boundary conditions (see Sect. \ref{sec:periodi_bc}) \\
 \qquad \qquad \qquad $\cdot$ Apply oxygen consumption (Algorithm \ref{algoO2consumption}) \\
 \qquad \qquad $\ast$ Oxygen particles injection (see Sect. \ref{sec:artery_boundary}) \\
 \qquad \qquad $\ast$ Update time from $t$ to  $t + \Delta t^n$ \\
 $\textbf{End Do}$ 
 \vspace{0.3cm}
 \caption{Flowchart of numerical algorithm}
 \label{alg:alg1}
\end{algorithm}

\begin{table}
\begin{center}
\begin{tabular}{|l|l|l|l|}
\hline
Quantity                                           & Symbol              &   Value            & Unit            \\
\hline
{\bf Finite-element-method for blood flow}         &                     &                    &                 \\
\hline
Mesh size in $x$-direction                         &$\Delta x$           & $1.25$             & $\mu$m          \\
\hline
Mesh size in $y$-direction                         &$\Delta y$           & $1.25$             & $\mu$m          \\
\hline
{\bf SPH particle method for oxygen concentration} &                     &                    &                 \\
\hline
Particle ``mass''                                  &$m$                & $1.0$              & $-$             \\
Smoothing parameter                                &$\eta$               & $5.0$              & $\mu$m          \\
CFL parameter                                      &$C$                  & $0.45$             &$ -$             \\
\hline
{\bf Point Poisson process for capillary creation} &                     &                    &                 \\
\hline
Number of sample points per time step              &$N_c$                & $10^5$             & $-$             \\
\hline
\end{tabular}
\caption{\it Numerical parameters}
\label{numericalParameters}
\end{center}
\end{table}


\section{Results}
\label{sect:results}



\subsection{Simulations for Geometry $\Omega_1$}

In this section we consider Geometry $\Omega_1$ which mimics the branching of a new capillary network from a blood vessel in the plane perpendicular to this blood vessel (see Fig. \ref{fig:bio_geometries}). We explore the influences of the different creation/deletion mechanisms of capillary elements introduced in Sect. \ref{subsec:capillaries} on the network structure. All the parameters used are those given in Tables \ref{TableOfParameters} and \ref{numericalParameters}.


\subsubsection{Including all capillary creation/deletion mechanisms}  
\label{subsubsec:all_capillary}

\begin{figure}
  \begin{center}
    \includegraphics[trim={1.85cm 18cm 10.3cm 1.35cm}, width=1\textwidth, clip]{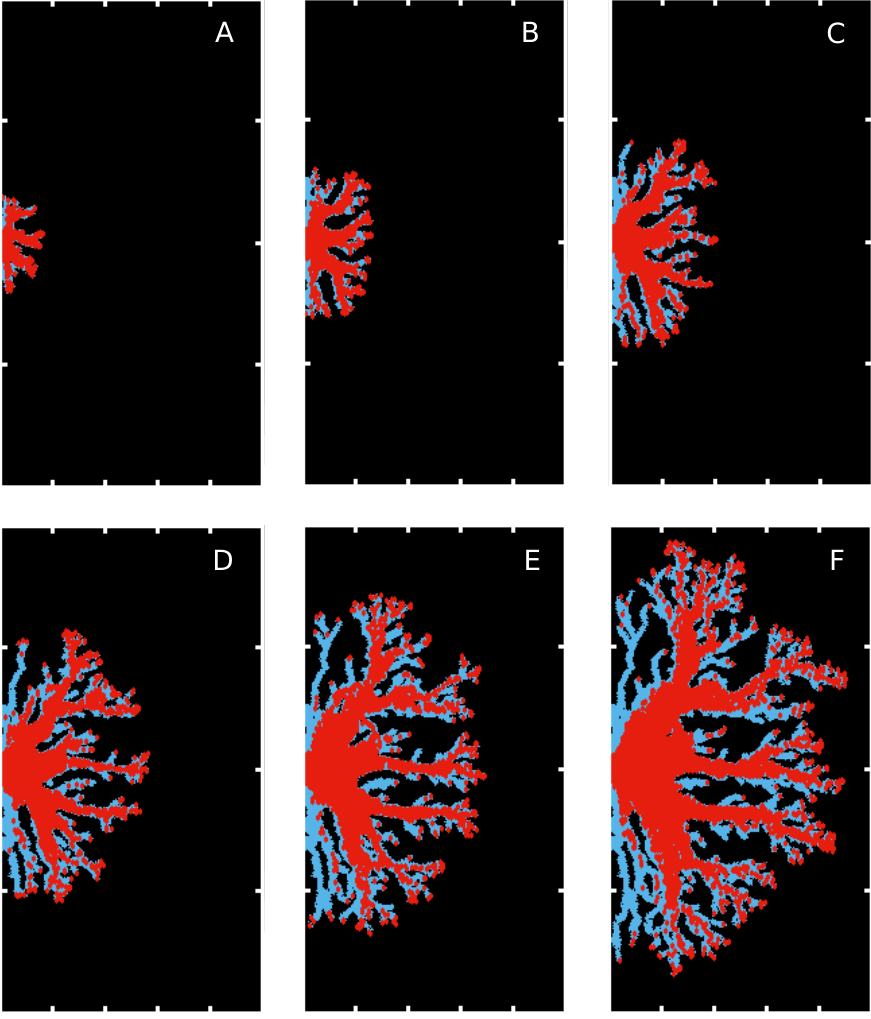}
    \caption{\it Positions of oxygen particles (red spots) and of capillary elements (blue rods) in the rectangular domain 
    $\Omega_1$ for a realization of the model. As red spots overlay the blue rods, capillary elements lying below the red oxygen 
    particles are present although not seen. All the creation/deletion mechanisms of capillary elements are turned 'On'. 
    All the parameter used are those given in Tables \ref{TableOfParameters} and \ref{numericalParameters}. Pictures (A) to (F) 
    give snapshots at increasing times: $2$ min (A), $4$ min (B), $6$ min (C), $8$ min (D), $10$ min (E), $12$ min (F) after initialization.}
\label{fig:all_mech}
  \end{center}
\end{figure}

\begin{figure}
  \begin{center}
    \includegraphics[trim={1.55cm 17.5cm 9cm 0.1cm}, width=1.02\textwidth, clip]{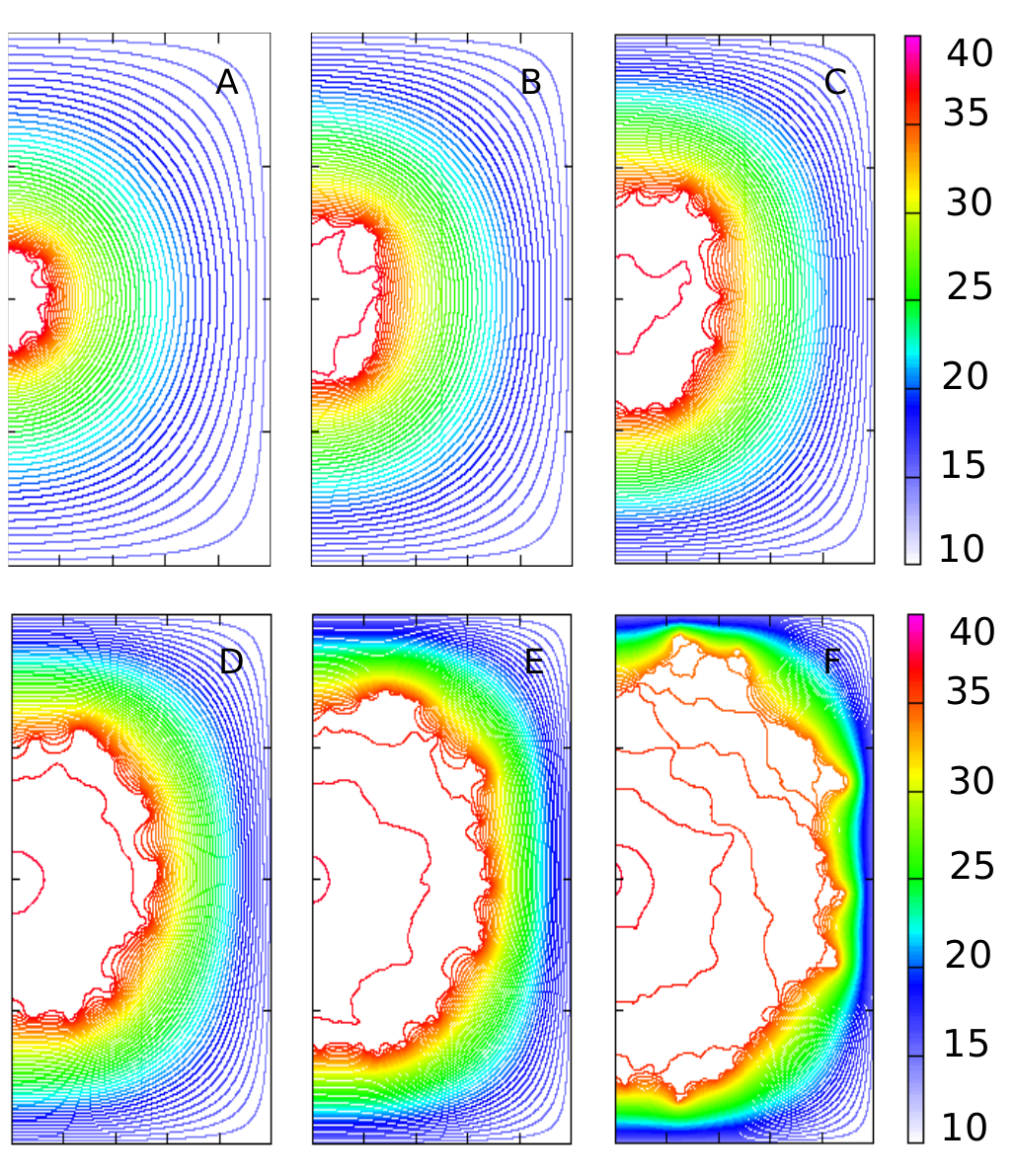}
    \caption{\it Isolines and heatmap of the pressure $p$ in the rectangular domain $\Omega_1$ for the same realization of the model as in Fig. 
    \ref{fig:all_mech}. All the creation/deletion mechanisms of capillary elements are turned 'On'. All the parameters used are those given in 
    Tables \ref{TableOfParameters} and \ref{numericalParameters}. Pictures (A) to (F) give snapshots at increasing times: $2$ min (A), $4$ min (B), 
    $6$ min (C), $8$ min (D), $10$ min (E), $12$ min (F) after initialization. The units are given in mmHg.}
    \label{fig:pressure_plots}
  \end{center}
\end{figure}

\begin{figure}
  \begin{center}
    \includegraphics[trim={2.1cm 16.3cm 8.5cm 1cm}, width=1.02\textwidth, clip]{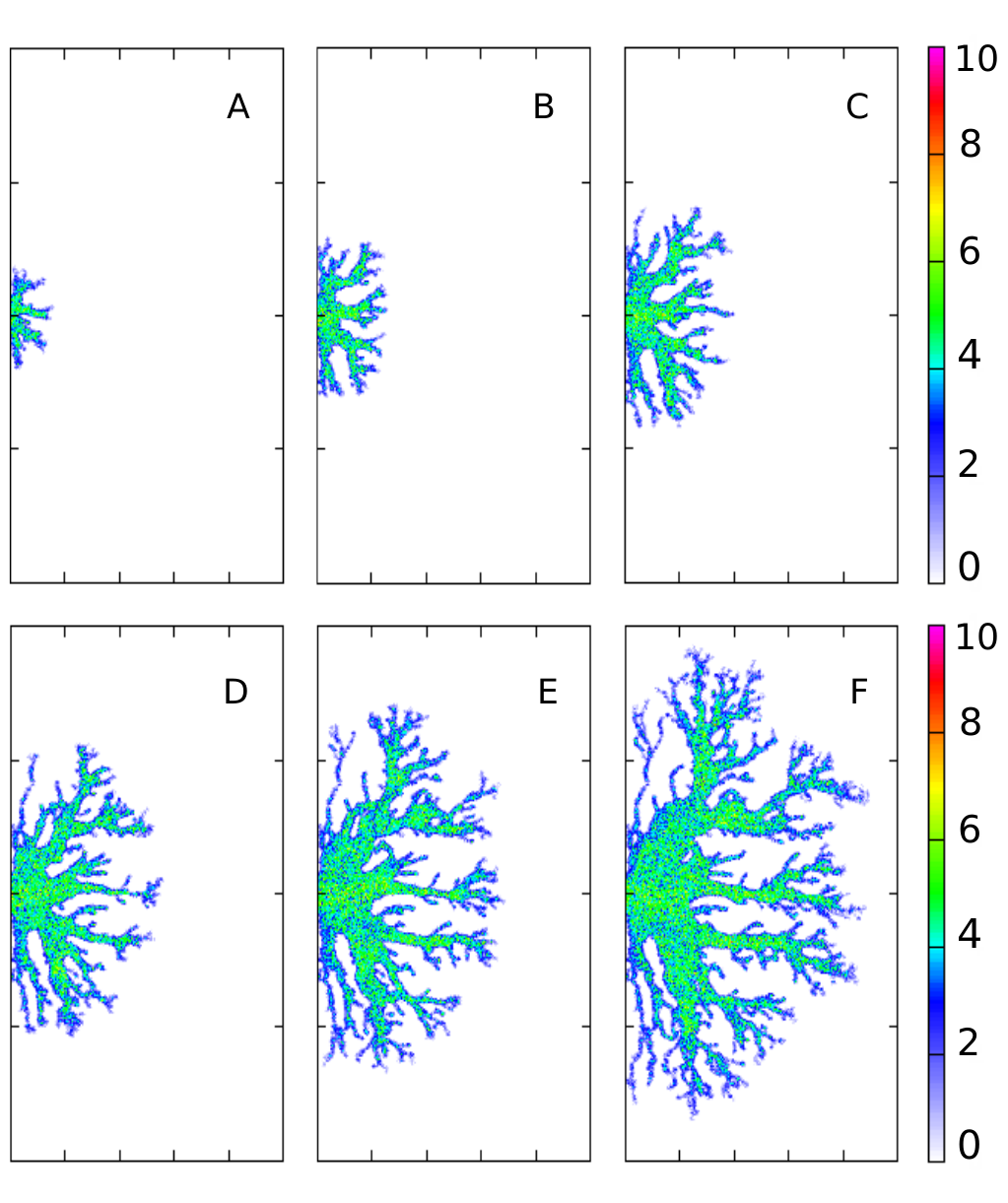}
    \caption{\it Heatmap of the Frobenius norm $\gamma$ of the hydraulic conductivity tensor $\textbf{K}$ in the rectangular domain $\Omega_1$ 
    for the same realization of the model as in Fig. \ref{fig:all_mech}. All the creation/deletion mechanisms of capillary elements are turned 'On'. 
    All the parameter used are those given in Tables \ref{TableOfParameters} and \ref{numericalParameters}. Pictures (A) to (F) give snapshots at 
    increasing times: $2$ min (A), $4$ min (B), $6$ min (C), $8$ min (D), $10$ min (E), $12$ min (F) after initialization. The units are given in
    $10^5$ $\mu$m$^2 / (mmHg \, \, min)$}
    \label{fig:averagedNetwork}
  \end{center}
\end{figure}

We first turn 'On' all the mechanisms described in Sect. \ref{subsec:capillaries} i.e. creation of capillary elements by oxygen gradient, reinforcement by blood flow and by WSS as well as capillary pruning. Since all these processes are of stochastic nature, different realizations of the same model will not give the same result. However, they are qualitatively similar. The result of one given realization is displayed in Fig. \ref{fig:all_mech}. On this figure, the rectangle represents the domain $\Omega_1$ i.e. corresponds to $1$ mm in length in the horizontal direction and $2$ mm in the vertical direction. The positions of the oxygen particles in this domain are represented by red spots and those of the capillary elements by tiny blue rods. As red spots overlay the blue rods, capillary elements lying below the red oxygen particles are present although not seen. For the same realization, the isolines and heatmap of the pressure $p$ in $\Omega_1$ are shown in Fig. \ref{fig:pressure_plots} and a heat map of the Frobenius norm $\gamma$ of the conductivity matrix $\textbf{K}$ is displayed in Fig. \ref{fig:averagedNetwork} (see Section \ref{Subsec:cap_pruning} for the definition of the Frobenius norm). In both figures, the colour code stands from blue (low value of $p$ or $\gamma$) to red (high values) through green and yellow (intermediate values). The six pictures (A) to (F) in these figures show various stages of the evolution of the network, $2$ min (A), $4$ min (B), $6$ min (C), $8$ min (D), $10$ min (E), $12$ min (F) after initialization. The initial condition, not depicted, corresponds to no oxygen and no capillary element at all, i.e. an empty domain for Fig. \ref{fig:all_mech}, regularly spaced concentric pressure isolines emanating from the middle of the left boundary in Fig. \ref{fig:pressure_plots} and a constant value equal to $k_h$ for Fig \ref{fig:averagedNetwork}.

As the pressure drops (see isolines in Fig. \ref{fig:pressure_plots}) between the portion $\Gamma_{1,D}$ of the left boundary (see Fig. \ref{fig:stress} for the nomenclature) representing a blood vessel, and the top, right and bottom boundaries $\Gamma_{2,D} \cup \Gamma_{3,D} \cup \Gamma_{4,D}$ where uptake of blood happens, there is a flow of blood from the former to the latter. Oxygen, which can enter the domain through $\Gamma_{1,D}$, is transported by blood flow. Blood flow and oxygen transport trigger the formation of capillaries through the various creation/deletion mechanisms described in Sect. \ref{subsec:capillaries}. This results in the initiation of a capillary network emanating from $\Gamma_{1,D}$ which gradually develops radially towards the surrounding boundaries $\Gamma_{2,D} \cup \Gamma_{3,D} \cup \Gamma_{4,D}$ as seen in Fig. \ref{fig:all_mech}. Capillaries are made by the aggregation of many capillary elements along distinctive branches and contribute to a dramatic increase of the hydraulic conductivity along these branches as seen on Fig. \ref{fig:averagedNetwork}. In Fig. \ref{fig:all_mech} and \ref{fig:averagedNetwork} we observe spontaneous branch sprouting, the merging of existing branches (forming so called 'anastomoses') and the spontaneous emergence of capillary tortuosity, which are characteristic features of actual vascular networks (see for instance \cite{G+1974} or \cite{M2015}). Fig. \ref{fig:averagedNetwork} which displays the norm of the conductivity matrix $\textbf{K}$ informs us on the density of capillary elements. In particular, the big red spot of oxygen particles in the trunk of the network seen in Fig. \ref{fig:all_mech} overlays a uniform bed of capillary elements (hidden by the red spot) as indicated by the uniform green color in the same region in Fig. \ref{fig:averagedNetwork}. The creation of a capillary network instead of the expansion of a homogeneous cloud of capillary elements is due to an instability which results from a positive feedback loop between the flows of blood and oxygen on the one hand and capillary creation on the other hand. The precise role of each of these mechanisms will be investigated below. 

The pressure seems to remain constant and equal to its boundary value on $\Gamma_{1,D}$ throughout the region covered by the capillary network and which roughly corresponds to its convex hull (see Fig. \ref{fig:pressure_plots}). Indeed, due to the very large hydraulic conductivity along the branches of the network, there is almost no pressure drop between $\Gamma_{1,D}$ and the tip of the branches. Even if the hydraulic conductivity between the branches is much lower, there is no reason for the pressure to drop significantly between these branches. So, all the pressure drop between $\Gamma_{1,D}$ and the outer domain boundary occurs between the boundary of the convex hull of the network (which we will call the 'envelope' of the network) and the outer boundary. As the envelope of the network moves towards the outer boundary, the pressure isolines become closer and closer to each other, indicating a dramatically increasing pressure gradient (see e.g. \ref{fig:pressure_plots} (F)). When a network branch touches the outer boundary (not depicted here), there is a sudden 'short circuit' of the pressure difference and the flow becomes virtually infinite, which does not make biological sense. So, the model currently cannot describe the last phase of the network expansion, when it connects to the outer boundary. Various mechanisms which may take place when the network transitions from an expansion phase to an established phase are being investigated.

The results show that the network evolves over a time scale of the order of $10$ minutes. This is too fast and experimental results from e.g. \cite{K+2011} suggest that the right time-scale would rather be $10$ hours. So, the network dynamics is between $1$ and $2$ orders of magnitude faster in the model than in reality. The issue there is a numerical issue. There is a time-step limitation for the stability of the SPH resolution of the oxygen concentration equation. It requires that a particle does not move more than a smoothing length $\eta$ during one time step. The particle velocity is roughly that of the blood velocity. A slower capillary creation/deletion dynamics would not dramatically affect this velocity, and thus, the stability constraint. So, the time-step would be roughly similar to that used in the current simulations. But the simulation would have to run for longer times, with 10 to 100 times more time-steps than in the current simulations. This would lead to unaffordable simulation times. Resorting to implicit particle methods would be a possible cure. Some implicit particle methods have been proposed in the context of plasma physics~\cite{langdon1983direct} but they are cumbersome and would require to be adapted and validated for the present case. Alternately, one could use the stationary form of the oxygen concentration equation. Indeed, the time-scale for equilibration of the oxygen concentration is very fast compared to the speed of network evolution. Therefore, we could safely assume an adiabatic evolution of the oxygen concentration, where it would instantaneously adjust to the new network structure exactly like the blood flow does in the current version of the model. The issue here is the resolution of a degenerate stationary convection-diffusion-reaction equation where the concentration takes zero values in large parts of the simulation domain. There is no established method for this kind of problem and there are indeed potential issues such as non-uniqueness of the solutions. This is why, in the present state, we have used the SPH method and 'accelerated' the evolution of the network to make solutions computable.

\subsubsection{Influence of individual capillary creation/deletion mechanisms}  
\label{subsubsec_influence}

\begin{figure}
  \begin{center}
    \includegraphics[trim={2cm 7cm 6cm 5cm}, width=0.94\textwidth, clip]{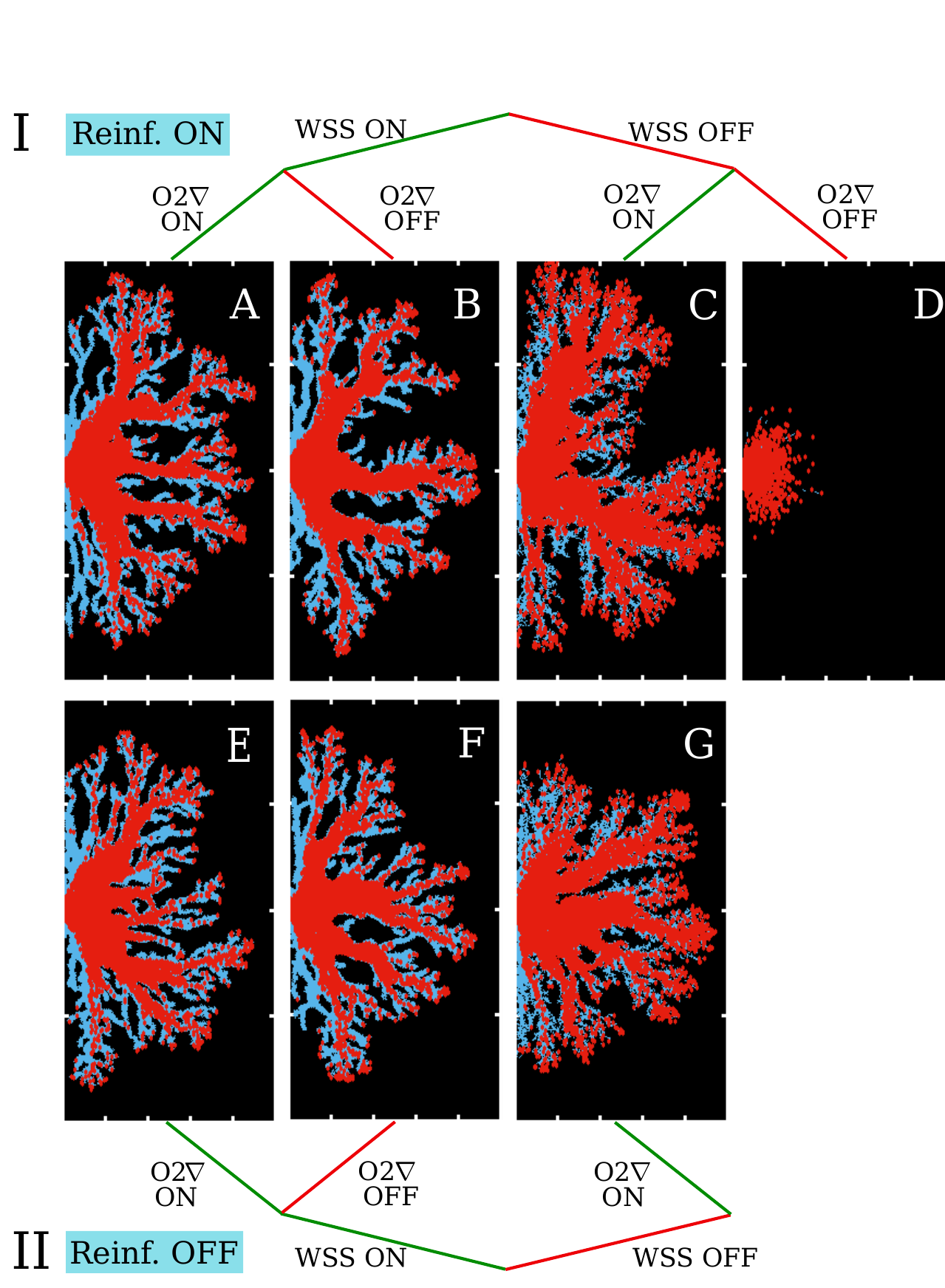}
    \caption{\it Two binary decision trees, placed back-to-back. The upper one (I) includes capillary creation by reinforcement while the lower one (II) excludes it. Each tree successively includes or excludes  capillary creation by WSS and oxygen gradient (noted $O2\nabla$). At the end of each branch, a typical realization of the model with corresponding inclusion/exclusion of  the mechanism is shown. The picture shows the positions of the oxygen particles (red spots) and those of the capillary elements (tiny blue rods). The times for each of the snapshots are the following: 12 min (A), 12 min (B), 19.5 min (C), 30 min (D), 12 min (E), 12 min (F) and 19.5 min (G), after initialization.}
\label{fig:main_plot}
  \end{center}
\end{figure}

In this section, we attempt to analyze the roles of individual creation/deletion mechanisms. In Fig. \ref{fig:main_plot} we display a set of simulations where the various creation mechanisms are turned 'On' or 'Off'. When a mechanism is turned 'Off', this simply means that the corresponding Poisson parameter $\nu^*$ is set to zero, while when it is turned 'On', $\nu^*$ is set to the value given in Table \ref{TableOfParameters}. In  Fig. \ref{fig:main_plot} we have kept the deletion (capillary pruning) mechanism (see Sect. \ref{Subsec:cap_pruning}) 'On'. If we turn this  mechanism 'Off' (not shown in the text but a video of this situation is provided in the supplement), the magnitude of (the Frobenius norm of) the conductivity tensor $\mathbf{K}$ reaches higher values in the middle of the branches, without this having a perceivable influence on the network morphology itself, except for the dynamics being slightly quicker (by about $10 \, \%$) than if the pruning mechanism is turned 'On'.

Fig. \ref{fig:main_plot} takes the form of two binary trees placed back-to-back, the upper one (I) corresponding to the reinforcement mechanism turned 'On' and the lower one (II) to that mechanism turned 'Off'. Within these trees, each branching corresponds to the choice of one of the two other mechanisms (creation by oxygen gradient or by WSS) being 'On' or 'Off'. The first branching (the topmost one for (I) and bottommost one for (II)) corresponds to choosing whether creation by WSS is 'On' (this is depicted by a green edge in the tree) or 'Off' (red edge). Then the second branching (the bottommost one for (I) and the topmost one for (II)) chooses whether creation by oxygen gradient (referred to as '$O2\nabla$' on the picture) is 'On' (greed edge) or 'Off' (red edge). Each leaf of the tree is a typical result of the model in the situation corresponding to the various mechanisms being turned 'On' or 'Off' according to the path followed in the tree. Like in Fig. \ref{fig:all_mech}, the picture shows the positions of the oxygen particles (red spots) and those of the capillary elements (tiny blue rods). We refer to the previous section for a detailed description.

First we examine what happens when only one of the creation mechanisms is turned 'On', the other ones being turned 'Off'. 
\begin{itemize}
\item {\bf Creation of capillary elements by oxygen gradien 'On', the other mechanisms 'Off':} this corresponds to Fig. \ref{fig:main_plot} (G). We notice a fully developed network. So, creation by Oxygen gradient alone provides enough positive feedback to trigger an instability. Let us describe a possible mechanism for this. First, due to the large hydraulic conductivity in the branches, the blood flow velocity there is large. Neglecting oxygen diffusivity, consumption and time variation of the oxygen concentration which are believed to play minor roles here, Eq. \eqref{O2PDE} roughly reduces to $u \cdot \nabla_x \rho = 0$, showing that oxygen concentration is nearly constant in the branch and equal to that in the inflow boundary $\Gamma_{1,D}$, i.e. $\rho_0$. Ahead of the branch tip, the hydraulic conductivity drops, and so does the blood velocity and consequently the oxygen concentration, which cannot be transported by the blood flow any more. So, the creation mechanisms detects a large oxygen gradient at this place, oriented in the direction along the branch. Therefore, it triggers the creation of a new capillary element ahead of the branch tip in the direction of the branch, thereby increasing its length. However, there are also oxygen gradients across the boundaries of the branches. This results in capillary creation across the branches and contributes to branch widening and the initiation of many new small branches across the main branch. Indeed, we can see from Fig. \ref{fig:main_plot} (G) that the trunk and main branches of the network are thick and that branches terminate in a sponge-like structure. Several anastomoses can also be seen. It takes about $20$ min to observe a fully developed network. 
\item {\bf Creation of capillary elements by WSS 'On', the other mechanisms 'Off':} this corresponds to Fig. \ref{fig:main_plot} (F). Again, the WSS mechanism alone is able to generate a full network. Here, the mechanism of network formation is a bit more mysterious as WSS creates capillaries normal to the main blood flow. So, a new capillary created at a branch tip will pursue the branch in the direction normal to the existing branch. However, the next capillary will be created in the direction normal to the direction of the previous capillary and thus again in the direction of the main branch. In this way the branch advances a bit like a sailing boat forced to tack in headwind. Indeed, the observation of the videos shows that the motion of the branch tip is undulatory while it is straighter in the previous case. Behind the tip, the branch consolidates into larger and straighter channels than in the previous case. The branch width is very large at the trunk and decreases quickly in the secondary branches. Obviously WSS also favors sprouting but most of these branches are not pursued because they are not used by blood flow which finds an easier way towards the outer boundary through the main branches. The formation of the network is faster than in the previous case, with a fully developed network after only $12$ min. 
\item {\bf Creation of capillary elements by reinforcement 'On', the other mechanisms 'Off':} this corresponds to Fig. \ref{fig:main_plot} (D). Here, we cannot see much of network formation. Rather, we observe the expansion of oxygen and capillary elements in a diffusive way from the inflow boundary $\Gamma_{1,D}$. The reinforcement mechanism does not carry enough positive feedback to trigger a significant instability at least within the $30$ min time of the simulation (compared with the $12$ min which are sufficient for a fully developed network with WSS). 
\end{itemize}

From this analysis, we can infer that the mechanisms that influence the network structure the most are the creation by oxygen gradient and by WSS. The creation by reinforcement seems to play a minor role. Let us examine this question more closely by comparing what happens when we turn the reinforcement 'On' and 'Off' with all other parameters unchanged. This corresponds to comparing homologous pictures in the upper (I) and lower (II) binary trees of Fig. \ref{fig:main_plot}, i.e. (A) with (E), (B) with (F) and (C) with (G) (we leave (D) aside because in this case, if we turn reinforcement 'Off', there is no capillary creation mechanism at all and nothing happens). 

\begin{itemize}
\item {\bf Creation of capillary elements by WSS and oxygen gradients both 'On'; comparison between reinforcement mechanism 'On' and 'Off':} this corresponds to comparing Figs. \ref{fig:main_plot} (A) and (E). Fig. \ref{fig:main_plot} (A) is the same as Fig. \ref{fig:all_mech} (F) but is reproduced again for the sake of comparison. The network structures in the two plots are fairly different. While in the presence of reinforcement (Fig. \ref{fig:main_plot} (A)) there are thick branches with significant sprouting near their tip, in the absence of it (Fig. \ref{fig:main_plot} (E)) branches are thinner and there is much less sprouting at the tip. In the latter case, there are more thin branches directly branching off the trunk of the network. Also, the network seems less developed in the absence of reinforcement, suggesting that network formation is slightly slower. The reinforcement mechanism consolidates small branches thereby giving them more chances to grow. When reinforcement is absent, branching looks more difficult. 

\item {\bf Creation of capillary elements by WSS 'On' and oxygen gradient 'Off'; comparison between reinforcement mechanism 'On' and 'Off':} this corresponds to comparing  Figs. \ref{fig:main_plot} (B) and (F). Roughly speaking the same comparisons as in the previous case can be made here, although the difference is less striking.

\item {\bf Creation of capillary elements by oxygen gradient 'On' and WSS 'Off'; comparison between reinforcement mechanism 'On' and 'Off':} this corresponds to comparing Figs. \ref{fig:main_plot} (C) and (G). Here the two network structures are qualitatively almost indistinguishable, except for the extension of the network, which is larger when reinforcement is on, than when it is off. 
\end{itemize}

So reinforcement seems to have little yet perceivable influence on the network structure. In general it gives slightly more regular shape of branches so, we are going to keep it in the next round of comparisons, thereby discarding the lower tree (II) of Figs. \ref{fig:main_plot}. Assuming that we include reinforcement  and looking only at the upper tree (I), we are going to compare the cases where both WSS and oxygen gradient creation are 'On' to the cases of either one or the other is 'On', i.e. compare Figs. \ref{fig:main_plot} (A), (B) and (C) with one another

\begin{itemize}
\item {\bf Comparison between creation of capillary elements by WSS and oxygen gradient both 'On' to cases where either one is 'On' and the other is 'Off' (creation by reinforcement 'On'):} this corresponds to comparing Figs. \ref{fig:main_plot} (A), (B) and (C). We can readily single out Fig. \ref{fig:main_plot} (C) (where WSS is absent) for the sponge-like structure of the network already noticed above. The WSS mechanism present in Figs. \ref{fig:main_plot} (A) and (B) provides a neater network with a well defined branching structure. This may be related to the way the branch progresses by creating a wider capillary path due to the undulatory motion of its tip as discussed above. On the other hand, in terms of homogeneity of the oxygen perfusion, the structure shown in Fig. \ref{fig:main_plot} (C) seems to be more satisfactory as a wider surface of the tissue seems to have access to oxygen. On the other hand, Figs. \ref{fig:main_plot} (A) and (B) look quite similar. A closer inspection shows that less branches are generated and the generated ones are thicker in the absence of the oxygen gradient mechanism (Fig. \ref{fig:main_plot} (B)) than when it is present (Fig. \ref{fig:main_plot} (A)). Overall, when  all mechanisms are present, the network seems to be better balanced, with a gradual decrease of the thickness of the branches when going from the trunk to the periphery. For this reason, we conclude that all mechanisms seem to concur to the formation of a harmonious network. 
\end{itemize}


\subsubsection{Robustness of the simulations with respect to mesh size}  

\begin{figure}
  \begin{center}
    \includegraphics[trim={1.5cm 19.5cm 1cm 1.8cm}, width=1.04\textwidth, clip]{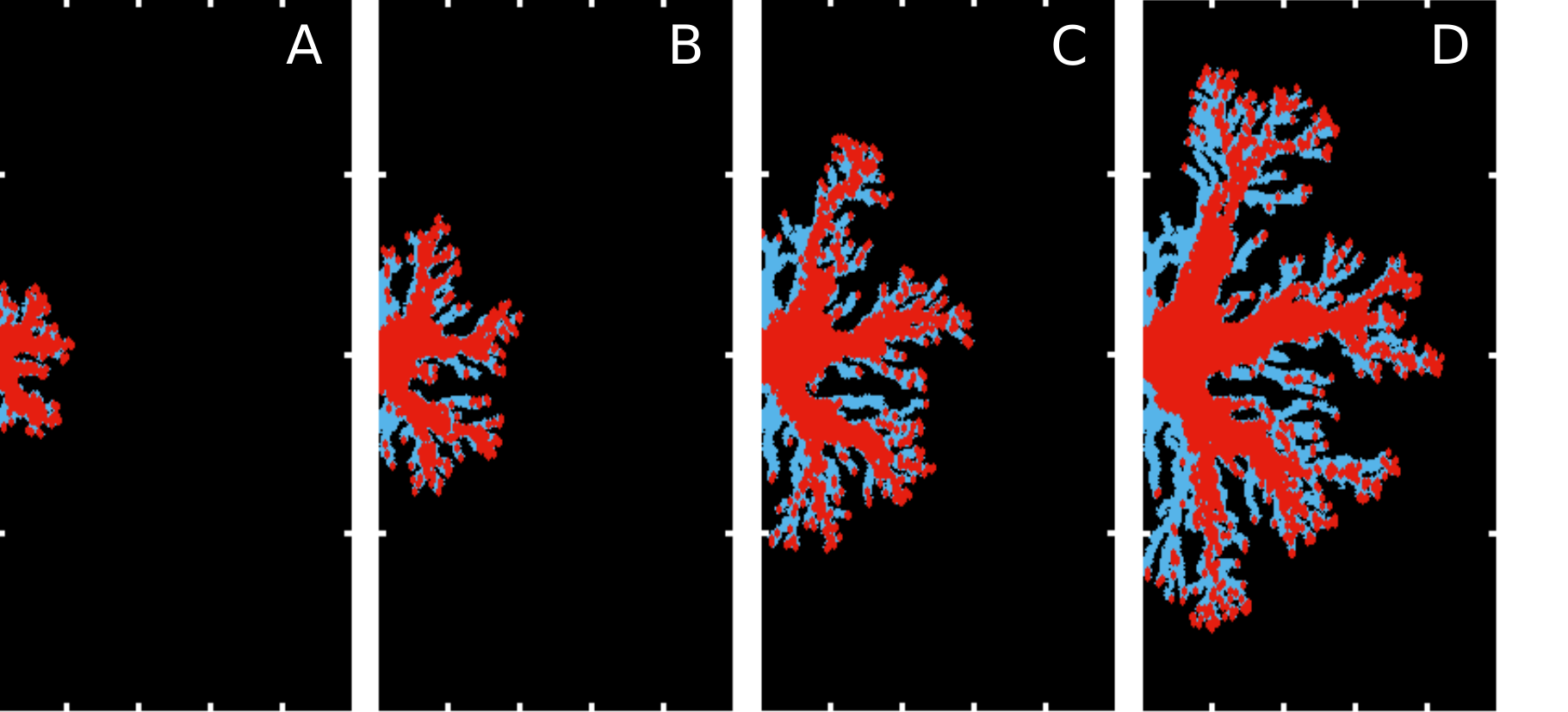}
    \caption{\it Positions of the oxygen particles (red spots) and of the capillary elements (tiny blue rods) in the domain $\Omega_1$ (see caption of Fig. \ref{fig:all_mech} for details) for a realization with  mesh-size $\Delta x = \Delta y = 5/8$, the other parameters in Tables \ref{TableOfParameters} and \ref{numericalParameters} being unchanged. Pictures (A) to (D) give snapshots at increasing times: $2.5$ min (A), $5$ min (B), $7.5$ min (C), $10$~min~(D) after initialization.}
    \label{fig:octa_mesh}
  \end{center}
\end{figure}

We have realized that too coarse a mesh-size for the finite-element method may impair the quality of the results. For all the simulations shown so far, we took $\Delta x = \Delta y = 1.25 \, \mu$m as mesh-size  (see Table \ref{numericalParameters}). This mesh-size resolves the scale of each capillary (tubes of length $15 \, \mu $m and width $4 \, \mu$m). By taking smaller mesh sizes we did not notice any significant difference. Coarser mesh sizes on the other hand gave significant differences which we attributed to a bad spatial resolution. In Fig. \ref{fig:octa_mesh} 
we display the results of a realization taking mesh size $\Delta x = \Delta y= 0.625$, i.e. half the size of the previous simulations, with all other parameters in Tables \ref{TableOfParameters} and \ref{numericalParameters} being the same as before. Figs. \ref{fig:octa_mesh} (D) is to be compared with Fig. \ref{fig:all_mech} (E) roughly corresponding to the same time. We do observe some differences due to stochastic nature of the model, which never provides exactly the same results twice. But qualitatively, the features of the network are comparable, with a gradual decrease of the thickness of the branches from trunk to tip. Hence, we concluded that the mesh size $\Delta x = \Delta y = 1.25 \, \mu$m  is a good compromise to resolve the small scales without excessive computer time.

\subsection{Simulations for geometry $\Omega_2$}

We now consider Geometry $\Omega_2$ which represents the branching of a new capillary network from a blood vessel in the plane parallel to itself (see Fig. \ref{fig:bio_geometries}). It is also relevant for experimental situations such as \cite{grogan2018importance, MKA1982}. We refer to Sect. \ref{subsec:geom} for the description of the geometrical setting and the boundary conditions. 

\begin{figure}
  \begin{center}
    \includegraphics[trim={1.1cm, 18.85cm, 9cm, 1cm}, width=1\textwidth, clip]{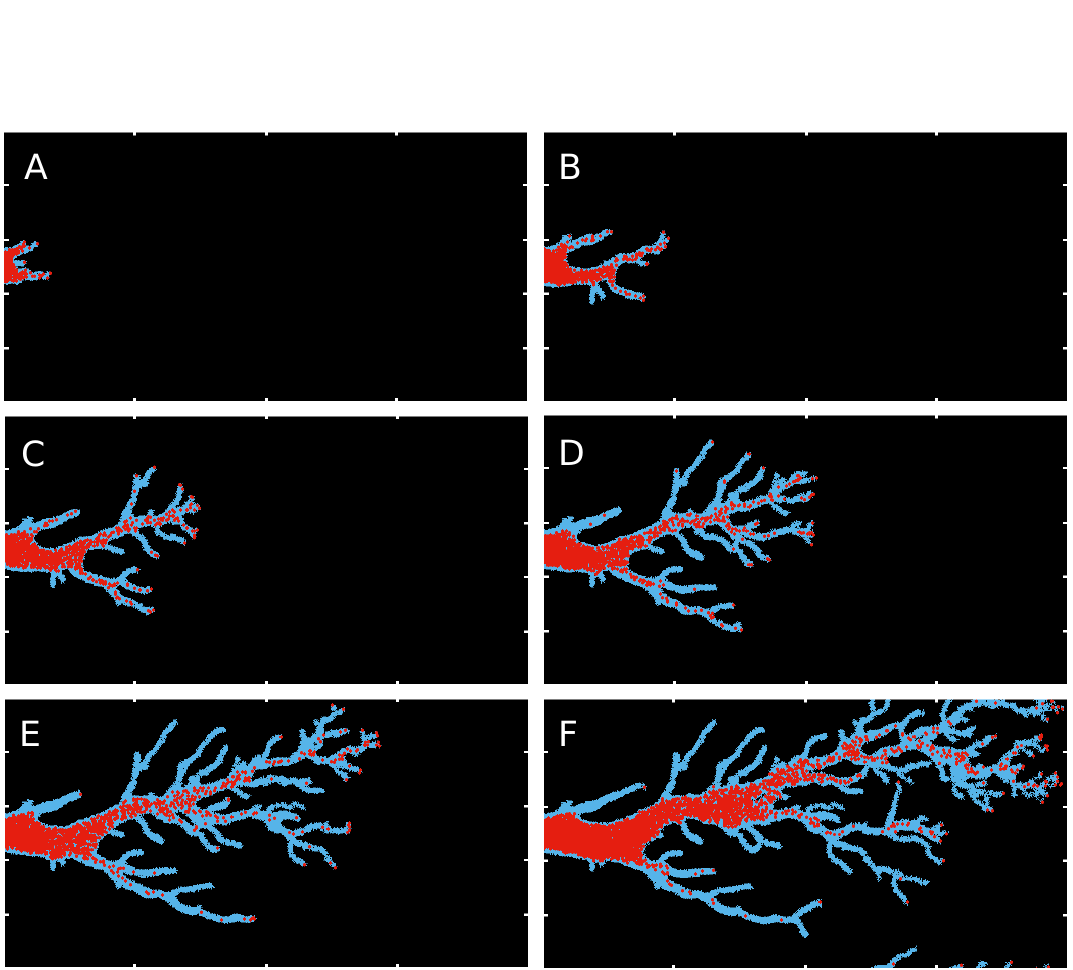}
    \caption{\it 
Positions of the oxygen particles (red spots) and of the capillary elements (tiny blue rods) in the domain $\Omega_2$ (see caption of Fig. \ref{fig:all_mech} for details) for a realization of the model. All the creation/deletion mechanisms of capillary elements are turned 'On'. All the parameters used are those given in Tables \ref{TableOfParameters} and \ref{numericalParameters}. Pictures (A) to (F) give snapshots at increasing times: $6.8$ min (A), $13.6$ min (B), $20.4$ min (C), $27.2$ min (D), $34$ min (E), $40.8$ min (F) after initialization.}
\label{fig:all_mech_geom2}
  \end{center}
\end{figure}

A realization of the model with all capillary element creation/deletion mechanisms turned 'On' and with parameters given in Tables \ref{TableOfParameters} and \ref{numericalParameters} is shown in Fig. \ref{fig:all_mech_geom2}. The graphics are the same as in Fig. \ref{fig:all_mech}, i.e.: (i) the rectangle represents the domain $\Omega_2$ which is $2$ mm long in the horizontal direction and $1$ mm long in the vertical direction and (ii) the red spots are the positions of the oxygen particles and the tiny blue rods depict the capillary elements. Again, the red spots overlay the blue rods, so there are capillary elements below the red oxygen dots even if they are not apparent. We remind that the boundary conditions along the horizontal boundaries are periodic. Fig. \ref{fig:all_mech_geom2} shows six different snapshots of the network evolution, $6.8$ min (A), $13.6$ min (B), $20.4$ min (C), $27.2$ min (D), $34$ min (E), $46.8$ min (F) after the initial time. The same instability as in geometry $\Omega_1$ takes place and contributes to initiate the capillary network from the injection boundary $\Gamma_{1,D}^{\mbox{\scriptsize mid}}$. Because of the presence of the WSS capillary creation mechanisms, the network branches are very distinct and of fairly constant width between two bifurcations. There is a gradual decrease of the branch width from trunk to branch tip as already observed in Fig. \ref{fig:all_mech}. There are several anastomoses. The periodic boundary conditions along the horizontal boundaries prevent the generation of a pressure gradient in the vertical direction. The blood flow is predominantly horizontal, from left to right and does not provide any drive for the development of network branches in the vertical direction. Therefore, most of the network branches develop horizontally. We also observe on Fig. \ref{fig:all_mech_geom2} (F) that the early made branches (those bifurcating away from the trunk close to the injection region) do not seem to be used by oxygen particles. This is because most of the blood flows along the central trunk due to its high hydraulic conductivity until it reaches the tip of the trunk. There, the pressure gradient is large because of the short gap between the tip and the right-hand outflow boundary. It outcompetes the low hydraulic conductivity in the gap and produces a large blood flow resulting in a large perfusion in the whole trunk. This large blood flow drags the oxygen particles away from the side branches which, after some time, become depopulated. In actual tissues, unused capillaries are pruned after some time. Therefore, the long term structure of the network would gradually become reduced to the mere trunk itself and would be close to the structures shown in \cite{grogan2018importance} or in \cite{MKA1982, PP1991,PK1989}. We note that the time scale is of the order of one hour to see a fully developed network, and is too fast by at least one order of magnitude compared to actual observations. We refer to Sect. \ref{subsubsec:all_capillary} for the rationale behind this discrepancy. 

\section{Discussion}
\label{sect:discussion}

In this model, the emergence of blood capillary networks relies on the positive feedback between blood circulation and oxygen transport on the one hand and capillary formation on the other hand. It is the first of this kind which gives a central role to the circulation of blood. As new capillary branches are formed, they elongate or are reinforced as a consequence of this positive feedback. Both the network topology (i.e. its connectivity) and its geometry (e.g. the branch widths, the branching angles, etc.) are emerging properties, i.e. they are not directly encoded in the capillary creation rules but rather emerge from the interactions between the various entities of the model. This is made possible by the main originality of the model: the description of the network through elementary entities, the capillary elements, which are not required to connect to one another. Rather, the connectivity is recovered when the characteristics of the tissue, its hydraulic conductivity, is computed through the summation of the individual contributions of each capillary element. In this way, the capillary elements are free to appear in the adequate locations and directions in response to  environmental cues, such as the local oxygen gradient or shear stress tensor. This flexible approach, which does not require to keep track of the network connectivity has first been proposed in the context of ant trail formation in \cite{BDM2013}.

There is a vast literature about angiogenesis and it is not feasible to compare our model with all previous approaches. However, most of them focus on the migration of EC as the main driving mechanism of capillary network formation (see e.g. \cite{balding1985mathematical, BJJ2009, byrne1995mathematical, DM2013, MC+2006, OAMB2009, pillay2017modeling, T+2014, T+2011} and the review \cite{SBP2013}). More importantly, most of them let the tip cell play a special role. The mechanism is that an EC, in response to a chemical signal emitted by a tumour or a hypoxic region, starts to move upwards the gradient of this chemical. Doing so, it lays down a chemical trail like a slime trail left behind by a snail (hence the nickname 'snail-trail model' for this kind of models). This chemical trail is then followed by other EC that form the stem of the blood vessel. An early model of this type is \cite{balding1985mathematical},  further elaborated in \cite{byrne1995mathematical, pillay2017modeling}. In the multidimensional cell-based versions of this model the rules for branching and creation of a new stem are often postulated with characteristics borrowed from the observations. Hence, the resulting patterns cannot be qualified as emergent in that they are directly resulting from the rules imposed to the agents. Importantly, our model does not give any special role to the tip capillary element. All capillary elements obey the same rules and this is enough to produce a blood vessel. To this extent our model is more parsimonious than the snail trail model. Likewise, there are no special instructions to tell the branches to bifurcate and how they should bifurcate. The resulting patterns emerge in a non-direct way from the individual rules. Finally while in most previously cited models, the driving force of network development is cell motion upwards a chemotactic gradient, in ours, the leading role is taken by the blood flow. Likely, both phenomena are important and the two types of mechanisms should be combined to accurately describe the biological phenomenology. Indeed, there is evidence of both EC motion during angiogenesis and sprouting of new blood vessels through WSS generated by blood flow.

Another viewpoint developed in the literature is that of the optimality of the network with respect to some cost functional. This approach dates back to \cite{murray1926physiological}. Recently, inspired by this approach and the discrete network models of \cite{hu2013adaptation}, a series of papers derive and analyse continuum models of network formation \cite{A+2017, haskovec2015mathematical, H+2016}. These are PDE models for a vector quantity akin to a magnetization in ferromagnetism and which encompasses both the density and orientation of the network. This quantity co-evolves with the hydraulic conductivity of the porous medium generating a positive feedback loop in a very similar fashion as in our model. There are however two important features that are needed to generate a network in these models: network diffusion on the one hand, and more importantly a nonlinear loss term interpreted as the metabolic cost of maintenance of the network on the other hand. In particular it is shown in \cite{H+2016} that there is no network generation unless a specific power law range is used for this nonlinear loss term. Our model does not involve any network diffusion (however, one could view the noise carried by the stochastic processes as a proxy for network diffusion). As reported at the beginning of Sect. \ref{subsubsec_influence}, it is insensitive to the presence of a nonlinear capillary pruning process, itself being analog to the metabolic cost term of \cite{H+2016}. It is not surprising that continuum models do not generate instabilities in situations where microscopic models do (see another example in \cite{BDM2013}). Noise involved in particle models is a powerful instability trigger which is absent from continuum models. This also brings some questions about the validity of the continuum models, although some rigorous results have been proved \cite{haskovec2018rigorous}.

An analogy worth being made is with Diffusion-Limited Aggregation (DLA) models \cite{amitrano1986growth, hastings1998laplacian, herrmann1986geometrical} which have been used to model electrical breakdown \cite{pietronero1984stochastic}. In the DLA process, a network is triggered by the large gradient of a potential function satisfying Laplace's equation and feedbacks on this potential by changing the boundary conditions. It could be seen as related to our model through a singular limit in which the capillary element conductivity would be infinite. Our model bears also analogies with canalization models in leaf venation processes (see \cite{mitchison1980model, mitchison1981polar} and the review \cite{rolland2005reviewing}). These models describe how auxin transport (a hormone essential for plant development) shapes the vein network of leaves. Similarly to our model, they are based on a diffusion equation in which the diffusion coefficient evolves self-consistently with the flux of auxin. This generates a positive feedback triggering an instability. In another domain, geosciences, models of landscape evolution are based on a similar feedback between water flux and soil elevation \cite{chen2014equations, chen2014landscape} although in this case, soil elevation feedbacks into the advection part of the water height equation rather than in its diffusion part.

Obviously there are many directions by which the model could be improved to better account for biological observations. We have already commented on the need to include growth factors such as VEGF and other signaling chemical species as well as to allow for capillary element mobility and chemiotactic sensing ability. This would require the addition of another convection-diffusion-reaction equations describing the transport of these chemicals and its coupling with the capillary creation processes. Another aspect is active migration of EC from existing vessel walls to form new blood vessels. Currently capillary elements appear from nowhere and are immobile. In reality, EC are recruited among those lining the existing blood vessels and move in procession (the snail trail) to form new blood vessels. This active migration could be easily added to the model. In this motion EC many be impeached or guided by the structure of the ECM, a process named haptotaxis. In doing so, EC may also contribute to remodel the ECM (see \cite{hillen2006m, MC+2006, painter2009modelling} for models of haptotaxis). In \cite{P+2017}, the self-organization of ECM and functional cells is modeled using a strategy similar to that presented here and the two models could be coupled to offer a comprehensive description of capillary formation in a developing tissue. The restriction to two-dimensions should be removed as biological reality is fully three-dimensional. Here, the real issue is computational efficiency and fast resolution methods should be sought. The time-scale problem already pointed out in Sect. \ref{sect:results} must also be addressed, by e.g. directly looking for a stationary solution of the oxygen transport equation rather than using its time-dependent version as in the current model. This requires the development of robust stationary solvers able to deal with regions of zero oxygen density. One can also dispute the validity of Darcy's law in the tissue. One improvement would be to consider Brinkman's equation, i.e. adding the influence of blood viscosity. This would allow for a larger span of possible boundary conditions for the blood velocity and the possible emergence of more complex flow patterns. An extra step in increasing complexity would be distinguishing between blood in the capillaries and the interstitial fluid outside. This is needed if one wants to take specific account of oxygen transport across the capillary walls \cite{penta2015multiscale}. More generally, tissues have poroelastic character (i.e. there are interactions between tissue deformations and blood flow) which might be interesting to take into account.

Moving towards identifying which structures are biologically relevant, there is a need for network shape quantification. One possibility would be to skeletonize the capillary network and extract quantifiers such as statistics of branch lengths, branch widths and branching angles at junctions. Other shape quantifiers could involve the surface ratio between the network and its convex envelope for instance. Once a set of relevant quantifiers are selected, a parametric exploration of the model could be attempted. However this would first require significant algorithmic improvements as discussed below. Due to the large dimension of parameter space, sensitivity analysis would be needed. After these prerequisites, it would be possible to compare the model with biological images on which the same quantifiers would have been calculated. In doing this model calibration, the use of recent machine-learning techniques may prove necessary. Of course, these comparisons will be limited by the two-dimension character of the model, which makes the development of a three-dimensional model all the more necessary. Among algorithmic improvements needed for the model to be of practical use, the first one is the development of a faster elliptic solver. Due to the simple domain shape, a Fast Fourier Transform (FFT) solver could be an option, provided attention is paid to the aliasing problem in view of the large spatial inhomogeneity of the problem \cite{bardos2015stability}. A possible remedy to that large inhomogenity could be multi-scale elliptic methods \cite{efendiev2009multiscale}. The acceptance/rejection for the capillary creation/deletion process could be improved by the use of Markov Chain Monte Carlo methods such as the Metropolis-Hastings algorithm. As pointed out above, the resolution of stationary oxygen transport would be necessary to recover time-scales compatible with observations. Such algorithmic improvements will not resolve the computational complexity bottleneck facing simulations of large tissues or organs. Here a paradigmatic shift is necessary and consists of resorting to a continuum model for the capillary network as well. A rigorous passage from the discrete model to the continuum one is however quite involved (see e.g. \cite{haskovec2018rigorous}) and faces conceptual problems, such as the greater difficulty to generate networks with a continuum model as discussed above. Without solving these difficulties, any phenomenological continuum model of capillary network would be subject to caution.  

As summarized here this new network formation model opens many different exciting research avenues. It offers a new paradigm for capillary network creation by placing the flow of blood at the central place in the process. This paper provides a proof of concept of this approach and elaborates a road map by which the model can be gradually improved towards a fully fledged simulator of blood capillary network formation. Such simulator would have huge potential for biological or clinical applications in cancer, wound healing, tissue engineering and regeneration. Besides biological or clinical sciences applications the approach could also be adapted to plant biology (for leaf venation or root formation), physics (lightnings of thunder) or engineering (dielectric breakdown).


\begin{appendices}

\section{Finite Element Method for blood flow}
\label{appendix:fem}

\subsection{Weak formulation and approximation}\label{sec:weak_for}
Let us first consider the elliptic problem \eqref{eq:elliptic} posed on $\Omega_1$ with boundary conditions \eqref{eq:dir1}, \eqref{eq:dir2}, \eqref{eq:dir3}. Let $\mathcal{T}^h$ be a decomposition of the domain $\Omega$ into non-overlapping identical rectangle elements with mesh size $\Delta x$ and $\Delta y$ in the $x$ and $y$ directions and let $h = \max\{\Delta x, \Delta y\}$. Let $\mathcal{V}_h$ be the conforming finite element space associated with the partition $\mathcal{T}^h$, i.e. $\mathcal{V}^h := \{ v \in \mathcal{C} ( \bar \Omega) : v|_{\mathcal R} \in \mathcal{Q}_1 \text{ for all } \mathcal{R} \in \mathcal{T}^h \}$, where $\mathcal{Q}_1$ denotes the space of polynomials of degree $1$ in each direction in ${\mathcal R}$ and ${\mathcal C} ( \bar \Omega)$ is the space of continuous functions defined on the closure $\bar \Omega$ of $\Omega$. Introduce the decomposition of the boundary $\partial \Omega = \Gamma_{N} \cup \Gamma_{D}$ with  $\Gamma_{D} = \Gamma_{1,D} \cup \Gamma_{2,D} \cup \Gamma_{3,D} \cup \Gamma_{4,D}$ and $\Gamma_{N} = \Gamma_{1,N} \cup \Gamma_{2,N}$. The set of test functions is given by $\mathcal{V}^h_0 := \{ u \in \mathcal{V}^h \, : \, u|_{\Gamma_D}=0\}$. Multiplying \eqref{eq:elliptic} by a test function $v \in \mathcal{V}^h_0$, integrating over $\Omega$ and using Green's formula we obtain
\begin{equation}
\int_{\Omega} \textbf{K} \nabla_x p \cdot \nabla_x v \, \text{d} x - \int_{\partial \Omega} (\textbf{K} \nabla_x p \cdot \hat{n})v \, \text{d} \sigma= 0. 
\label{eq:integral}
\end{equation}
Thanks to the assumptions on the test function and the boundary condition \eqref{eq:dir3}, the second term of the left-hand-side of \eqref{eq:integral} is zero. Let $\{ \varphi_1, \varphi_2, \ldots, \varphi_N \}$ be a basis of $\mathcal{V}^h_0$ and introduce $p_{\mbox{\scriptsize lift}}$ the function of $\mathcal{V}^h$ which interpolates the Dirichlet boundary values \eqref{eq:dir1}, \eqref{eq:dir2} of $p$ at the nodes on $\Gamma_D$. We introduce  $\tilde p = p - p_{\mbox{\scriptsize lift}}$ and denote by $\tilde p^h$ an approximation of $\tilde p$ in $\mathcal{V}^h_0$ satisfying  
\begin{equation}\label{eq:discrete}
 \int_\Omega ( \textbf{K} \nabla_x \tilde p^h) \cdot \nabla_x \varphi_j \d x = - \int_\Omega ( \textbf{K} \nabla_x p_{\mbox{\scriptsize lift}}) \cdot \nabla_x \varphi_j \d x, \quad \forall j \in \{1, \ldots, N \} \, .
\end{equation}
Introducing $A = ( A_{ij})_{i,j = 1, \ldots, N}$ the $N \times N$ symmetric positive-definite matrix whose entries are 
$A_{ij} = \int_\Omega ( \textbf{K} \nabla_x \varphi_i) \cdot \nabla_x \varphi_j \d x$,  equation \eqref{eq:discrete} gives rise
to the linear system 
\begin{equation}\label{eq:linsys}
 A \textbf{p} = b 
\end{equation}
where $b = (b_i)_{i = 1, \ldots, N}$ is the vector with entries 
$b_i = - \int_\Omega ( \textbf{K} \nabla_x \varphi_i) \cdot \nabla_x p_{\mbox{\scriptsize lift}} \d x$ and where 
$\textbf{p} = (p_i)_{i = 1, \ldots, N}$ is the vector of unknown values $p_i$ such that $\tilde p^h = \sum_{i=1}^N p_i \, \varphi_i$. 
To solve~\eqref{eq:linsys} we use a preconditioned conjugate gradient method using the inverse of the diagonal of $A$ as a 
preconditioner (see \cite{BD2008}). In order to compute an approximation for each of the entries of the matrix $A$
we approximate $\textbf{K}(x)$ in each element $\mathcal{R}$ of the triangulation $\mathcal{T}^h$ by a 
constant function $\widetilde{\textbf{K}}$ equal to the average of the values of $\textbf{K}$ at the vertices of $\mathcal{R}$.
With this approximation we compute the local stiffness matrix $A^{\mathcal{R}}$ in the element $\mathcal{R}$ given as (where the nodes are ordered in counterclockwise order starting from the lower left corner):

$$
A^{\mathcal{R}} = 
\begin{pmatrix}
    \displaystyle \frac{ \widetilde{k_1} + \widetilde{k_3}}{ 3} + \frac{ \widetilde{k_2}}{ 2} & \displaystyle - \frac{ \widetilde{k_1}}{ 3} + \frac{ \widetilde{k_3}}{ 6} & \displaystyle - \frac{\widetilde{k_1} + \widetilde{k_3}}{6} - \frac{ \widetilde{k_2}}{ 2} & \displaystyle \frac{ \widetilde{k_1}}{ 6} - \frac{ \widetilde{k_3}}{3}  \\[10pt]
    \displaystyle - \frac{ \widetilde{k_1}}{ 3} + \frac{ \widetilde{k_3}}{ 6}                 & \displaystyle \frac{\widetilde{k_1} + \widetilde{k_3}}{ 3} - \frac{ \widetilde{k_2}}{ 2}                                                                              & \displaystyle \frac{ \widetilde{k_1}}{ 6} - \frac{ \widetilde{k_3}}{ 3}  & \displaystyle - \frac{ \widetilde{k_1} + \widetilde{k_3}}{ 6} + \frac{ \widetilde{k_2}}{ 2} \\[10pt]
    \displaystyle - \frac{\widetilde{k_1} + \widetilde{k_3}}{6} - \frac{ \widetilde{k_2}}{ 2} & \displaystyle \frac{ \widetilde{k_1}}{ 6} - \frac{ \widetilde{k_3}}{3} & \displaystyle \frac{ \widetilde{k_1} + \widetilde{k_3}}{ 3} + \frac{ \widetilde{k_2}}{2} & \displaystyle - \frac{ \widetilde{k_1}}{ 3} + \frac{ \widetilde{k_3}}{6} \\[10pt]
    \displaystyle \frac{ \widetilde{k_1}}{ 6} - \frac{ \widetilde{k_3}}{3}                    & \displaystyle - \frac{ \widetilde{k_1} + \widetilde{k_3}}{ 6} + \frac{ \widetilde{k_2}}{ 2} & \displaystyle - \frac{ \widetilde{k_1}}{ 3} + \frac{ \widetilde{k_3}}{6} & \displaystyle \frac{ \widetilde{k_1} + \widetilde{k_3}}{ 3} - \frac{ \widetilde{k_2}}{2}
 \end{pmatrix}
 \, .
$$
\vspace{0.4cm}

For Geometry 2, we proceed following a similar procedure as for Geometry 1 taking into account the periodic boundary conditions. For more details on the theory and implementation of the finite-element method we refer to \cite{BS2007, G2006}.


\subsection{Velocity computation}

We compute the gradient of the pressure using a second order finite difference method and for the sake of completeness we state the formulas below. For the partial derivatives with respect to $x$ and assuming that $( x_i, y_j)$ is a node not lying on the boundary of $\Omega$ (where $\Omega$ is $\Omega_1$ or $\Omega_2$) we use the centered difference scheme
\begin{equation}
\frac{\partial p( x_i, y_j)}{\partial x} \approx \frac{p(x_{i+1}, y_{j}) - p(x_{i-1}, y_j)}{2 \Delta x} \, .
\label{centeredFDM}
\end{equation}
For nodes on the left boundary of $\Omega$ we use a forward finite difference scheme
\begin{equation}
\frac{\partial p(x_i, y_j)}{\partial x} \approx \frac{-3p(x_{i}, y_j) + 4p(x_{i+1}, y_j) - p(x_{i+2}, y_j)}{2\Delta x} \, ,
\label{forwardFDM}
\end{equation}
and a symmetric formula for nodes on the right boundary. The partial derivatives with respect to $y$ are computed in a similar fashion.


\section{Particle approximation of the oxygen flow}
\label{appendix:sph}

We use a splitting method to solve \eqref{O2PDE}. In the first splitting step, we use the Smoothed Particle Hydrodynamics (SPH) method to solve for the advection and diffusion terms, i.e. the left-hand side of \eqref{O2PDE}. In the second step, we solve for the reaction term i.e. the right-hand side of \eqref{O2PDE}.


\subsection{SPH approximation of the convection-diffusion step}

Here, we give the details for the first step, i.e. we assume $\beta = 0$ throughout this section. Supposing for a while that the vector field $\textbf{v}$ in \eqref{eq:v} is given and smooth, then the density 
\begin{equation}
\rho(x,t) = m \sum_{\ell=1}^L \delta(x- \textbf{Y}_\ell(t)),
\label{eq:particleapprox}
\end{equation}
 where $\delta$ is the Dirac delta at $0$ and $m>0$ is any positive constant, is a solution of Eq. \eqref{O2PDE} if and only if $\textbf{Y}_\ell(t)$ satisfies the following ODE:
\begin{equation}
\frac{ \d \textbf{Y}_\ell}{ \d t} = \textbf{v} (\textbf{Y}_\ell(t),t)  = \textbf{u}( \textbf{Y}_\ell(t), t) - \textbf{D}( \textbf{Y}_\ell(t),t) \frac{\nabla_x \rho}{ \rho + \widetilde{ \rho}}(\textbf{Y}_\ell(t),t). 
\label{particleDynamics}
\end{equation}
A measure satisfying \eqref{eq:particleapprox}, \eqref{particleDynamics} is called a particle solution,  $\textbf{Y}_\ell(t)$ is the position of the $\ell$-th particle at time $t$ and $m$ is the particle mass (here chosen identical for all particles). There is flexibility in choosing $m$ and the initial particle positions $\textbf{Y}_\ell(0)$ which is used to best approximate initial and boundary conditions. However, as such, the formula does not make sense. Indeed, $\rho$ being a sum of Dirac deltas, the right-hand side of \eqref{particleDynamics} is not defined. The SPH methodology consists of introducing a mollifier kernel $W$: $x \in \Omega \mapsto W(x,\eta) \in [0,\infty)$ where $\eta>0$ satisfying $\int_{\Omega} W(x,\eta) \, dx = 1$ and $W(\cdot,\eta) \to \delta$ as $\eta \to 0$ in the distributional sense. Then $\rho$ can be approximated by 
\begin{equation}
\rho(x,t) \approx \rho^\eta(x,t) := m \sum_{\ell=1}^L  \, W(x - \textbf{Y}_\ell(t), \eta) \, ,
\label{densitySPH}
\end{equation}
which is now a smooth function. Thus, it can be composed with a nonlinear function $U$: $r \in [0,\infty) \mapsto U(r) \in {\mathbb R}$ giving 
\begin{equation}
U(\rho)(x,t) \approx U(\rho)^\eta(x,t):= U \big( m \sum_{\ell=1}^L  \, W(x - \textbf{Y}_\ell(t), \eta) \big) \, ,
\label{eq:Urho}
\end{equation}
and it can be differentiated:
\begin{equation}
\nabla_x \rho(x,t) \approx (\nabla_x \rho)^\eta (x,t):=m \, \sum_{\ell=1}^{ L} \nabla_x W(x - \textbf{Y}_\ell(t), \eta) \, .
\label{gradientSPH}
\end{equation}
This strategy is at the core of the SPH method \cite{monaghan1992smoothed}. Here, we use this strategy to replace \eqref{particleDynamics} by
\begin{equation}\label{eq:O2_particle}
 \frac{ \d \textbf{Y}_\ell}{ \d t} = \textbf{V}_\ell(t):= \textbf{u}({\textbf{Y}}_\ell(t),t) - \textbf{D}({\textbf{Y}}_\ell(t),t) \,  \frac{(\nabla_x \rho)^\eta ({\textbf{Y}}_\ell(t),t)}{\rho^\eta ({\textbf{Y}}_\ell(t),t)+ \widetilde{\rho}},
\end{equation}
which is now completely well-defined. In the present context of diffusion equations, this method is called the diffusion velocity method and was first introduced in \cite{DM1990}. Several choices for the mollifier kernel $W$ are possible depending on the context. We have chosen the {\it poly6} kernel for its simplicity and its reliability in the numerical solution of the Navier-Stokes equations (see \cite{M2003}). This kernel is defined in $\R^2$ by
\begin{equation}
W( x, \eta) = 
\begin{cases}
\displaystyle \frac{4}{\pi \eta^8} \left (\eta^2 - | x |^2 \right)^3 \quad \quad \hbox{ if } 0 \leq | x | \leq \eta \,,\\
\\
0 \quad \quad \quad \quad \quad \quad \quad \quad \quad \quad \hbox{otherwise} \, .\\
\end{cases}
\label{kernel}
\end{equation}
We solve \eqref{eq:O2_particle} using the forward Euler scheme 
\begin{equation}
\textbf{Y}_\ell^{n+1} = \textbf{Y}_\ell^n + \Delta t^n \textbf{V}_\ell^n
\label{updatePosition}
\end{equation}
where $\textbf{Y}_\ell^n$; $\textbf{V}^n_\ell$ are approximations of $\textbf{Y}_\ell (t^n)$, $\textbf{V}_\ell (t^n)$ and where $t^{n}$ is the $n$-th time discretization point.  The time-step $\Delta t^n= t^{n+1} - t^n$ is chosen respecting the CFL condition
\begin{equation}
\Delta t^n = C \frac{ \eta}{\mbox{max}_\ell | \textbf{V}_\ell^n |}
\label{CFL}
\end{equation}
to preserve the stability of the scheme,  with $C \leq 1/2$ (see Table \ref{TableOfParameters}). In addition, in order to have a good acceptance-rejection sampling for the creation of capillaries, we also impose the constraint $\Delta t^n \leq 0.01$. The blood velocity $\textbf{u}({\textbf{Y}}_\ell(t),t)$ is obtained through a linear interpolation from the values of $\textbf{u}$ given by the finite element calculation at the nodes of the finite element mesh. The diffusivity matrix $\textbf{D}$ is updated at the end of the capillary creation-removal process (described below) through \eqref{eq:tissue2}. For better efficiency, it is first computed at the nodes of the finite-element mesh. Then, the diffusion coefficient $\textbf{D}({\textbf{Y}}_\ell(t),t)$ is recovered by linear interpolation exactly like the blood velocity $\textbf{u}$.


\subsection{Death process for the reaction step}

Now, we consider the right-hand side of \eqref{O2PDE}, i.e. we assume $\textbf{v}=0$ in \eqref{O2PDE}. We approximate this equation by a simple death process on the particle approximation \eqref{eq:particleapprox}. This step is described in Algorithm \ref{algoO2consumption} below. 

\begin{algorithm}
$$
\begin{cases}
  \hbox{For } \ell = 1, \ldots, L \\
  \qquad \hbox{Compute }\beta(\rho_\ell^n) = \frac{\beta_{\mbox{\scriptsize sat}}}{\rho_\ell^n + K_m} \mbox{ with } \rho_\ell^n = \rho^\eta(\textbf{Y}_\ell^n, t^n) \\
  \qquad \hbox{Pick randomly } p\in[0,1] \hbox{ according to a uniform law}\\
  \qquad \hbox{If } (p < 1 - e^{-\beta(\rho_\ell^n) \Delta t^n} ) \hbox{ then}\\
  \qquad \qquad \hbox{Remove particle }\ell \\
  \qquad \hbox{End if} \\
  \hbox{End for}
\end{cases}
$$
\caption{Algorithm for the particle death process modelling oxygen consumption (assuming time is equal to $t^n$)}
\label{algoO2consumption}
\end{algorithm}


\subsection{Boundary conditions}

\subsubsection{Periodic boundary conditions}
\label{sec:periodi_bc}

These are boundary conditions \eqref{eq:C2_rho-per} in Case 2 of Sect. \ref{subsubsect:boundary_conditions}. If the $y$-coordinate of a particle exits the range $[0,L_y)$, it is changed to $y + k L_y$ where $k$ is the unique integer in ${\mathbb Z}$ such that $y + k L_y \in [0,L_y)$. The $x$-coordinate is unchanged.


\subsubsection{Outflow boundary conditions}
\label{sec:outflow_boundary}

As described in Sect. \ref{subsubsect:boundary_conditions}, the boundary conditions \eqref{eq:dir2_rho} and \eqref{eq:C2_rho_outflow} guarantee that the oxygen flow is outgoing across the corresponding boundaries. They are taken into account by removing the particles that exit the domain through these boundaries. 


\subsubsection{Blood vessel boundary condition}
\label{sec:artery_boundary}

These are conditions \eqref{eq:dir1_rho} and \eqref{eq:C2_rho0}. We define a ghost domain $\Omega(L_{\hbox{\scriptsize ghost}}) = [-L_{\hbox{\scriptsize ghost}},0] \times [L_{\mbox{\scriptsize min}},L_{\mbox{\scriptsize max}}]$, where $L_{\hbox{\scriptsize ghost}}$ is changed dynamically as explained below. The ghost domain intersects $\Omega$ along the boundaries where the Dirichlet conditions \eqref{eq:dir1_rho} or \eqref{eq:C2_rho0} need to be enforced, i.e. $\Gamma_{1,D}$ (Case 1) or $\Gamma_{1,D}^{\mbox{\scriptsize mid}}$ (Case 2), which here, for the ease of notation will be collectively denoted by $\Gamma_{1,D}$.
Supposing that in $\Omega(L_{\hbox{\scriptsize ghost}})$ there is a density $\rho_0$ of oxygen particles and that their velocity in the $x$-direction is $\bar{v}$, their flux $j$ across $\Gamma_{1,D}$ is then $j = \rho_0 \bar v$. A sensible estimate of $\bar{v}$ is given by the blood velocity $\textbf{u}$ along $\Gamma_{1,D}$. Thus, we take $\bar{v}$ equal to the average of $\textbf{u}$ along $\Gamma_{1,D}$ (in practice, we take the average of $\textbf{u}$ at the finite-element nodes along $\Gamma_{1,D}$). But this flux $j$ is also given by $j = N / ( L \Delta t)$ where $N$ is the number of oxygen particles (of mass $m=1$) crossing $\Gamma_{1,D}$ during a small time interval $\Delta t$ and $L = L_{\mbox{\scriptsize max}} - L_{\mbox{\scriptsize min}}$ is the length of $\Gamma_{1,D}$. We can then estimate the number $N^n$ of particles to be created during time step $\Delta t^n$ as $N^n = \rho_0 \, L \, \Delta t^n \, \bar v^n$ (where $\bar v^n$ is the estimate of $\bar v$ at time $t^n$). We take $L_{\hbox{\scriptsize ghost}}^n$ (the width of the ghost domain at time $t^n$) to be $L_{\hbox{\scriptsize ghost}}^n = \Delta t^n \, \bar v^n$ (we recall that $\Delta t^{(n)}$ is determined by \eqref{CFL}) and put $N^n$ particles randomly uniformly in the corresponding ghost domain $\Omega(L_{\hbox{\scriptsize ghost}}^n)$. During the time step $\Delta t^n$, they are moved with velocity $\bar v^n$ and end up inside the actual domain $\Omega$. In this way, we ensure that the flux of particles entering the domain corresponds to the density $\rho_0$. As $N^n$ is in general not an integer, we insert $\floor*{N^n}$ particles if $N^n \geq 1$ (where $\floor*{N}$ denotes the greatest integer smaller than or equal to $N$). If $N^n < 1$, we pick $r$ in $[ 0, 1]$ with uniform probability and insert one particle in $\Omega(L_{\hbox{\scriptsize ghost}}^n)$ randomly uniformly if $r \leq N^n $ and create no particle otherwise. In the unlikely event where there are no particles in the computational domain $\Omega$, \eqref{CFL} cannot be use. Instead, we set 
$ \Delta t =  C h / \bar v $, where $C$ is the same CFL number as that used in \eqref{CFL}.


\subsubsection{No oxygen boundary condition}

This is condition \ref{eq:dir3_rho} along $\Gamma_{1,N} \cup \Gamma_{2,N}$ in Case 1 of Sect. \ref{subsubsect:boundary_conditions}. We use the same methodology as in Sect. \ref{sec:artery_boundary} but with $\rho_0=0$, simply meaning that we inject no particle across this boundary.


\section{Acceptance-Rejection sampling method for the capillary creation}
\label{appendix:acceptance}

\subsection{Creation of new capillaries and modification of the tissue}

We use an acceptance-rejection method in order to sample the spatio-temporal Poisson processes involved in the creation of capillary elements. Set time equal to $t^n$ and $\Delta t^n$ given by \eqref{CFL}. We first choose the total number of sampling points to be $N_c$ (see Table \ref{TableOfParameters} for the value used in our simulations and Sect. \ref{subsec:numerics} for a discussion of this choice). We set $S = \frac{|\Omega|}{N_c}$ where $|\Omega|$ is the area of the full domain $\Omega$ and we define $\textbf{u}^\perp ( \textbf{X})$ to be the rotation by $90^\circ$ in counterclockwise direction of $\textbf{u} (\textbf{X})$.  The algorithms for capillary element creation by oxygen gradient, by reinforcement or by WSS, (see Sect. \ref{subsec:capillaries}) are respectively given at Algorithms \ref{algonuc}, \ref{algonuf}, \ref{algonuwss}.

\begin{algorithm}[ht!]
$$
\begin{cases}
  \hbox{For } j=1, \ldots, N_c \hbox{ Do:} \\
  \hbox{\qquad Pick up randomly a position } \textbf{X} \in \Omega \\
  \hbox{\qquad Compute $\rho(\textbf{X})$ and $\nabla_x \rho(\textbf{X})$ using  formulas (\ref{densitySPH}) and (\ref{gradientSPH})} \\
  \hbox{\qquad Compute the Poisson intensity $\nu_c (\textbf{X})$ using formula (\ref{nuc})}  \\
  \hbox{\qquad Pick up randomly } p \in [0,1] \hbox{ (uniform law)} \\
  \hbox{\qquad If } p < (1 - e^{-\nu_c S \Delta t^{(n)}}) \hbox{ then} \\
  \hbox{\qquad \qquad Create a capillary at position } \textbf{X} \hbox{ with direction } \displaystyle \frac{\nabla_x \rho(\textbf{X})}{| \nabla_x \rho(\textbf{X})) |} \\
  \hbox{\qquad End if} \\
  \hbox{End Do}
\end{cases}
$$
\caption{Algorithm for capillary element creation along the gradient of oxygen concentration (see \ref{Subsec:creation_gradient})}
\label{algonuc}
\end{algorithm}

\begin{algorithm}[ht!]
$$
\begin{cases}
  \hbox{For } j=1, \ldots, N_c \hbox{ Do:} \\
  \hbox{\qquad Pick up randomly a position } \textbf{X} \\
  \hbox{\qquad Compute the local blood flow and density, }\textbf{u}(\textbf{X}) \hbox{ and } \rho(\textbf{X}), \hbox{ respectively} \\
  \hbox{\qquad Compute Poisson intensity $\nu_f(\rho, \textbf{u})$ using (\ref{nuf})} \\
  \hbox{\qquad Pick up randomly } p \in [0,1] \hbox{ (uniform law)} \\
  \hbox{\qquad If } p < (1 - e^{-\nu_f S \Delta t^{(n)}}) \hbox{ then} \\
  \hbox{\qquad\qquad Create a capillary at position } \textbf{X} \hbox{ with direction } \displaystyle \frac{ \textbf{u} (\textbf{X})}{| \textbf{u} (\textbf{X})) |} \\
  \hbox{\qquad End if} \\
  \hbox{End Do}
\end{cases}
$$
\caption{Algorithm for capillary element creation by reinforcement along blood flow (see \ref{Subsec:reinforcement})}
\label{algonuf}
\end{algorithm}

\begin{algorithm}[ht!]
$$
\begin{cases}
  \hbox{For } j=1, \ldots, N_c \hbox{ Do:} \\
  \hbox{\qquad Pick up randomly a position } \textbf{X} \\
  \hbox{\qquad Compute }\sigma(\textbf{X}) \hbox{ according to \eqref{eq:sigma}} \\
  \hbox{\qquad Compute the leading eigenvalue }\lambda\hbox{ of }\sigma(\textbf{X}) \\
  \hbox{\qquad Compute the Poisson coefficient $\nu_w (\lambda)$ using (\ref{eq:wall})} \\
  \hbox{\qquad Pick up randomly } p \in [0,1] \\
  \hbox{\qquad If } p < (1 - e^{-\nu_w S \Delta t^{(n)}}) \hbox{ then} \\
  \hbox{\qquad \qquad Create a capillary at position } \textbf{X} \hbox{ with direction } \displaystyle \frac{ \textbf{u}^\perp ( \textbf{X})}{| \textbf{u}^\perp ( \textbf{X}) |} \\
  \hbox{\qquad End if} \\
  \hbox{End Do}
\end{cases}
$$
\caption{Algorithm for capillary element creation by WSS (see \ref{Subsec:wss})}
\label{algonuwss}
\end{algorithm}


\subsection{Removal of unused capillaries}

The capillary pruning algorithm is a Poisson time process only, not a spatio-temporal Poisson process. So we do not need the area discretization parameter $S$. The capillary pruning algorithm is given in Algorithm \ref{algonur}.

\begin{algorithm}[ht!]
$$
\begin{cases}
  \hbox{Loop over all the capillaries of the network} \\
  \qquad \hbox{Get the capillary position $\textbf{X}$} \\
  \qquad \hbox{Compute $\gamma(\textbf{X})$, the Frobenius norm of $\textbf{K}(\textbf{X})$}\\ 
  \qquad \hbox{Compute the Poisson intensity $\nu_r$ using formula (\ref{nur})} \\
  \qquad \hbox{Pick up randomly } p \in [0,1] \hbox{ (uniform law)} \\
  \qquad \hbox{If } p < (1 - e^{-\nu_r \Delta t^{(n)}}) \hbox{ then}\\
  \qquad \qquad \hbox{Remove capillary} \\
  \qquad \hbox{End if} \\
  \hbox{End loop} 
\end{cases}
$$
\caption{Algorithm for capillary pruning (see \ref{Subsec:cap_pruning})}
\label{algonur}
\end{algorithm}


\section{Description of the videos}

Videos can be found on \href{https://doi.org/10.6084/m9.figshare.c.4287575.v1}{https://doi.org/10.6084/m9.figshare.c.4287575.v1}

\smallskip
\noindent
The videos have been obtained by running the model with the parameters set to the values given in Tables \ref{TableOfParameters} and \ref{numericalParameters} and with the following choices for the geometry and the capillary element creation/pruning mechanisms:

\begin{itemize}
\item[1)] Movies (a)-(d): Geometry $\Omega_1$. All the capillary element creation/pruning mechanisms ``ON'', corresponding to Figs. \ref{fig:all_mech}-\ref{fig:averagedNetwork} and \ref{fig:main_plot}(A). \\
\underline{Movie (a)} Positions of oxygen particles (red spots) and capillary elements (blue rods), corresponding to Fig. \ref{fig:all_mech}. \\
\underline{Movie (b)} Isolines and heatmap of the pressure $p$, corresponding to Fig. \ref{fig:pressure_plots}. \\
\underline{Movie (c)} Heatmap of the Frobenius norm of the hydraulic conductivity tensor, corresponding to Fig. \ref{fig:averagedNetwork}. \\
\underline{Movie (d)} Same as movie (a) but with mesh size divided by $2$, corresponding to Fig.~\ref{fig:octa_mesh}. 
\item[2)] Movies (e)-(j): Geometry $\Omega_1$. Positions of oxygen (red spots) and capillary elements (blue rods) with  some capillary creation/pruning mechanisms turned off (corresponding to Figs. \ref{fig:main_plot}(B) to (G)).\\
\underline{Movie (e)} (corresponds to Fig.\ref{fig:main_plot}(B))  \\
WSS  “on”,  oxygen  gradient  “off”,  reinforcement  “on”  and  pruning “on”. \\
\underline{Movie (f)} (corresponds to Fig.\ref{fig:main_plot}(C))  \\
WSS  “off”,  oxygen  gradient  “on”,  reinforcement  “on”  and  pruning “on”. \\
\underline{Movie (g)} (corresponds to Fig.\ref{fig:main_plot}(D))  \\
WSS  “off”,  oxygen  gradient  “off”,  reinforcement  “on”  and  pruning “on”. \\
\underline{Movie (h)} (corresponds to Fig.\ref{fig:main_plot}(E))  \\
WSS  “on”,  oxygen  gradient  “on”,  reinforcement  “off”  and  pruning “on”.\\
\underline{Movie (i)} (corresponds to Fig.\ref{fig:main_plot}(F))  \\
WSS  “on”,  oxygen  gradient  “off”,  reinforcement  “off”  and  pruning “on”.\\
\underline{Movie (j)} (corresponds to Fig.\ref{fig:main_plot}(G))  \\
WSS  “off”,  oxygen  gradient  “on”,  reinforcement  “off”  and  pruning“on”.\\
\item[3)] Movie (k):  Geometry $\Omega_1$.   Positions of oxygen (red spots) and capillary elements (blue rods) with all the capillary creation mechanisms “on” and pruning turned “off” (Correspondong figure not shown in the text).
\item[4)] Movie (l):  Geometry $\Omega_2$. Positions of oxygen (red spots) and capillary elements  (blue  rods)  with  all  the  capillary  creation/pruning  mechanisms turned “on” (corresponding to Fig. \ref{fig:all_mech_geom2}).
\end{itemize}


\end{appendices}


\begin{thebibliography}{10}

\bibitem{A+2017}
{\sc G.~Albi, M.~Burger, J.~Haskovec, P.~Markowich, and M.~Schlottbom}, {\em
  Continuum modeling of biological network formation}, in Active Particles,
  Volume 1, Springer, 2017, pp.~1--48.

\bibitem{amitrano1986growth}
{\sc C.~Amitrano, A.~Coniglio, and F.~Di~Liberto}, {\em Growth probability
  distribution in kinetic aggregation processes}, Phys. Rev. Lett., 57 (1986),
  p.~1016.

\bibitem{balding1985mathematical}
{\sc D.~Balding and D.~McElwain}, {\em A mathematical model of tumour-induced
  capillary growth}, J. Theoret. Biol., 114 (1985), pp.~53--73.

\bibitem{bardos2015stability}
{\sc C.~Bardos and E.~Tadmor}, {\em Stability and spectral convergence of
  fourier method for nonlinear problems: on the shortcomings of the $2/3$
  de-aliasing method}, Numer. Math., 129 (2015), pp.~749--782.

\bibitem{BJJ2009}
{\sc A.~L. Bauer, T.~L. Jackson, and Y.~Jiang}, {\em Topography of
  extracellular matrix mediates vascular morphogenesis and migration speeds in
  angiogenesis}, PLoS computational biology, 5 (2009), p.~e1000445.

\bibitem{BDM2013}
{\sc E.~Boissard, P.~Degond, and S.~Motsch}, {\em Trail formation based on
  directed pheromone deposition}, J. Math. Biol., 66 (2013), pp.~1267--1301.

\bibitem{BS2007}
{\sc S.~Brenner and R.~Scott}, {\em The mathematical theory of finite element
  methods}, vol.~15, Springer Science \& Business Media, 2007.

\bibitem{byrne1995mathematical}
{\sc H.~Byrne and M.~Chaplain}, {\em Mathematical models for tumour
  angiogenesis: numerical simulations and nonlinear wave solutions}, Bull.
  Math. Biol., 57 (1995), pp.~461--486.

\bibitem{CJ2000}
{\sc P.~Carmeliet and R.~K. Jain}, {\em Angiogenesis in cancer and other
  diseases}, Nature, 407 (2000), p.~249.

\bibitem{chen2014equations}
{\sc A.~Chen, J.~Darbon, G.~Buttazzo, F.~Santambrogio, and J.-M. Morel}, {\em
  On the equations of landscape formation}, Interfaces Free Bound., 16 (2014),
  pp.~105--136.

\bibitem{chen2014landscape}
{\sc A.~Chen, J.~Darbon, and J.-M. Morel}, {\em Landscape evolution models: a
  review of their fundamental equations}, Geomorphology, 219 (2014),
  pp.~68--86.

\bibitem{curcio2014kinetics}
{\sc E.~Curcio, A.~Piscioneri, S.~Morelli, S.~Salerno, P.~Macchiarini, and
  L.~De~Bartolo}, {\em Kinetics of oxygen uptake by cells potentially used in a
  tissue engineered trachea}, Biomaterials, 35 (2014), pp.~6829--6837.

\bibitem{BD2008}
{\sc G.~Dahlquist and {\AA}.~Bj{\"o}rck}, {\em Numerical methods in scientific
  computing, volume i}, Society for Industrial and Applied Mathematics, 8
  (2008).

\bibitem{DM2013}
{\sc J.~T. Daub and R.~M. Merks}, {\em A cell-based model of
  extracellular-matrix-guided endothelial cell migration during angiogenesis},
  Bull. Math. Biol., 75 (2013), pp.~1377--1399.

\bibitem{DM1990}
{\sc P.~Degond and F.-J. Mustieles}, {\em A deterministic approximation of
  diffusion equations using particles}, SIAM J. Sci. Stat. Comput., 11 (1990),
  pp.~293--310.

\bibitem{efendiev2009multiscale}
{\sc Y.~Efendiev and T.~Y. Hou}, {\em Multiscale finite element methods: theory
  and applications}, vol.~4, Springer Science \& Business Media, 2009.

\bibitem{F+2005}
{\sc I.~Fischer, J.-P. Gagner, M.~Law, E.~W. Newcomb, and D.~Zagzag}, {\em
  Angiogenesis in gliomas: biology and molecular pathophysiology}, Brain
  pathology, 15 (2005), pp.~297--310.

\bibitem{F1995}
{\sc J.~Folkman}, {\em Angiogenesis in cancer, vascular, rheumatoid and other
  disease}, Nature medicine, 1 (1995), p.~27.

\bibitem{F2017}
{\sc R.~L. Fournier}, {\em Basic transport phenomena in biomedical
  engineering}, CRC press, 2017.

\bibitem{G+2014}
{\sc P.~A. Galie, D.-H.~T. Nguyen, C.~K. Choi, D.~M. Cohen, P.~A. Janmey, and
  C.~S. Chen}, {\em Fluid shear stress threshold regulates angiogenic
  sprouting}, Proc. Natl. Acad. Sci. USA, 111 (2014), pp.~7968--7973.

\bibitem{garipcan2011image}
{\sc B.~Garipcan, S.~Maenz, T.~Pham, U.~Settmacher, K.~D. Jandt, J.~Zanow, and
  J.~Bossert}, {\em Image analysis of endothelial microstructure and
  endothelial cell dimensions of human arteries--a preliminary study}, Advanced
  engineering materials, 13 (2011), pp.~B54--B57.

\bibitem{G+1974}
{\sc M.~A. Gimbrone~Jr, R.~S. Cotran, S.~B. Leapman, and J.~Folkman}, {\em
  Tumor growth and neovascularization: an experimental model using the rabbit
  cornea}, Journal of the National Cancer Institute, 52 (1974), pp.~413--427.

\bibitem{G2006}
{\sc M.~S. Gockenbach}, {\em Understanding and implementing the finite element
  method}, vol.~97, Siam, 2006.

\bibitem{GP2000}
{\sc D.~Goldman and A.~S. Popel}, {\em A computational study of the effect of
  capillary network anastomoses and tortuosity on oxygen transport}, J.
  Theoret. Biol., 206 (2000), pp.~181--194.

\bibitem{G+2016}
{\sc J.~A. Gonz{\'a}lez, F.~J. Rodr{\'\i}guez-Cort{\'e}s, O.~Cronie, and
  J.~Mateu}, {\em Spatio-temporal point process statistics: a review}, Spat.
  Stat., 18 (2016), pp.~505--544.

\bibitem{grogan2018importance}
{\sc J.~A. Grogan, A.~J. Connor, J.~M. Pitt-Francis, P.~K. Maini, and H.~M.
  Byrne}, {\em The importance of geometry in the corneal micropocket
  angiogenesis assay}, PLoS computational biology, 14 (2018), p.~e1006049.

\bibitem{haskovec2018rigorous}
{\sc J.~Haskovec, L.~M. Kreusser, and P.~Markowich}, {\em Rigorous continuum
  limit for the discrete network formation problem}, arXiv preprint
  arXiv:1808.01526,  (2018).

\bibitem{haskovec2015mathematical}
{\sc J.~Haskovec, P.~Markowich, and B.~Perthame}, {\em Mathematical analysis of
  a pde system for biological network formation}, Comm. Partial Differential
  Equations, 40 (2015), pp.~918--956.

\bibitem{H+2016}
{\sc J.~Haskovec, P.~Markowich, B.~Perthame, and M.~Schlottbom}, {\em Notes on
  a pde system for biological network formation}, Nonlinear Anal., 138 (2016),
  pp.~127--155.

\bibitem{hastings1998laplacian}
{\sc M.~B. Hastings and L.~S. Levitov}, {\em Laplacian growth as
  one-dimensional turbulence}, Phys. D, 116 (1998), pp.~244--252.

\bibitem{herrmann1986geometrical}
{\sc H.~J. Herrmann}, {\em Geometrical cluster growth models and kinetic
  gelation}, Physics Reports, 136 (1986), pp.~153--224.

\bibitem{hillen2006m}
{\sc T.~Hillen}, {\em M 5 mesoscopic and macroscopic models for mesenchymal
  motion}, J. Math. Biol., 53 (2006), pp.~585--616.

\bibitem{hu2013adaptation}
{\sc D.~Hu and D.~Cai}, {\em Adaptation and optimization of biological
  transport networks}, Phys. Rev. Lett., 111 (2013), p.~138701.

\bibitem{I+1997}
{\sc S.~Ichioka, M.~Shibata, K.~Kosaki, Y.~Sato, K.~Harii, and A.~Kamiya}, {\em
  Effects of shear stress on wound-healing angiogenesis in the rabbit ear
  chamber}, Journal of Surgical Research, 72 (1997), pp.~29--35.

\bibitem{K+2008}
{\sc H.~Kang, K.~J. Bayless, and R.~Kaunas}, {\em Fluid shear stress modulates
  endothelial cell invasion into three-dimensional collagen matrices}, American
  Journal of Physiology-Heart and Circulatory Physiology, 295 (2008),
  pp.~H2087--H2097.

\bibitem{K+2011}
{\sc R.~Kaunas, H.~Kang, and K.~J. Bayless}, {\em Synergistic regulation of
  angiogenic sprouting by biochemical factors and wall shear stress}, Cellular
  and molecular bioengineering, 4 (2011), pp.~547--559.

\bibitem{K+2005}
{\sc B.~Kaur, F.~W. Khwaja, E.~A. Severson, S.~L. Matheny, D.~J. Brat, and
  E.~G. Van~Meir}, {\em Hypoxia and the hypoxia-inducible-factor pathway in
  glioma growth and angiogenesis}, Neuro-oncology, 7 (2005), pp.~134--153.

\bibitem{langdon1983direct}
{\sc A.~B. Langdon, B.~I. Cohen, and A.~Friedman}, {\em Direct implicit large
  time-step particle simulation of plasmas}, J. Comput. Phys., 51 (1983),
  pp.~107--138.

\bibitem{MN1998}
{\sc M.~Matsumoto and T.~Nishimura}, {\em Mersenne twister: a 623-dimensionally
  equidistributed uniform pseudo-random number generator}, ACM Transactions on
  Modeling and Computer Simulation (TOMACS), 8 (1998), pp.~3--30.

\bibitem{MC+2006}
{\sc S.~R. McDougall, A.~R. Anderson, and M.~A. Chaplain}, {\em Mathematical
  modelling of dynamic adaptive tumour-induced angiogenesis: clinical
  implications and therapeutic targeting strategies}, J. Theoret. Biol., 241
  (2006), pp.~564--589.

\bibitem{mitchison1980model}
{\sc G.~Mitchison}, {\em A model for vein formation in higher plants}, Proc. R.
  Soc. Lond. B, 207 (1980), pp.~79--109.

\bibitem{mitchison1981polar}
\leavevmode\vrule height 2pt depth -1.6pt width 23pt, {\em The polar transport
  of auxin and vein patterns in plants}, Phil. Trans. R. Soc. Lond. B, 295
  (1981), pp.~461--471.

\bibitem{monaghan1992smoothed}
{\sc J.~J. Monaghan}, {\em Smoothed particle hydrodynamics}, Annual review of
  astronomy and astrophysics, 30 (1992), pp.~543--574.

\bibitem{M2003}
{\sc M.~M{\"u}ller, D.~Charypar, and M.~Gross}, {\em Particle-based fluid
  simulation for interactive applications}, in Proceedings of the 2003 ACM
  SIGGRAPH/Eurographics symposium on Computer animation, Eurographics
  Association, 2003, pp.~154--159.

\bibitem{M2015}
{\sc W.~L. Murfee}, {\em Implications of fluid shear stress in capillary
  sprouting during adult microvascular network remodeling}, Mechanobiology of
  the Endothelium,  (2015), p.~166.

\bibitem{murray1926physiological}
{\sc C.~D. Murray}, {\em The physiological principle of minimum work: I. the
  vascular system and the cost of blood volume}, Proc. Natl. Acad. Sci. USA, 12
  (1926), pp.~207--214.

\bibitem{MKA1982}
{\sc V.~Muthukkaruppan, L.~Kubai, and R.~Auerbach}, {\em Tumor-induced
  neovascularization in the mouse eye}, Journal of the National Cancer
  Institute, 69 (1982), pp.~699--708.

\bibitem{OAMB2009}
{\sc M.~R. Owen, T.~Alarc{\'o}n, P.~K. Maini, and H.~M. Byrne}, {\em
  Angiogenesis and vascular remodelling in normal and cancerous tissues}, J.
  Math. Biol., 58 (2009), p.~689.

\bibitem{painter2009modelling}
{\sc K.~Painter}, {\em Modelling cell migration strategies in the extracellular
  matrix}, J. Math. Biol., 58 (2009), p.~511.

\bibitem{PP1991}
{\sc S.~Paku and N.~Paweletz}, {\em First steps of tumor-related
  angiogenesis.}, Laboratory investigation; a journal of technical methods and
  pathology, 65 (1991), pp.~334--346.

\bibitem{P+2011}
{\sc J.~Y. Park, J.~B. White, N.~Walker, C.-H. Kuo, W.~Cha, M.~E. Meyerhoff,
  and S.~Takayama}, {\em Responses of endothelial cells to extremely slow
  flows}, Biomicrofluidics, 5 (2011), p.~022211.

\bibitem{PK1989}
{\sc N.~Paweletz and M.~Knierim}, {\em Tumor-related angiogenesis}, Critical
  reviews in oncology/hematology, 9 (1989), pp.~197--242.

\bibitem{penta2015multiscale}
{\sc R.~Penta, D.~Ambrosi, and A.~Quarteroni}, {\em Multiscale homogenization
  for fluid and drug transport in vascularized malignant tissues}, Math. Models
  Methods Appl. Sci., 25 (2015), pp.~79--108.

\bibitem{P+2017}
{\sc D.~Peurichard, F.~Delebecque, A.~Lorsignol, C.~Barreau, J.~Rouquette,
  X.~Descombes, L.~Casteilla, and P.~Degond}, {\em Simple mechanical cues could
  explain adipose tissue morphology}, J. Theoret. Biol., 429 (2017),
  pp.~61--81.

\bibitem{PH2009}
{\sc L.-K. Phng and H.~Gerhardt}, {\em Angiogenesis: a team effort coordinated
  by notch}, Developmental cell, 16 (2009), pp.~196--208.

\bibitem{pietronero1984stochastic}
{\sc L.~Pietronero and H.~Wiesmann}, {\em Stochastic model for dielectric
  breakdown}, J. Stat. Phys., 36 (1984), pp.~909--916.

\bibitem{pillay2017modeling}
{\sc S.~Pillay, H.~M. Byrne, and P.~K. Maini}, {\em Modeling angiogenesis: A
  discrete to continuum description}, Phys. Rev. E, 95 (2017), p.~012410.

\bibitem{R1997}
{\sc W.~Risau}, {\em Mechanisms of angiogenesis}, Nature, 386 (1997), p.~671.

\bibitem{rolland2005reviewing}
{\sc A.-G. Rolland-Lagan and P.~Prusinkiewicz}, {\em Reviewing models of auxin
  canalization in the context of leaf vein pattern formation in arabidopsis},
  The Plant Journal, 44 (2005), pp.~854--865.

\bibitem{S+2012}
{\sc M.~Schneider, J.~Reichold, B.~Weber, G.~Sz{\'e}kely, and S.~Hirsch}, {\em
  Tissue metabolism driven arterial tree generation}, Medical image analysis,
  16 (2012), pp.~1397--1414.

\bibitem{SBP2013}
{\sc M.~Scianna, C.~Bell, and L.~Preziosi}, {\em A review of mathematical
  models for the formation of vascular networks}, J. Theoret. Biol., 333
  (2013), pp.~174--209.

\bibitem{S+2013}
{\sc T.~W. Secomb, J.~P. Alberding, R.~Hsu, M.~W. Dewhirst, and A.~R. Pries},
  {\em Angiogenesis: an adaptive dynamic biological patterning problem}, PLoS
  computational biology, 9 (2013), p.~e1002983.

\bibitem{SP1996}
{\sc T.~C. Skalak and R.~J. Price}, {\em The role of mechanical stresses in
  microvascular remodeling}, Microcirculation, 3 (1996), pp.~143--165.

\bibitem{SF2007}
{\sc M.~A. Swartz and M.~E. Fleury}, {\em Interstitial flow and its effects in
  soft tissues}, Annu. Rev. Biomed. Eng., 9 (2007), pp.~229--256.

\bibitem{takahashi1966oxygen}
{\sc G.~Takahashi, I.~Fatt, and T.~Goldstick}, {\em Oxygen consumption rate of
  tissue measured by a micropolarographic method}, The Journal of general
  physiology, 50 (1966), pp.~317--335.

\bibitem{T+2014}
{\sc L.~Tang, A.~L. van~de Ven, D.~Guo, V.~Andasari, V.~Cristini, K.~C. Li, and
  X.~Zhou}, {\em Computational modeling of 3d tumor growth and angiogenesis for
  chemotherapy evaluation}, PloS one, 9 (2014), p.~e83962.

\bibitem{T+2011}
{\sc R.~D. Travasso, E.~C. Poir{\'e}, M.~Castro, J.~C. Rodrguez-Manzaneque, and
  A.~Hern{\'a}ndez-Machado}, {\em Tumor angiogenesis and vascular patterning: a
  mathematical model}, PloS one, 6 (2011), p.~e19989.

\bibitem{williams1988dynamic}
{\sc S.~Williams, S.~Wasserman, D.~Rawlinson, R.~Kitney, L.~Smaje, and
  J.~Tooke}, {\em Dynamic measurement of human capillary blood pressure},
  Clinical science, 74 (1988), pp.~507--512.

\bibitem{X+2014}
{\sc Y.~Xiong, P.~Yang, R.~L. Proia, and T.~Hla}, {\em Erythrocyte-derived
  sphingosine 1-phosphate is essential for vascular development}, The Journal
  of clinical investigation, 124 (2014), pp.~4823--4828.

\end{thebibliography}
\end{document}